\documentclass[12pt,epsf]{article}
\usepackage{graphicx}
\usepackage{amsmath}
\usepackage{amssymb}
\usepackage{amsfonts}
\usepackage{setspace}
\usepackage{slashed}
\usepackage{geometry}
\DeclareMathAlphabet{\nrmcal}{OMS}{cmsy}{m}{n}
\usepackage{calrsfs}

\usepackage{xcolor}

\usepackage{cancel}

\newcommand{\modulo}[1]{\overset{\textrm{\tiny$\left(#1\right)$}}{=}}
\newcommand{\nodulo}[1]{\overset{\textrm{\tiny$\left(#1\right)$}}{\neq}}

\begin{document}
%%%%%%%%%%%%%%%%%%%%%%%%%%%%%%%%%%%%%%%%%%%

\def\a{\alpha}
\def\b{\beta}
\def\c{\varepsilon}
\def\d{\delta}
\def\e{\epsilon}
\def\f{\phi}
\def\g{\gamma}
\def\h{\theta}
\def\k{\kappa}
\def\l{\lambda}
\def\m{\mu}
\def\n{\nu}
\def\p{\psi}
\def\q{\partial}
\def\r{\rho}
\def\s{\sigma}
\def\t{\tau}
\def\u{\upsilon}
\def\v{\varphi}
\def\w{\omega}
\def\x{\xi}
\def\y{\eta}
\def\z{\zeta}
\def\D{\Delta}
\def\G{\Gamma}
\def\H{\Theta}
\def\L{\Lambda}
\def\F{\Phi}
\def\P{\Psi}
\def\S{\Sigma}

\def\o{\over}
\def\beq{\begin{eqnarray}}
\def\eeq{\end{eqnarray}}
\newcommand{\gsim}{ \mathop{}_{\textstyle \sim}^{\textstyle >} }
\newcommand{\lsim}{ \mathop{}_{\textstyle \sim}^{\textstyle <} }
\newcommand{\vev}[1]{ \left\langle {#1} \right\rangle }
\newcommand{\bra}[1]{ \langle {#1} | }
\newcommand{\ket}[1]{ | {#1} \rangle }
\newcommand{\EV}{ {\rm eV} }
\newcommand{\KEV}{ {\rm keV} }
\newcommand{\MEV}{ {\rm MeV} }
\newcommand{\GEV}{ {\rm GeV} }
\newcommand{\TEV}{ {\rm TeV} }
\def\diag{\mathop{\rm diag}\nolimits}
\def\Spin{\mathop{\rm Spin}}
\def\SO{\mathop{\rm SO}}
\def\O{\mathop{\rm O}}
\def\SU{\mathop{\rm SU}}
\def\U{\mathop{\rm U}}
\def\Sp{\mathop{\rm Sp}}
\def\SL{\mathop{\rm SL}}
\def\tr{\mathop{\rm tr}}
\def\mpl{M_{PL}}

\def\IJMP{Int.~J.~Mod.~Phys. }
\def\MPL{Mod.~Phys.~Lett. }
\def\NP{Nucl.~Phys. }
\def\PL{Phys.~Lett. }
\def\PR{Phys.~Rev. }
\def\PRL{Phys.~Rev.~Lett. }
\def\PTP{Prog.~Theor.~Phys. }
\def\ZP{Z.~Phys. }

%%%%%%%%%%%%%%%%%%%%%%%%%%%%%%%%%%%%%%%%%%%%%%%%%%%%%%%%%%%%%%%%%%%%%%%%%%%%%%%%%%%%%%%%%%%%%%%%%%%%

\baselineskip 0.7cm

\begin{titlepage}

\begin{flushright}
IPMU 13-0152\\
ICRR 657-2013-6
\end{flushright}

\vskip 1.35cm
\begin{center}
\begin{Large}\textbf{The Peccei-Quinn Symmetry\\
from a Gauged Discrete} \boldmath{$R$} \textbf{Symmetry}\end{Large}

\vskip 1.2cm
Keisuke Harigaya$^1$, Masahiro Ibe$^{2,1}$, Kai Schmitz$^{1}$ and Tsutomu T. Yanagida$^1$
\vskip 0.4cm
$^1${\it Kavli IPMU (WPI), University of Tokyo, Kashiwa 277-8583, Japan}\\
$^2${\it ICRR, University of Tokyo, Kashiwa 277-8582, Japan}
\vskip 1.5cm

\end{center}

\begin{abstract}
\onehalfspacing
The axion solution to the strong $CP$ problem
calls for an explanation as to why the Lagrangian should be invariant
under the global Peccei-Quinn symmetry, $U(1)_{\textrm{PQ}}$,
to such a high degree of accuracy.
In this paper, we point out that the $U(1)_{\textrm{PQ}}$ can indeed survive as an
accidental symmetry in the low-energy effective theory, if the standard model gauge
group is supplemented by a gauged and discrete $R$ symmetry, $Z_{N}^R$, forbidding all
dangerous operators that explicitly break the Peccei-Quinn symmetry.
In contrast to similar approaches, the requirement that the $Z_{N}^R$ symmetry
be anomaly-free
\textit{forces} us, in general, to extend the supersymmetric standard model
by new matter multiplets.
Surprisingly, we find a large landscape of viable scenarios that all
individually fulfill the current experimental constraints on the QCD vacuum
angle as well as on the axion decay constant.
In particular, choosing the number of additional multiplets appropriately, the order
$N$ of the $Z_N^R$ symmetry can take any integer value larger than $2$.
This has interesting consequences with respect to possible solutions of the
$\mu$ problem, collider searches for vector-like
quarks and axion dark matter.
\end{abstract}

\end{titlepage}

\setcounter{page}{2}

%%%%%%%%%%%%%%%%%%%%%%%%%%%%%%%%%%%%%%%%%%%%%%%%%%%%%%%%%%%%%%%%%%%%%%%%%%%%%%%%%%%%%%%%%%%%%%%%%%%%

\tableofcontents

%%%%%%%%%%%%%%%%%%%%%%%%%%%%%%%%%%%%%%%%%%%%%%%%%%%%%%%%%%%%%%%%%%%%%%%%%%%%%%%%%%%%%%%%%%%%%%%%%%%%

\section{Introduction}

%%%%%%%%%%%%%%%%%%%%%%%%%%%%%%%%%%%%%%%%%%%%%%%%%%%%%%%%%%%%%%%%%%%%%%%%%%%%%%%%%%%%%%%%%%%%%%%%%%%%

The Peccei-Quinn (PQ) symmetry, $U(1)_{\textrm{PQ}}$, provides us with a very attractive
mechanism to solve the strong $CP$ problem in quantum chromodynamics
(QCD)~\cite{Peccei:1977hh,Peccei:1977ur}.
Up to now, a convincing explanation for the origin of the PQ symmetry is, however,
still pending, since it is a global symmetry and any global symmetry is
believed to be broken by quantum gravity
effects~\cite{Kamionkowski:1992mf,Barr:1992qq,Holman:1992us}.
In order for the PQ symmetry to accidentally survive in the low-energy effective
theory, one thus has to arrange for a sufficient suppression of all unwanted
operators that explicitly break it.
The tight experimental upper bound on the
QCD vacuum angle, $\bar{\theta} \lesssim 10^{-10}$~\cite{Baker:2006ts},
necessitates in particular that this suppression be extremely efficient.
One natural way to protect the PQ symmetry is to invoke some gauge symmetry
that accidentally forbids all the operators that would break it too severely.
In this paper, we point out that, in the context of the supersymmetric
standard model, the role of this protective gauge symmetry could be played by
a gauged discrete $R$ symmetry, $Z_{N}^R$.

%%%%%%%%%%%%%%%%%%%%%%%%%%%%%%%%%%%%%%%%%%%%%%%%%%%%%%%%%%%%%%%%%%%%%%%%%%%%%%%%%%%%%%%%%%%%%%%%%%%%

Given only the particle content of the minimal supersymmetric standard model (MSSM),
any $Z_{N}^R$ symmetry, except for $Z_{3}^R$ and $Z_{6}^R$, is anomalously broken by
$SU(3)_C$ and $SU(2)_L$ instanton effects~\cite{Lee:2011dya,Evans:2011mf}.%
\footnote{We restrict ourselves to generation-independent $Z_N^R$ symmetries, where $N > 2$,
that commute with $SU(5)$ and do not consider anomaly cancellation via the Green-Schwarz
mechanism~\cite{Green:1984sg} coming from string theory, cf.\ Sec.~\ref{subsec:MSSM}.}
On the supposition that a different $Z_{N}^R$ symmetry, other than $Z_{3}^R$ or $Z_{6}^R$, might
account for the protection of the $U(1)_{\textrm{PQ}}$, we are hence \textit{naturally} led to
introduce an extra matter sector canceling the MSSM contributions to the $Z_{N}^R$ anomalies.
For a particular value of $N$ as well as  $k$ additional pairs of vector-quark
superfields charged under the MSSM gauge group, the requirement that the shift in the QCD vacuum
angle induced by PQ-breaking operators be less than $10^{-10}$ then implies
an upper bound on the axion decay constant $f_a$.
By identifying those extensions of the MSSM that yield an upper bound on $f_a$
above the astrophysical lower bound of $f_a \gtrsim 10^9\,\textrm{GeV}$~\cite{Raffelt:2006cw},
we are thus able to single out the values of $N$ and $k$ that are phenomenologically viable.
Surprisingly, for each integer value of $N$ larger than $2$, a variety
of $k$ values is admissible.
Here, $k$ can in particular always be chosen such that the unification of the
gauge coupling constants still occurs at the perturbative level.
Moreover, for $k= 5,6$ and $k\geq 8$, it is possible to protect the PQ symmetry
by means of a $Z_4^R$ symmetry.
As we will discuss, this is an especially interesting case, since a $Z_4^R$ may not only
explain the origin of the PQ symmetry, but at the same time
also allow for a simple solution of the MSSM $\mu$ problem.

%%%%%%%%%%%%%%%%%%%%%%%%%%%%%%%%%%%%%%%%%%%%%%%%%%%%%%%%%%%%%%%%%%%%%%%%%%%%%%%%%%%%%%%%%%%%%%%%%%%%

The very idea to protect the PQ symmetry against gravity effects by means of
a gauge symmetry is, of course, not new.
Many authors have, for instance, considered extensions of the standard model
gauge group $G_{\textrm{SM}} = SU(3)_C \times SU(2)_L \times U(1)_Y$
by some \textit{continuous} symmetry.
Early examples of such attempts include models based on the gauge group
$G_{\textrm{SM}} \times U(1)'$~\cite{Barr:1992qq} or on the group
$E_6 \times U(1)'$~\cite{Holman:1992us}.
Also extensions of the gauge group by a continuous \textit{and}
a discrete symmetry, such as
$G_{\textrm{SM}} \times SU(4) \times Z_N$~\cite{Dine:1992vx},
$SU(3)_C \times SU(3)_L \times U(1)_Y \times Z_{13} \times Z_2$~\cite{Dias:2002gg} or
$SU(5) \times SU(N) \times Z_N$~\cite{Carpenter:2009zs},
have been studied in the literature.
Likewise, next to these field-theoretic models, string constructions have been shown
to give rise to accidental PQ symmetries.
By compactifying the heterotic string on Calabi-Yau manifolds~\cite{Lazarides:1985bj}
or on $Z_6$-II orbifolds~\cite{Choi:2009jt}, it is, for example, feasible
to retain accidental global symmetries
in the low-energy effective theory as remnants of exact \textit{stringy} discrete symmetries.
All of these approaches, however, rely on rather speculative assumptions about the
UV completion of the standard model (SM).
In particular, they require in many cases an \textit{ad hoc}
extension of the particle content of the
standard model that is motivated  by the intention to eventually end up with
a global PQ symmetry in the first place.
In view of this situation, it is thus of great interest to assess what a \textit{minimal}
extension of the standard model or the MSSM would look like that still accomplishes a
successful protection of the PQ symmetry.
The model presented in Ref.~\cite{Dias:2002hz} might, for instance, be considered
a step into this direction.
It forgoes any additional continuous symmetry, but only
extends $G_{\textrm{SM}}$ by a discrete $Z_{13}\times Z_3$.
Still, it comes with a multi-Higgs sector that, while being certainly interesting from
a phenomenological point of view, lacks a decisive reason for its origin from
a fundamental perspective.

%%%%%%%%%%%%%%%%%%%%%%%%%%%%%%%%%%%%%%%%%%%%%%%%%%%%%%%%%%%%%%%%%%%%%%%%%%%%%%%%%%%%%%%%%%%%%%%%%%%%

Now, invoking nothing but a discrete $Z_{N}^R$ symmetry in order to protect the
PQ symmetry rests, by contrast, on a very sound conceptional footing.
A discrete $R$ symmetry is an often important and sometimes even imperative
ingredient to model building and phenomenology in supersymmetry (SUSY).
It allows for a solution to the $\mu$ problem~\cite{Giudice:1988yz,Yanagida:1997yf,Dine:2009swa},
prevents too rapid proton decay~\cite{Dimopoulos:1981zb,Sakai:1981pk} and forbids a constant
term in the superpotential of order the Planck scale which, in scenarios of low-scale
SUSY breaking, would otherwise result in a huge negative
cosmological constant~\cite{Izawa:1997he}.
The existence of an $R$ symmetry and its potential spontaneous or explicit
breaking is furthermore closely linked to the spontaneous breaking of
SUSY, irrespectively of whether our present non-supersymmetric
vacuum corresponds to a true~\cite{Nelson:1993nf} or merely metastable
ground state~\cite{Intriligator:2006dd}.
Finally, it is interesting to observe that higher-dimensional supergravity
theories such as superstring theory always feature an $R$ symmetry, which might
be naturally broken down to its discrete subgroup $Z_N^R$ upon the compactification of the
extra dimensions~\cite{Imamura:2001es}.
This last point may again be regarded to be rather speculative, but
it does not alter the fact that discrete $R$ symmetries surely play an preeminent role
among all conceivable symmetries by which $G_{\textrm{SM}}$ could possibly be extended.
In this sense, the main result of this paper is that nothing but
the arguably simplest and most natural extra gauge symmetry,
namely a gauged and discrete $R$ symmetry $Z_N^R$, could be responsible for shielding
the PQ symmetry from the dangerous effects of gravity.

%%%%%%%%%%%%%%%%%%%%%%%%%%%%%%%%%%%%%%%%%%%%%%%%%%%%%%%%%%%%%%%%%%%%%%%%%%%%%%%%%%%%%%%%%%%%%%%%%%%%

After having outlined why we are particularly interested in enlarging
$G_{\textrm{SM}}$ by a gauged $Z_N^R$ symmetry, we shall present in the next section our
minimal extension of the MSSM and explain (i) how the colour and weak anomalies of the discrete
$Z_N^R$ symmetry force
us to introduce new matter multiplets, (ii) how these new matter multiplets
acquire masses as well as (iii) how a $\mu$ term of order of the soft masses
can be generated dynamically.
In Secs.~\ref{sec:constraints},
we will then study the phenomenological constraints on our model and identify
the viable combinations of $N$ and $k$ along with upper and lower bounds
on the axion decay constant $f_a$.
Finally, we conclude with a summary of our model and a short overview
of its phenomenological implications.
Two appendices deal with the $R$ charges of the MSSM fields and a slight
modification of our model that manages to avoid the axion domain wall problem, respectively.

%%%%%%%%%%%%%%%%%%%%%%%%%%%%%%%%%%%%%%%%%%%%%%%%%%%%%%%%%%%%%%%%%%%%%%%%%%%%%%%%%%%%%%%%%%%%%%%%%%%%

\section{Minimal extension of the MSSM with a PQ symmetry}
\label{sec:model}

%%%%%%%%%%%%%%%%%%%%%%%%%%%%%%%%%%%%%%%%%%%%%%%%%%%%%%%%%%%%%%%%%%%%%%%%%%%%%%%%%%%%%%%%%%%%%%%%%%%%

We shall now demonstrate how an anomaly-free discrete $R$ symmetry $Z_N^R$ in combination
with an extra matter sector automatically gives rise to a global PQ symmetry.
As a preparation, let us first summarize our conventions and assumptions regarding
the MSSM sector.

%%%%%%%%%%%%%%%%%%%%%%%%%%%%%%%%%%%%%%%%%%%%%%%%%%%%%%%%%%%%%%%%%%%%%%%%%%%%%%%%%%%%%%%%%%%%%%%%%%%%

\subsection{Supersymmetric Standard Model Sector}

%%%%%%%%%%%%%%%%%%%%%%%%%%%%%%%%%%%%%%%%%%%%%%%%%%%%%%%%%%%%%%%%%%%%%%%%%%%%%%%%%%%%%%%%%%%%%%%%%%%%

We take the renormalizable MSSM superpotential to be of the following form,
\begin{align}
W_{\textrm{MSSM}} = h_{ij}^u \, \mathbf{10}_i \mathbf{10}_j H_u +
h_{ij}^d \, \mathbf{5}_i^* \mathbf{10}_j H_d +
h_{ij}^\nu \, \mathbf{5}_i^* \mathbf{1}_j H_u +
\frac{1}{2} M_i \mathbf{1}_i \mathbf{1}_i \,,
\label{eq:WMSSM}
\end{align}
where we have arranged the MSSM chiral quark and lepton superfields into $SU(5)$ multiplets,
$\mathbf{10} = \left(q,u^c,e^c\right)$ and $\mathbf{5}^* = (d^c, \ell)$.
Throughout this paper, we shall assume that the tiny masses of
the SM neutrinos are accounted for by the seesaw mechanism~\cite{Minkowski:1977sc}.
That is why we have also introduced neutrino singlet fields, $\mathbf{1} = (n^c)$,
in Eq.~\eqref{eq:WMSSM}, next to the actual matter content of the MSSM.%
\footnote{As this sometimes falls victim to bad jargon, we emphasize that
the fermions contained in $n^c$ are \textit{left-handed}. In fact, they are
the hermitian conjugates of the right-handed neutrinos required for the seesaw mechanism.}
Moreover, $H_u$ and $H_d$ is the usual pair of MSSM Higgs doublets,
$h^u$, $h^d$ and $h^\nu$ are Yukawa matrices and $M$ denotes the diagonalized
Majorana mass matrix for the heavy neutrinos involved in the seesaw mechanism.
$i$ and $j$ finally label the three different generations of quarks and leptons,
\textit{i.e.}\ $i,j = 1,2,3$.

%%%%%%%%%%%%%%%%%%%%%%%%%%%%%%%%%%%%%%%%%%%%%%%%%%%%%%%%%%%%%%%%%%%%%%%%%%%%%%%%%%%%%%%%%%%%%%%%%%%%

We assume the MSSM quark and lepton fields to be unified in $SU(5)$ representations
in order to allow for an embedding of the MSSM into a grand unified theory (GUT).
Note, however, that taking $SU(5)$ alone to be the full GUT gauge group is problematic.
The minimal supersymmetric $SU(5)$ GUT model~\cite{Dimopoulos:1981zb,Sakai:1981gr}
namely fails to give GUT-scale masses to the coloured Higgs triplets that are expected
to pair up with the MSSM Higgs doublets in complete $SU(5)$ multiplets.
This results in too rapid proton decay and represents what is known as the infamous
doublet-triplet splitting problem~\cite{Witten:1981nf}.
In addition to that, the standard way to break $SU(5)$ to $G_{\textrm{SM}}$ by means
of a $\mathbf{24}$-plet is not compatible with the assumption of an unbroken
$R$ symmetry below the GUT scale.\footnote{If we managed to break $SU(5)$ without breaking the
$R$ symmetry, we would be left with potentially interesting or dangerous
$G_{\textrm{SM}}$-charged exotics whose masses would only
receive soft SUSY-breaking contributions~\cite{Fallbacher:2011xg}.}
Because of that, we shall assume that $SU(5)$ is merely a proper subgroup of the
full GUT group, $SU(5) \subset G_{\textrm{GUT}}$.
An attractive possibility in this context is unification based on the product
group $SU(5) \times U(3)_H$, which can be formulated in an $R$-invariant
fashion~\cite{Izawa:1997he,Kurosawa:2001iq},
while solving the doublet-triplet splitting problem in a natural way~\cite{Yanagida:1994vq}.

%%%%%%%%%%%%%%%%%%%%%%%%%%%%%%%%%%%%%%%%%%%%%%%%%%%%%%%%%%%%%%%%%%%%%%%%%%%%%%%%%%%%%%%%%%%%%%%%%%%%

Finally, we point out that we define the MSSM to conserve matter parity, $P_M$,
so as to forbid all dangerous baryon and lepton number-violating operators in
the renormalizable superpotential.
This renders the actual gauge group of the MSSM slightly larger than the
one of the standard model, $G_{\textrm{MSSM}} = G_{\textrm{SM}} \times P_M$.
One possibility to account for the origin of matter parity is to interpret it
as the remnant discrete subgroup of a local $U(1)_{B-L}$ symmetry that is spontaneously
broken above the electroweak scale~\cite{Martin:1992mq}.
Here, $B$$-$$L$ stands for the difference between baryon number $B$ and lepton $L$.
Assuming the presence of an additional Abelian factor $U(1)_X$ in the GUT gauge group
orthogonal to $SU(5)$, it can be expressed in terms of
the Abelian GUT charge $X$ and the weak hypercharge $Y$ through the relation
$X +4Y = 5\,(B$$-$$L)$, cf.\ also Appendix~\ref{app:MSSMRcharges}.

%%%%%%%%%%%%%%%%%%%%%%%%%%%%%%%%%%%%%%%%%%%%%%%%%%%%%%%%%%%%%%%%%%%%%%%%%%%%%%%%%%%%%%%%%%%%%%%%%%%%

\subsection[Extra matter sector required by a non-anomalous $Z_N^R$ symmetry]
{Extra matter sector required by a non-anomalous \boldmath{$Z_N^R$} symmetry}
\label{subsec:MSSM}

%%%%%%%%%%%%%%%%%%%%%%%%%%%%%%%%%%%%%%%%%%%%%%%%%%%%%%%%%%%%%%%%%%%%%%%%%%%%%%%%%%%%%%%%%%%%%%%%%%%%

As outlined in the introduction, a discrete $R$ symmetry
$Z_N^R$ represents a unique choice when considering possible extensions
of the MSSM gauge group.
We now perform just such an extension, such that
the full gauge group $G$ of our model also features a $Z_N^R$ factor.
A priori, we allow $N$, the order of the $Z_N^R$ symmetry, to
take any integer value larger than $2$.
We disregard the case $N=2$ since a $Z_2^R$ symmetry, \textit{i.e.}\ $R$ parity,
is not an $R$ symmetry in the actual sense.
By including a Lorentz rotation, it can always be reformulated as an
ordinary $Z_2$ parity~\cite{Dine:2009swa}.
On top of that, given only a $Z_2^R$ symmetry, we would also be unable
to forbid a constant term in the superpotential, which would result
in a  cosmological constant of order the Planck scale.
On the other hand, we point out that, in the case of even $N$,
the $Z_N^R$ symmetry contains $R$ parity as a subgroup, $Z_N^R \supset Z_2^R$
for $N = 4,6,8,..$.
Depending on the details of the $R$ charge assignments to the particles
of our model, this $R$ parity coincides in some cases with the ordinary
matter parity $P_M$.
In these cases, we then do not need to additionally impose matter parity by hand,
as it is already included in the $Z_N^R$ factor of the gauge group.
In all other cases, we rely on the assumption that a spontaneously broken
$U(1)_{B-L}$ gauge symmetry gives rise to matter parity at low energies.
In summary, the gauge group of our model is, hence, given by
\begin{align}
G = SU(3)_C \times SU(2)_L \times U(1)_Y \times
\begin{cases}
Z_N^R \times P_M & ;\:\: Z_N^R\not\supset P_M \\
Z_N^R & ; \:\: Z_N^R \supset P_M
\end{cases} \,.
\end{align}

%%%%%%%%%%%%%%%%%%%%%%%%%%%%%%%%%%%%%%%%%%%%%%%%%%%%%%%%%%%%%%%%%%%%%%%%%%%%%%%%%%%%%%%%%%%%%%%%%%%%

\subsubsection[Gauge anomalies of the $Z_N^R$ symmetry]
{Gauge anomalies of the \boldmath{$Z_N^R$} symmetry}

%%%%%%%%%%%%%%%%%%%%%%%%%%%%%%%%%%%%%%%%%%%%%%%%%%%%%%%%%%%%%%%%%%%%%%%%%%%%%%%%%%%%%%%%%%%%%%%%%%%%

We attribute the origin of the $Z_N^R$ factor in the gauge group
to the presence of a continuous gauged $R$ symmetry at high energies,
after the breaking of which $Z_N^R$ remains as a discrete subgroup.
Thus being part of the gauge group, it is crucial that the 
$Z_N^R$ symmetry be anomaly-free.
The relevant anomaly cancellation conditions are those related
to the colour as well as to the weak anomaly of the $Z_N^R$, \textit{i.e.}\
the $Z_{N}^R\left[SU(3)_C\right]^2$ and the $Z_{N}^R\left[SU(2)_L\right]^2$
anomaly, respectively.
The anomaly coefficients for these
two anomalies, $\nrmcal{A}_R^{(C)}$ and $\nrmcal{A}_R^{(L)}$,
are given by~\cite{Kurosawa:2001iq,Ibanez:1992ji}
\begin{align}
\nrmcal{A}_R^{(C)} = 6 + N_g \, (3 \,r_{\mathbf{10}} + r_{\mathbf{5}^*} - 4) \,,\quad
\nrmcal{A}_R^{(L)} = 4 + N_g \, (3 \,r_{\mathbf{10}} + r_{\mathbf{5}^*} - 4) +
(r_{H_u} + r_{H_d} - 2) \,.
\label{eq:ARCARL}
\end{align}
Here, $r_{\mathbf{10}}$, $r_{\mathbf{5}^*}$, $r_{\mathbf{1}}$, $r_{H_u}$ and $r_{H_d}$
denote the $R$ charges of the MSSM matter multiplets and Higgs doublets and
$N_g = 3$ is the number of fermion generations in the MSSM.%
\footnote{The authors of Ref.~\cite{Evans:2011mf} have recently made the interesting observation
that $N_g \geq 3$ is a \textit{necessary condition} for consistently extending
the MSSM gauge group by an anomaly-free discrete $R$ symmetry $Z_N^R$ with $N > 2$.}
Note that we have assumed the $R$ charges of the matter fields to be generation-independent.
Otherwise, \textit{i.e.}\ in the case of generation-dependent $R$ charges, the $R$ symmetry
would suppress some of the entries in the Yukawa matrices $h^u$ and $h^d$ too heavily.
We also remark that the $R$ charges are normalized such that
the anti-commuting superspace coordinate $\theta$ carries $R$ charge $r_\theta = 1$.
By choosing a different value for the $R$ charge of $\theta$, say, $r_\theta' \neq 1$,
we always have the option to collectively rescale all $R$ charges by the common
factor $r_\theta'/r_\theta$.

%%%%%%%%%%%%%%%%%%%%%%%%%%%%%%%%%%%%%%%%%%%%%%%%%%%%%%%%%%%%%%%%%%%%%%%%%%%%%%%%%%%%%%%%%%%%%%%%%%%%
 
Besides the colour and the weak anomaly, all
further anomalies involving at least one $Z_N^R$ factor also have to vanish
in order to render the $Z_N^R$ symmetry fully anomaly-free.
The anomalies non-linear in $Z_N^R$,
such as $\left[Z_N^R\right]^3$ or $\left[Z_N^R\right]^2 U(1)_Y$,
are, however, sensitive to heavy, fractionally
charged states at high energies~\cite{Banks:1991xj}.
Similarly, the gravitational anomaly, $Z_N^R \left[\textrm{gravity}\right]^2$,
also receives contributions from light sterile fermions
as well as from hidden-sector fermions
acquiring large masses of order the SUSY-breaking scale
in the course of spontaneous SUSY breaking~\cite{Lee:2011dya}.
All of these anomalies hence highly depend on the particle spectrum in the UV
and, thus, do not allow us to derive further constraints on our model.
In general, the $Z_N^R \left[U(1)_Y\right]^2$ anomaly also
does not yield a useful condition because the SM hypercharge
is not quantized~\cite{Banks:1991xj,Ibanez:1991hv}.
Only if the GUT group is semi-simple, such that the normalization of the hypercharge
is dictated by the gauge structure, the $Z_N^R \left[U(1)_Y\right]^2$ anomaly
provides a meaningful constraint on the $Z_N^R$ symmetry as well as on the set of particles
charged under it.%
\footnote{We mention in passing that neither
of the previously discussed GUT gauge groups, \textit{i.e.}\
neither $SU(3) \times U(3)_H$ nor $SU(5) \times U(1)_X$, is semi-simple.
Assuming one of these two groups to correspond to the GUT
gauge group, we are hence not able
to make use of the anomaly cancellation condition
for the $Z_N^R \left[U(1)_Y\right]^2$ anomaly.}

%%%%%%%%%%%%%%%%%%%%%%%%%%%%%%%%%%%%%%%%%%%%%%%%%%%%%%%%%%%%%%%%%%%%%%%%%%%%%%%%%%%%%%%%%%%%%%%%%%%%

Obviously, we only included the contributions from the MSSM sector to
the anomaly coefficients in Eq.~\eqref{eq:ARCARL}.
$\nrmcal{A}_R^{(C)}$ and $\nrmcal{A}_R^{(L)}$ could, however, still receive
corrections $\Delta\nrmcal{A}_R^{(C)}$ and $\Delta\nrmcal{A}_R^{(L)}$
due to new coloured or weakly interacting fermions with masses
at or above the electroweak scale.
This extra matter would need to be assembled in complete $SU(5)$ multiplets
in order not to spoil the unification of the gauge coupling constants.
Consequently, extra fermions ought to equally contribute to
$\nrmcal{A}_R^{(C)}$ and $\nrmcal{A}_R^{(L)}$, such that
the corresponding corrections are equal to each other,
$\Delta\nrmcal{A}_R = \Delta\nrmcal{A}_R^{(C)} = \Delta\nrmcal{A}_R^{(L)}$,
and such that the difference between $\nrmcal{A}_R^{(C)}$ and $\nrmcal{A}_R^{(L)}$
ends up being independent of the properties of the extra matter sector.
A minimal necessary condition for rendering the $Z_N^R$ symmetry anomaly-free
is hence that
\begin{align}
\nrmcal{A}_R^{(L)} - \nrmcal{A}_R^{(C)} = r_{H_u} + r_{H_d} - 4 \modulo{N} 0 \,,
\label{eq:anomaly-free}
\end{align}
where we have introduced the symbol $\modulo{N}$ as a shorthand notation
to denote equality modulo $N$,
\begin{align}
a \modulo{N} b \quad\Leftrightarrow\quad a \textrm{ mod } N = b \textrm{ mod } N
\quad\Leftrightarrow\quad \exists! \: \ell \in \mathbb{Z}: \: a = b + \ell\,N \,.
\label{eq:modulo}
\end{align}
The condition in Eq.~\eqref{eq:anomaly-free}
is equivalent to $r_{H_u} + r_{H_d} \modulo{N} 4$.
We therefore see that an anomaly-free $Z_N^R$ symmetry automatically
suppresses the $\mu$ term for the MSSM Higgs doublets.

%%%%%%%%%%%%%%%%%%%%%%%%%%%%%%%%%%%%%%%%%%%%%%%%%%%%%%%%%%%%%%%%%%%%%%%%%%%%%%%%%%%%%%%%%%%%%%%%%%%%

\subsubsection[Constraints on the $R$ charges of the MSSM fields]
{Constraints on the \boldmath{$R$} charges of the MSSM fields}

%%%%%%%%%%%%%%%%%%%%%%%%%%%%%%%%%%%%%%%%%%%%%%%%%%%%%%%%%%%%%%%%%%%%%%%%%%%%%%%%%%%%%%%%%%%%%%%%%%%%

Next to Eq.~\eqref{eq:anomaly-free}, the requirement that the first two
terms in the superpotential $W_{\textrm{MSSM}}$, cf.\ Eq.~\eqref{eq:WMSSM}, be in accordance
with the $Z_N^R$ symmetry provides us with two further constraints
on the $R$ charges of the MSSM fields,
\begin{align}
2 r_{\mathbf{10}} + r_{H_u} \modulo{N} 2 \,, \quad 
r_{\mathbf{5}^*} + r_{\mathbf{10}} + r_{H_d} \modulo{N} 2 \,.
\label{eq:yukawacond}
\end{align}
The combination of all three conditions then implies
$3r_{\mathbf{10}} + r_{\mathbf{5}}^* \modulo{N} 0$,
which automatically forbids the dangerous dimension-5 operator
$\mathbf{10}\,\mathbf{10}\,\mathbf{10}\,\mathbf{5}^*$ in the
superpotential, which would otherwise induce too rapid proton decay~\cite{Sakai:1981pk}.
Together with matter parity, the anomaly-free $Z_N^R$ symmetry
thus bans all baryon and lepton number-violating operators
up to dimension 5 except for the operator
$\mathbf{5}^* H_u\, \mathbf{5}^*H_u$, which we, of course, want to
retain to be able to explain the small neutrino masses~\cite{Weinberg:1979sa}.
Finally, in the seesaw extension of the MSSM, we also have to ensure
that the last two terms in $W_{\textrm{MSSM}}$ respect the $Z_N^R$ symmetry,
which translates into
\begin{align}
r_{\mathbf{5}*} + r_{\mathbf{1}} + r_{H_u} \modulo{N} 2 \,,\quad
2  r_{\mathbf{1}} \modulo{N} 2 \,.
\label{eq:neutrinocond}
\end{align}
Here, as for the second condition, we have assumed zero $R$ charge for
the Majorana neutrino mass $M$, which is to say that we consider
its origin to be independent of the mechanism responsible for the
spontaneous breaking of $R$ symmetry.

%%%%%%%%%%%%%%%%%%%%%%%%%%%%%%%%%%%%%%%%%%%%%%%%%%%%%%%%%%%%%%%%%%%%%%%%%%%%%%%%%%%%%%%%%%%%%%%%%%%%

In conclusion, we find that extending the particle content of the MSSM by three neutrino
singlets, the five $R$ charges $r_{\mathbf{10}}$, $r_{\mathbf{5}^*}$, $r_{\mathbf{1}}$,
$r_{H_u}$ and $r_{H_d}$ are determined by the five conditions
in Eqs.~\eqref{eq:anomaly-free}, \eqref{eq:yukawacond} and \eqref{eq:neutrinocond}.
However, due to the fact that all of these conditions only constrain the MSSM
$R$ charges up to integer multiples of $N$, they do not suffice to
fix the values of $r_{\mathbf{10}}$, $r_{\mathbf{5}^*}$, $r_{\mathbf{1}}$,
$r_{H_u}$ and $r_{H_d}$ uniquely.
Instead, for each value of $N$, there exist exactly ten
different possibilities to assign $R$ charges to the MSSM fields.
In Appendix~\ref{app:MSSMRcharges}, we derive and discuss
these solutions in more detail.
In particular, we show that, for any given value of $N$,
the different $R$ charge assignments are related to each other
by gauge transformations.
First of all, in consequence of the $SU(5)$ invariance of the MSSM Lagrangian,
the ten solutions split into two equivalence classes
of respectively five solutions.
As shown in Appendix~\ref{app:MSSMRcharges}, these two classes are generated
by the action of $Z_5$ transformations on the following two $R$ charge assignments,
\begin{align}
r_{\mathbf{10}} \modulo{N} \frac{1}{5} + \ell \,\frac{N}{2} \,,\quad
r_{\mathbf{5}^*} \modulo{N} -\frac{3}{5} + \ell \,\frac{N}{2}\,, \quad
r_{\mathbf{1}} \modulo{N} 1 + \ell \,\frac{N}{2} \,, \quad
r_{H_u} \modulo{N} 2 - \frac{2}{5} \,, \quad r_{H_d} \modulo{N} 2 + \frac{2}{5} \,,
\label{eq:MSSMRcharges}
\end{align}
where $\ell=0,1$ and where $Z_5 \subset SU(5)$ is the center of $SU(5)$.
Furthermore, if matter parity stems from a $U(1)_X$ symmetry that is part of the gauge
group at high energies, \textit{i.e.}\ if $P_M \subset U(1)_X$, these two
solutions are in turn related to each other by a $P_M$ transformation, such that
eventually all ten $R$ charge assignments end up being physically equivalent.
On the other hand, if matter parity is a subgroup of the $Z_N^R$ symmetry,
\textit{i.e.}\ if $P_M \subset Z_N^R$, the two solutions in Eq.~\eqref{eq:MSSMRcharges}
cannot be related to each other and we are left with two inequivalent classes
of solutions.
Last but not least, we remark that, for all values of $\ell$ and $N$, all of the $R$
charges in Eq.~\eqref{eq:MSSMRcharges}, expect for $r_{\mathbf{1}}$ in some cases,
are fractional.
In Appendix~\ref{app:MSSMRcharges}, we however show that, for each
$N \neq 5,10,15,..$, there exists at least one $R$ charge assignment
that is equivalent to one of the two assignments in Eq.~\eqref{eq:MSSMRcharges}
and which only involves integer-valued $R$ charges. 
But not only that, we also demonstrate that, in a $U(1)_X$-invariant extension
of our model, all $R$ charges in Eq.~\eqref{eq:MSSMRcharges} can always be
rendered integer-valued by means of a $U(1)_X$ transformation.

%%%%%%%%%%%%%%%%%%%%%%%%%%%%%%%%%%%%%%%%%%%%%%%%%%%%%%%%%%%%%%%%%%%%%%%%%%%%%%%%%%%%%%%%%%%%%%%%%%%%

\subsubsection{Anomaly cancellation owing to new matter fields}
\label{subsubsec:anomaly}

%%%%%%%%%%%%%%%%%%%%%%%%%%%%%%%%%%%%%%%%%%%%%%%%%%%%%%%%%%%%%%%%%%%%%%%%%%%%%%%%%%%%%%%%%%%%%%%%%%%%

Irrespectively of the concrete $R$ charges in Eq.~\eqref{eq:MSSMRcharges},
the anomaly constraint in Eq.~\eqref{eq:anomaly-free} in combination with
the two conditions in Eq.~\eqref{eq:yukawacond} immediately implies for
the anomaly coefficients in Eq.~\eqref{eq:ARCARL}
\begin{align}
\nrmcal{A}_R^{(C)} \modulo{N} \nrmcal{A}_R^{(L)} \modulo{N} 6  - 4 N_g  \modulo{N} -6 \,.
\label{eq:nonzeroA}
\end{align}
As this result does not rely on any of the two conditions in Eq.~\eqref{eq:neutrinocond},
it is independent of the fact that we extended the MSSM particle content by
three right-handed neutrinos.
It rather equally applies in the MSSM as well as in its seesaw extension.
But more importantly, it leads us to one of the key observations of this paper:
\textit{as long as the order $N$ of the
$Z_N^R$ symmetry is different from $N = 3$ or $N = 6$, we are
forced to introduce a new matter sector in order to cancel the MSSM contributions
to the colour and the weak $Z_N^R$ anomaly.
In this sense, the introduction of new coloured and weakly interacting
states in our model is not an ad hoc measure,
but rather a natural consequence of the requirement
of an anomaly-free discrete $R$ symmetry.}%
\footnote{Again, this statement can be defined down by allowing for anomaly
cancellation via the Green-Schwarz mechanism, in the case of which not only
$Z_3^R$ and $Z_6^R$ can be rendered anomaly-free solely within the MSSM, but also
$Z_4^R$, $Z_8^R$, $Z_{12}^R$ and $Z_{24}^R$~\cite{Lee:2011dya,Lee:2010gv}.
Moreover, it is worth noting that, in the context of a two-singlet
extension of the MSSM, the anomaly-free $Z_{24}^R$ symmetry can be
used to successfully protect the PQ symmetry~\cite{Lee:2011dya}.\smallskip}

%%%%%%%%%%%%%%%%%%%%%%%%%%%%%%%%%%%%%%%%%%%%%%%%%%%%%%%%%%%%%%%%%%%%%%%%%%%%%%%%%%%%%%%%%%%%%%%%%%%%

The simplest way to cancel the MSSM anomalies in Eq.~\eqref{eq:nonzeroA}
without spoiling the unification of the gauge coupling constants is
to introduce $k$ pairs of vector-quarks and anti-quarks,
$Q_i$ and $\bar{Q_i}$, where $i=1,..,k$, that respectively
transform in the $\mathbf{5}$ and $\mathbf{5}^*$ of $SU(5)$.%
\footnote{Transforming in the $\mathbf{5}$ and $\mathbf{5}^*$ of $SU(5)$, the new multiplets
$Q_i$ and $\bar{Q_i}$, of course, also contain lepton doublets.
From a phenomenological point of view and with regard to the PQ solution of the $CP$ problem,
these are however less interesting as compared to the corresponding quark triplets.
Because of that, we will
refer to $Q_i$ and $\bar{Q}_i$ as the new \textit{quark} and \textit{anti-quark}
superfields in the following.}
As they transform in complete $SU(5)$ multiplets, the extra quarks and
anti-quarks yield equal non-MSSM contributions $\Delta\nrmcal{A}_R^{(C)}$
and $\Delta\nrmcal{A}_R^{(L)}$ to the colour and weak anomaly coefficients
of the $Z_N^R$ asymmetry.
According to Eq.~\eqref{eq:nonzeroA}, we must require that
\begin{align}
\Delta\nrmcal{A}_R^{(C)}= \Delta\nrmcal{A}_R^{(L)} =
k \, (r_Q + r_{\bar{Q}} - 2 ) \modulo{N} 6 \,, \quad
r_{Q\bar{Q}} = r_Q + r_{\bar{Q}} \modulo{N} 2 + \frac{1}{k} \left(6 + \ell_Q N\right)
\,, \quad \ell_Q \in \mathbb{Z} \,,
\label{eq:QRcharges}
\end{align}
where $r_Q$ and $r_{\bar{Q}}$ denote the generation-independent $R$ charges
of the extra quarks and anti-quarks, respectively, and $r_{Q\bar{Q}}$ is
the common $R$ charge of the bilinear quark operators
$\left(Q\bar{Q}\right)_i = Q_i\bar{Q}_i$.
Just like all other $R$ charges, the $R$ charge $r_{Q\bar{Q}}$ is only defined up to
the addition of integer multiples of $N$.
Hence, all inequivalent solutions to the condition in Eq.~\eqref{eq:QRcharges}
lie in the interval $[0,N)$.
In addition, we observe that, varying $\ell_Q$ in integer steps,
the $R$ charge $r_{Q\bar{Q}}$ changes in steps of $\frac{N}{k}$.
Consequently, for each pair of values for $N$ and $k$, there are $k$ inequivalent
choices for $r_{Q\bar{Q}}$,
\begin{align}
r_{Q\bar{Q}} \modulo{N}  2 + \frac{1}{k} \left(6 + \ell_Q N\right) \,,\quad
\ell_Q = 0,..,k-1 \,.
%-\textrm{ceil}\left[\frac{6+2k}{N}\right], ..,
%-\textrm{floor}\left[\frac{6+2k}{N}\right] + k\,.
\label{eq:rQQ}
\end{align}

%%%%%%%%%%%%%%%%%%%%%%%%%%%%%%%%%%%%%%%%%%%%%%%%%%%%%%%%%%%%%%%%%%%%%%%%%%%%%%%%%%%%%%%%%%%%%%%%%%%%

A crucial implication of this result is that, in most cases, the extra quarks and anti-quarks
are massless as long as the $Z_N^R$ is unbroken.
Only for $r_{Q\bar{Q}} = 2$, a supersymmetric and $R$-invariant mass term is allowed for
the extra quark fields in the superpotential.
An $R$ charge $r_{Q\bar{Q}}$ of $2$ can, however, only be obtained in the case
of a $Z_3^R$ or a $Z_6^R$ symmetry,
\begin{align}
r_{Q\bar{Q}} = 2\:: \quad (N,\ell_Q) = (3,-2) \,,
\quad (N,\ell_Q) = (6,-1) \,, \quad k = 1,2,3,.. \,,
\label{eq:rQQ2}
\end{align}
which are just the two $Z_N^R$ symmetries that do not require an extension of the MSSM
particle content in the first place.
As soon as the introduction of a new matter sector is \textit{mandatory}
in order to render the $Z_N^R$ symmetry anomaly-free, the new quark fields
are therefore guaranteed to be massless.
This observation also reflects the self-consistency of our result
in Eq.~\eqref{eq:rQQ}.
If we had started with the requirement of non-zero contributions from the
new quarks to the $Z_N^R$ anomalies and we
had found that the new quarks could possibly be massive,
our derivation of $r_{Q\bar{Q}}$ would be faulty, since
massive quarks would not contribute to
the $Z_N^R$ anomalies to begin with.

%%%%%%%%%%%%%%%%%%%%%%%%%%%%%%%%%%%%%%%%%%%%%%%%%%%%%%%%%%%%%%%%%%%%%%%%%%%%%%%%%%%%%%%%%%%%%%%%%%%%

Of course, the requirement of massless quarks at high energies does not say anything
about the masses of the new quarks at low energies, where the $Z_N^R$ is spontaneously broken.
The spontaneous breaking of the $Z_N^R$ symmetry might,
in fact, even generate masses for the extra quarks
of the order of the gravitino mass~\cite{Inoue:1991rk},
cf.\ also Sec.~\ref{subsubsec:GMmechanism}.
A necessary condition for this to happen is that $r_{Q\bar{Q}} = 0$,
which can be fulfilled for each value of $N$ as long as $k$ and $\ell_Q$ are
chosen such that $-2k \modulo{N} 6$.
This means in particular that, for $k=1$ and $N=4,8$, the $R$ charge $r_{Q\bar{Q}}$
is always zero.
For $N=3,4,6,8$ and only one pair of extra quark fields, $k=1$, the generation
of a sufficiently large mass for the new quark flavour is therefore not an issue.
The new quark either exhibits a supersymmetric mass from the outset or
it acquires a mass in the course of $R$ symmetry breaking~\cite{Inoue:1991rk}.
For completeness, we also mention that, provided $r_{Q\bar{Q}} = 0$, the new quarks
could equally acquire masses of the order of the gravitino mass in the course of
spontaneous SUSY breaking via the Giudice-Masiero mechanism~\cite{Giudice:1988yz}.
A further necessary prerequisite in this case would then be that there exists a coupling
of the extra quark fields to the SUSY breaking sector in the K\"ahler potential.

%%%%%%%%%%%%%%%%%%%%%%%%%%%%%%%%%%%%%%%%%%%%%%%%%%%%%%%%%%%%%%%%%%%%%%%%%%%%%%%%%%%%%%%%%%%%%%%%%%%%

In our following analysis, we will disregard the two exceptional choices
for $N$ and $\ell_Q$ in Eq.~\eqref{eq:rQQ2} as well as all combinations of $N$, $k$
and $\ell_Q$ yielding $r_{Q\bar{Q}} = 0$.
Instead, we shall focus on $R$ charges $r_{Q\bar{Q}}$ that imply vanishing masses for the
new quarks and anti-quarks before and after $R$ symmetry breaking as long as no further
fields are introduced.
The absence of a supersymmetric mass term is then equivalent to the
statement that the renormalizable superpotential of the extra quark sector vanishes completely.
This is because all SM singlets solely composed out of the fields $Q$ and $\bar{Q}$
must be combinations of the operator products $Q\bar{Q}$, $Q^5$ and $\bar{Q}^5$, such that
$Q\bar{Q}$ is the \textit{only} conceivable operator which could potentially show up in the
renormalizable superpotential.
At the renormalizable level, the global flavour symmetry of the extra
matter sector by itself, \textit{i.e.}\ neglecting its interactions with the
other sectors of our model for a moment, is therefore maximally large,
\begin{align}
U(k)_{Q} \times U(k)_{\bar{Q}} \cong
SU(k)_Q^V \times SU(k)_Q^A \times U(1)_Q^V \times U(1)_Q^A \,,
\label{eq:Qflavsym}
\end{align}
with $U(k)_{Q}$ and $U(k)_{\bar{Q}}$ accounting
for the flavour rotations of the left-chiral superfields $Q_i$ and $\bar{Q}_i$,
respectively.
As we will see in Secs.~\ref{subsec:PQsym}, the axial Abelian flavour symmetry $U(1)_Q^A$
will play an important role in the identification of the PQ symmetry.
Finally, we remark that higher-dimensional operators as well as couplings
of the new quarks and anti-quarks to other fields explicitly break the flavour symmetry.
In order not to spoil the PQ solution to the strong $CP$ problem,
these explicit breaking effects must be sufficiently suppressed by means of
a protective gauge symmetry.
We will return to this point in Sec.~\ref{sec:constraints}.

%%%%%%%%%%%%%%%%%%%%%%%%%%%%%%%%%%%%%%%%%%%%%%%%%%%%%%%%%%%%%%%%%%%%%%%%%%%%%%%%%%%%%%%%%%%%%%%%%%%%

\subsection{Extra singlet sector required to render the extra matter massive}
\label{subsec:singletsector}

%%%%%%%%%%%%%%%%%%%%%%%%%%%%%%%%%%%%%%%%%%%%%%%%%%%%%%%%%%%%%%%%%%%%%%%%%%%%%%%%%%%%%%%%%%%%%%%%%%%%

In the previous section, we have seen how the requirement of an anomaly-free
$Z_N^R$ symmetry forces us to extend the MSSM particle content by new
quark fields, $Q_i$ and $\bar{Q}_i$.
Except for some special cases, these quark fields are, however, massless as long
as the $Z_N^R$ symmetry is unbroken.
Extra massless coloured and weakly interacting particles are, of course, in conflict
with observations, which is why we have to extend our model once more, so as to provide
masses to the new quarks and anti-quarks.

%%%%%%%%%%%%%%%%%%%%%%%%%%%%%%%%%%%%%%%%%%%%%%%%%%%%%%%%%%%%%%%%%%%%%%%%%%%%%%%%%%%%%%%%%%%%%%%%%%%%

\subsubsection{Coupling of the extra matter fields to a new singlet sector}

%%%%%%%%%%%%%%%%%%%%%%%%%%%%%%%%%%%%%%%%%%%%%%%%%%%%%%%%%%%%%%%%%%%%%%%%%%%%%%%%%%%%%%%%%%%%%%%%%%%%

In order to generate sufficiently large
mass terms for the quark pairs $\left(Q\bar{Q}\right)_i$ in the
superpotential, we are in need of a SM singlet that
acquires a vacuum expectation value (VEV) at least above the electroweak scale.
No such singlet exists in the MSSM or its seesaw extension, so that we are required to
introduce another new field.
Let us refer to this field as $P$ and demand that it couples to the quark pairs
$\left(Q\bar{Q}\right)_i$ in the following way,%
\footnote{Note that the field $P$ might be part of the hidden sector responsible for
the spontaneous breaking of SUSY~\cite{Feldstein:2012bu}.}
\begin{align}
W_Q = \frac{1}{M_{\textrm{Pl}}^{n-1}} \sum_{i=1}^k \lambda_i\, P^n \,(Q\bar{Q})_i \,.
\label{eq:WQ}
\end{align}
Here, $M_{\textrm{Pl}} = (8\pi G)^{-1/2} = 2.44 \times 10^{18}\,\textrm{GeV}$
is the reduced Planck mass and the $\lambda_i$
denote dimensionless coupling constants, which we assume to be of $\nrmcal{O}(1)$.
The power $n$ can, \textit{a priori}, be any integer number, $n=1,2,..$.
Moreover, the coupling in Eq.~\eqref{eq:WQ} fixes the $R$ charge $r_P$
of the singlet field $P$.
In order to ensure that it is indeed allowed in the superpotential, we require that
\begin{align}
n\,r_P + r_{Q\bar{Q}} = 2 + \ell_P N \,, \quad
r_P \modulo{N} \frac{1}{n} \left(2 - r_{Q\bar{Q}} + \ell_P N\right)
\,, \quad \ell_P \in \mathbb{Z} \,.
\label{eq:rPcond}
\end{align}
Making use of our result for the $R$ charge $r_{Q\bar{Q}}$ for the quark pairs,
cf.\ Eq.~\eqref{eq:rQQ}, we then find
\begin{align}
r_P \modulo{N} - \frac{6}{nk} + (k \, \ell_P - \ell_Q)\frac{N}{nk} \,. \label{eq:rP}
\end{align}
Similarly as in the case of the extra quark fields, the $R$ charge $r_P$ is not uniquely
determined.
For each combination of values for $N$, $n$ and $k$, there are instead $nk$ inequivalent
solutions to the condition in Eq.~\eqref{eq:rPcond}.
These are all of the form
given in Eq.~\eqref{eq:rP}, with $(k \, \ell_P - \ell_Q) = 0,1,.., nk-1$.
%
%taking values in the following range,
% %
% \begin{align}
% (k \, \ell_P - \ell_Q) = \textrm{ceil}\left[\frac{6}{N}\right], ..,
% \textrm{floor}\left[\frac{6}{N}\right] + nk \,.
% \end{align}

%%%%%%%%%%%%%%%%%%%%%%%%%%%%%%%%%%%%%%%%%%%%%%%%%%%%%%%%%%%%%%%%%%%%%%%%%%%%%%%%%%%%%%%%%%%%%%%%%%%%

\subsubsection{Superpotential of the extra singlet sector}
\label{subsubsec:WP}

%%%%%%%%%%%%%%%%%%%%%%%%%%%%%%%%%%%%%%%%%%%%%%%%%%%%%%%%%%%%%%%%%%%%%%%%%%%%%%%%%%%%%%%%%%%%%%%%%%%%

So far, the field $P$ does not possess any
interactions that would endow it with a non-vanishing VEV.
We thus introduce another singlet field $X$ and couple it to the field $P$,
in order to generate a non-trivial $F$-term potential for the scalar component of $P$,
\begin{align}
W_P = \kappa \, X \left[\frac{\Lambda^2}{2} - f(P,..)\right] \,,
\label{eq:WPf}
\end{align}
where $\kappa$ is a coupling constant, $\Lambda$ denotes some
mass scale  and $f$ stands for a function of $P$ and probably
other fields.
We assume the scale $\Lambda$ to carry zero $R$ charge, which directly entails
that the singlet field $X$ and the function $f$ must have $R$ charges $2$ and $0$,
respectively.
Besides that, we also assume a value for $r_P$ such that
none of the operators $P$, $P^2$, $P^3$, $XP$, $XP^2$ and $X^2P$
is allowed in the superpotential $W_P$, \textit{i.e.}\
we require $r_P$ to fulfill all of the following relations at once,
\begin{align}
r_P \nodulo{N} 2 \,, \quad
2 r_P \nodulo{N} 2 \,, \quad 
3 r_P \nodulo{N} 2 \,, \quad 
2 r_P \nodulo{N} 0 \,, \quad 
r_P \nodulo{N} - 2 \,.
\label{eq:rPcond5}
\end{align}
As we will see shortly, these conditions ensure that $W_P$
ends up featuring a flat direction which can be identified with the axion
and its superpartners.

%%%%%%%%%%%%%%%%%%%%%%%%%%%%%%%%%%%%%%%%%%%%%%%%%%%%%%%%%%%%%%%%%%%%%%%%%%%%%%%%%%%%%%%%%%%%%%%%%%%%

Now, if the function $f$ were merely composed out of powers $P^m$ of the field $P$,
where $m = 3,4,..$, the $R$ charge of $f$ would only vanish for particular values of $r_P$ and
$m$ in the case of particular $Z_N^R$ symmetries.
We, however, wish to be able to give masses to the new quarks and anti-quarks,
irrespectively of the concrete value of $N$.
For that reason, we have to introduce a singlet field $\bar{P}$ carrying
the opposite $R$ charge of the field $P$,
\begin{align}
r_{\bar{P}} \modulo{N} -r_P \modulo{N} \frac{6}{nk} - (k \, \ell_P - \ell_Q)\frac{N}{nk} \,,
\label{eq:rPbar}
\end{align}
such that we are able to render the function $f$ an $R$ singlet by taking it
to be a function of the singlet pair $P\bar{P}$.
The superpotential in Eq.~\eqref{eq:WPf} can then be fixed to be of the following form,
\begin{align}
W_P = \kappa \, X \left[\frac{\Lambda^2}{2} - f(P\bar{P})\right] =
\kappa \, X \left(\frac{\Lambda^2}{2} - P\bar{P}\right) + .. \,,
\label{eq:WPQ}
\end{align}
with the dots after the plus sign indicating higher-dimensional non-renormalizable terms
and where, similarly as above, we have assumed that $r_{\bar{P}} = - r_P$ is
such that none of the operators $\bar{P}$, $\bar{P}^2$, $\bar{P}^3$, $X\bar{P}$, $X\bar{P}^2$
and $X^2\bar{P}$ is allowed in the superpotential $W_P$.
In addition to the five conditions in Eq.~\eqref{eq:rPcond5}, we therefore
also have to require that
\begin{align}
2 r_P \nodulo{N} - 2 \,, \quad  3 r_P \nodulo{N} - 2 \,.
\label{eq:rPcond2}
\end{align}

%%%%%%%%%%%%%%%%%%%%%%%%%%%%%%%%%%%%%%%%%%%%%%%%%%%%%%%%%%%%%%%%%%%%%%%%%%%%%%%%%%%%%%%%%%%%%%%%%%%%

In total, we hence impose seven conditions on the $R$ charge $r_P$, which,
depending on $N$, allow us to forbid as many as 14 different values for $r_P$.%
\footnote{To see this, note that all of our conditions can be written as
$r_P \neq a/q + \ell/qN$, where $q\in\left\{1,2,3\right\}$, $a\in\left\{-2,0,2\right\}$
and $\ell\in\mathbb{Z}$.
The number of different $r_P$ values forbidden by some condition therefore corresponds to
its value for $q$.}
We now also see that each of the combinations
of $N$, $\ell_Q$ and $k$ that either result
in $r_{Q\bar{Q}} = 0$ or $r_{Q\bar{Q}} = 2$ violates
exactly one of these conditions.
If $r_{Q\bar{Q}} = 0$, we know that $n \, r_P \modulo{N} 2$, such that either
$P$ or $P^2$ is allowed.
Similarly, $r_{Q\bar{Q}} = 2$ implies $n \, r_P \modulo{N} 0$, such that
$XP$ and/or $XP^2$ is allowed.
This means that, in those cases in which we do not depend on an extra singlet
sector to generate masses for the extra quarks, $Q_i$ and $\bar{Q}_i$,
we would not even succeed in doing so, if we attempted it nonetheless.
Finally, we emphasize that, by construction, $X \Lambda^2$ and $X P\bar{P}$
end up being the only renormalizable operators in $W_P$ that are compatible
with the $Z_N^R$ symmetry for any value of $N$.
In the following, we shall now show that the new singlet sector consisting of
the fields $X$, $P$ and $\bar{P}$ has the potential to accommodate the \textit{invisible}
axion and its superpartners and hence provide a solution of the strong $CP$ problem
via the PQ mechanism.

%%%%%%%%%%%%%%%%%%%%%%%%%%%%%%%%%%%%%%%%%%%%%%%%%%%%%%%%%%%%%%%%%%%%%%%%%%%%%%%%%%%%%%%%%%%%%%%%%%%%

\subsubsection{Identification of the PQ symmetry}
\label{subsec:PQsym}

%%%%%%%%%%%%%%%%%%%%%%%%%%%%%%%%%%%%%%%%%%%%%%%%%%%%%%%%%%%%%%%%%%%%%%%%%%%%%%%%%%%%%%%%%%%%%%%%%%%%

Evidently, the superpotential in Eq.~\eqref{eq:WPQ} exhibits a global $U(1)$ symmetry,
\textit{viz.}\ it is invariant under a global phase rotation of the fields $P$ and $\bar{P}$.
Let us refer to this symmetry as $U(1)_P$ and stipulate that
the two singlets $P$ and $\bar{P}$ respectively
carry charge $q_P^{(P)} = 1$ and $q_{\bar{P}}^{(P)} = - 1$
under it.
The $U(1)_P$ symmetry is explicitly broken by the coupling of the singlet operator
$P^n$ to the quark pairs $\left(Q\bar{Q}\right)_i$ in the superpotential in Eq.~\eqref{eq:WQ}.
At the same time, this coupling also breaks the $U(1)_Q^A$ symmetry in the extra quark sector.
Altogether, the coupling between the new quark sector and the new singlet sector
reduces the number of global Abelian symmetries from three to two,
\begin{align}
U(1)_P \times U(1)_Q^V \times U(1)_Q^A
\quad\rightarrow\quad U(1)_{\textrm{PQ}} \times U(1)_Q^V \,.
\label{eq:PQPQ}
\end{align}
The operators $\left(Q\bar{Q}\right)_i$ are invariant under $U(1)_Q^V$ transformations,
which is why the global vectorial symmetry in the quark sector survives the introduction of
the superpotential in Eq.~\eqref{eq:WQ}.
The other global symmetry leaving the coupling in Eq.~\eqref{eq:WQ} invariant
corresponds to some linear combination of
$U(1)_P$, $U(1)_Q^V$ and $U(1)_Q^A$.
It is this symmetry that we shall identify with the PQ symmetry.
In the remainder of this paper, we will now investigate under which circumstances it may
be successfully protected against the effects of higher-dimensional operators.

%%%%%%%%%%%%%%%%%%%%%%%%%%%%%%%%%%%%%%%%%%%%%%%%%%%%%%%%%%%%%%%%%%%%%%%%%%%%%%%%%%%%%%%%%%%%%%%%%%%%

Before continuing, let us, however, reiterate once more for clarity:
$U(1)_P$, $U(1)_Q^V$ and $U(1)_Q^A$ are accidental global symmetries of the new singlet
and quark sectors at the renormalizable level that arise due to our particular
choice of $R$ charges.
Neither of them manages to survive as an exact symmetry in the full low-energy
effective theory.
To begin with, the coupling between the two new sectors in Eq.~\eqref{eq:WQ}
breaks $U(1)_P \times U(1)_Q^V \times U(1)_Q^A$ to its subgroup
$U(1)_{\textrm{PQ}} \times U(1)_Q^V$.
This residual symmetry is, in turn, explicitly broken by other higher-dimensional
operators.
The dimension-6 operators $Q^5$ and $\bar{Q}^5$, for instance, explicitly break
the vectorial Abelian symmetry in the new quark sector.
The crucial question which we will have to address in the following therefore is
how severe the explicit breaking of the PQ symmetry turns out to be and whether
it remains sufficiently small enough, so that our model can still explain a
QCD vacuum angle $\bar{\theta}$ of less than $\nrmcal{O}\left(10^{-10}\right)$.

%%%%%%%%%%%%%%%%%%%%%%%%%%%%%%%%%%%%%%%%%%%%%%%%%%%%%%%%%%%%%%%%%%%%%%%%%%%%%%%%%%%%%%%%%%%%%%%%%%%%

Up to now, we are unable to specify the PQ charges of
the new quark and anti-quark fields separately, as the superpotential in Eq.~\eqref{eq:WQ}
only contains the quark product operators $\left(Q\bar{Q}\right)_i$.
Demanding that the PQ charges of the singlet fields $P$ and $\bar{P}$ coincide with
their $U(1)_P$ charges, all we can say is that the operators $\left(Q\bar{Q}\right)_i$
must carry a total PQ charge of $-n$.
For the time being, we may thus work with the following PQ charges,
\begin{align}
q_P = 1 \,, \quad q_{\bar{P}} = -1 \,, \quad q_Q \in \mathbb{R}
\,, \quad q_{\bar{Q}} = -n - q_Q \,, \quad q_{Q\bar{Q}} = -n \,.
\label{eq:qPPQQ}
\end{align}

%%%%%%%%%%%%%%%%%%%%%%%%%%%%%%%%%%%%%%%%%%%%%%%%%%%%%%%%%%%%%%%%%%%%%%%%%%%%%%%%%%%%%%%%%%%%%%%%%%%%

The PQ charges of the MSSM fields
$q_i$, where $i$ now runs over $i = q,\,u^c,\,d^c,\,\ell,\,e^c,\,n^c,\,H_u,\,H_d$,
are subject to constraints deriving from the Yukawa couplings in the superpotential
$W_{\textrm{MSSM}}$ in Eq.~\eqref{eq:WMSSM}.
The first two terms in $W_{\textrm{MSSM}}$ yield the following three conditions
\begin{align}
q_{u^c} + q_q + q_{H_u} = 0 \,, \quad
q_{d^c} + q_q + q_{H_d} = 0 \,, \quad
q_{e^c} + q_{\ell} + q_{H_d} = 0 \,,
\label{eq:PQconds}
\end{align}
the first two of which combine to give $q_{u^c} + q_{d^c} + 2 q_q + q_{H_u} + q_{H_d} = 0$.
As we will see in Sec.~\ref{subsec:muterm}, the PQ charges of the two MSSM Higgs doublets
must sum to zero, $q_{H_u} + q_{H_d} = 0$, implying that 
\begin{align}
q_{u^c} + q_{d^c} + 2 q_q = 0 \,.
\label{eq:totalq0}
\end{align}
The total PQ charge of all MSSM quark fields hence vanishes, such that the colour
anomaly of the PQ symmetry ends up receiving contributions only from the extra matter sector
and none from the MSSM sector, cf.\ also Eq.~\eqref{eq:U1PQanomaly} further below.

%%%%%%%%%%%%%%%%%%%%%%%%%%%%%%%%%%%%%%%%%%%%%%%%%%%%%%%%%%%%%%%%%%%%%%%%%%%%%%%%%%%%%%%%%%%%%%%%%%%%

Having derived this important result, we would still like to know which values
the MSSM PQ charges can actually take.
Forgetting for a moment about the neutrino singlets required for the seesaw mechanism,
the answer is clearly all values compatible with the three conditions in Eq.~\eqref{eq:PQconds}.
The PQ charges $q_i$ can then, for instance, be parametrized
in terms of $q_q,\,q_{\ell},\,q_{H_u} \in \mathbb{R}$.
Moreover, we note that in the course of electroweak symmetry breaking the
Yukawa couplings in $W_{\textrm{MSSM}}$
turn into mass terms for the MSSM matter fields,
breaking the PQ symmetry unless $q_{H_u} = - q_{H_d} = 0$.
In this particular case, the PQ symmetry can be identified as a linear combination of
$U(1)_B$ and $U(1)_L$, the global Abelian symmetries associated with baryon number
$B$ and lepton number $L$.
This result is a useful crosscheck, since $U(1)_B$ and $U(1)_L$ are
the \textit{unique} accidental global symmetries of the standard model.
In the seesaw extension of the MSSM, the conditions in Eq.~\eqref{eq:PQconds}
are supplemented by two further conditions
deriving from the last two terms in $W_{\textrm{MSSM}}$,
\begin{align}
q_{n^c} + q_{\ell} + q_{H_u} = 0 \,, \quad
2 q_{n^c} = 0 \,,
\end{align}
eliminating the PQ charge $q_{\ell}$ as a free parameter.
Upon extending the MSSM by three neutrino singlet fields, the PQ charges $q_i$
can therefore be parametrized by only two charges, $q_q,\,q_{H_u} \in \mathbb{R}$.
Setting $q_{H_u}$ to zero now renders the PQ symmetry proportional to
$U(1)_B$, which is, of course, expected, since the $U(1)_L$
is explicitly broken by the Majorana mass term in $W_{\textrm{MSSM}}$.
The only relation among the PQ charges $q_i$
relevant for our further analysis is Eq.~\eqref{eq:totalq0}.
Without loss of generality, we are thus free to take
$q_q$, $q_{\ell}$ and $q_{H_u}$ to be zero,
so that $q_i = 0$ for all fields $i$.
The field content of our model as well as our assignment of the PQ charges are hence
similar as in the KSVZ axion model proposed by Kim~\cite{Kim:1979if} as well as by
Shifman, Vainshtein and Zakharov~\cite{Shifman:1979if}.

%%%%%%%%%%%%%%%%%%%%%%%%%%%%%%%%%%%%%%%%%%%%%%%%%%%%%%%%%%%%%%%%%%%%%%%%%%%%%%%%%%%%%%%%%%%%%%%%%%%%

\subsubsection{Spontaneous breaking and colour anomaly of the PQ symmetry}

%%%%%%%%%%%%%%%%%%%%%%%%%%%%%%%%%%%%%%%%%%%%%%%%%%%%%%%%%%%%%%%%%%%%%%%%%%%%%%%%%%%%%%%%%%%%%%%%%%%%

In the true vacuum of the scalar potential corresponding to the superpotential $W_P$
in Eq.~\eqref{eq:WPQ}, the singlet field $X$
vanishes and the PQ symmetry is spontaneously broken,%
\footnote{Spontaneous $R$ symmetry breaking results in a tadpole term for $X$
in the scalar potential, $V \supset -\kappa\,m_{3/2}\,\Lambda^2 X$.
Besides that, $X$ also couples to other fields of our model, cf.\ Sec.~\ref{subsubsec:SMSSM},
such that its VEV eventually turns out to be of order the gravitino mass rather than zero,
$\left<X\right> \sim m_{3/2}$.}
\begin{align}
\langle X\rangle = 0 \,,\quad
\langle P\rangle = \frac{\Lambda}{\sqrt{2}} \,
\exp\left(\frac{A}{\Lambda}\right) \,,\quad
\langle\bar{P}\rangle = \frac{\Lambda}{\sqrt{2}} \,
\exp\left(-\frac{A}{\Lambda}\right) \,, \quad
\phi \subset A \,, \quad \phi = \frac{1}{\sqrt{2}}\left(b + i a\right) \,,
\label{eq:XPPvevs}
\end{align}
where the chiral superfield $A$ represents the axion multiplet, which consists of
the pseudo-scalar axion $a$, the scalar saxino $b$ and the fermionic axino $\tilde{a}$.
The various factors of $\sqrt{2}$ in Eq.~\eqref{eq:XPPvevs} serve two purposes.
First, they render the kinetic term of the axion canonically normalized;
second, they ensure that the scalar mass eigenstate that actually breaks
the PQ symmetry, $p_+ = \frac{1}{\sqrt{2}}\left(p + \bar{p}^*\right)$,
where $p$ and $\bar{p}$ are the complex scalars contained in $P$ and $\bar{P}$,
acquires a VEV $\langle p_+ \rangle = \Lambda$.
%
% As $P$ and $\bar{P}$ carry non-zero $R$ charges,
% the $Z_N^R$ symmetry also becomes spontaneously broken in the course of PQ breaking.
% %
% If the $Z_N^R$ contains a subgroup $Z_{N'}^R$, it is broken down
% to exactly this subgroup, provided that $\left|r_P\right| = N'$.
% %
% Otherwise, it is broken completely, leaving no remnant subgroup behind.
% %
% This latter case hence entails a simple solution to the domain wall problem
% that usually arises when considering the spontaneous breaking of
% discrete $R$ symmetries in the context of cosmology~\cite{Dine:2010eb}.

%%%%%%%%%%%%%%%%%%%%%%%%%%%%%%%%%%%%%%%%%%%%%%%%%%%%%%%%%%%%%%%%%%%%%%%%%%%%%%%%%%%%%%%%%%%%%%%%%%%%

Before continuing, we remark that, in the special case of a $Z_4^R$ symmetry,
also a cubic term in the singlet field $X$ is allowed in the superpotential $W_P$,
\begin{align}
N=4 \:: \quad W = \kappa X \left(\frac{\Lambda^2}{2} - P\bar{P}\right) - \lambda_X X^3 + .. \,,
\end{align}
where $\lambda_X$ is some dimensionless coupling constant of $\nrmcal{O}(1)$.
In this case, the field configuration in Eq.~\eqref{eq:XPPvevs} no longer
represents the unique vacuum of the scalar potential corresponding to $W_P$.
At $\langle P \rangle = \langle \bar{P} \rangle = 0$ and
$\langle X \rangle  = \sqrt{\kappa/(6\lambda_X)}\Lambda$, the scalar potential
exhibits another local minimum.
Because of the linear term in $X$ in the scalar potential,
$V \supset -\kappa\,m_{3/2}\,\Lambda^2 X$, this vacuum then has a negative
energy density, the absolute value of which is much larger than the energy
density of the PQ-breaking vacuum in Eq.~\eqref{eq:XPPvevs}.
There exists, however, no flat direction connecting the alternative vacuum
with our PQ-breaking vacuum, which is why we do not have to worry about the
stability of the latter one.
We merely have to assume that, in the course of the cosmological evolution,
our universe has settled in the vacuum in Eq.~\eqref{eq:XPPvevs} rather than
in the alternative vacuum.
In fact, this is a very plausible assumption, if we believe that the field
$X$ is stabilized at $\langle X \rangle = 0$ during inflation due to a large
Hubble-induced mass.

%%%%%%%%%%%%%%%%%%%%%%%%%%%%%%%%%%%%%%%%%%%%%%%%%%%%%%%%%%%%%%%%%%%%%%%%%%%%%%%%%%%%%%%%%%%%%%%%%%%%

In order to solve the strong $CP$ problem, it is necessary that the PQ symmetry
has a colour anomaly.
Thanks to our derivation of the PQ charges of all coloured matter fields in the previous
subsection, we are now able to calculate the anomalous divergence of the axial
PQ current $J_{\textrm{PQ}}^\mu$ and show that it is non-zero,
\begin{align}
\partial_\mu J_{\textrm{PQ}}^\mu = \nrmcal{A}_{\textrm{PQ}} \,\frac{\alpha_s}{8\pi}
\textrm{Tr} \left[ G_{\mu\nu} \tilde{G}^{\mu\nu}\right] \,,\quad
\nrmcal{A}_{\textrm{PQ}} = k \, q_{Q\bar{Q}} + q_{u^c} + q_{d^c} + 2q_q = - nk \,,
\label{eq:U1PQanomaly}
\end{align}
where we have introduced $\nrmcal{A}_{\textrm{PQ}}$ as the anomaly coefficient of
the $U(1)_{\textrm{PQ}}\left[SU(3)_C\right]^2$ anomaly.
This colour anomaly of the PQ symmetry induces an extra term in the effective
Lagrangian~\cite{Peccei:1977hh,Peccei:2006as},
\begin{align}
\mathcal{L}_{\textrm{QCD}}^{\textrm{eff}} \supset \left(\bar{\theta} -
\frac{a}{f_a}\right) \frac{\alpha_s}{8\pi}
\textrm{Tr} \left[ G_{\mu\nu} \tilde{G}^{\mu\nu}\right] \,,
\quad f_a = \frac{\sqrt{2}\Lambda}{\left|\nrmcal{A}_{\textrm{PQ}}\right|} \,,
\label{eq:LQCDeff}
\end{align}
with $f_a$ denoting the axion decay constant.
In consequence of this coupling of the axion $a$ to the gluon field strength $G_{\mu\nu}$,
an effective non-perturbative potential for the axion is generated,
\begin{align}
V_a^{\textrm{eff}} = \Lambda_{\textrm{QCD}}^4
\left[1 - \cos\left(\bar{\theta}-\frac{a}{f_a}\right)\right] \,,
\label{eq:Vaeff}
\end{align}
the minimum of which is located at $\left<a\right> = f_a \bar{\theta}$.
Shifting $a$ by its VEV $\left<a\right>$
then cancels the $\bar{\theta}$ term in Eq.~\eqref{eq:LQCDeff}, thereby
rendering the QCD Lagrangian $CP$-invariant.
Our singlet sector consisting of the fields
$X$, $P$ and $\bar{P}$ hence entails a manifestation of the PQ solution to
the strong $CP$ problem.

%%%%%%%%%%%%%%%%%%%%%%%%%%%%%%%%%%%%%%%%%%%%%%%%%%%%%%%%%%%%%%%%%%%%%%%%%%%%%%%%%%%%%%%%%%%%%%%%%%%%

An important detail to note is that it is the scale $f_a$, rather than $\Lambda$, which
determines the strength of all low-energy interactions of the axion~\cite{Georgi:1986df}.
This is also the reason why experimental constraints on the axion coupling are always formulated
as bounds on $f_a$ and not on $\Lambda$.
Requiring, for instance, that astrophysical objects such as supernovae or white dwarfs do
not lose energy too fast due to axion emission allows one to put a lower bound
of $\nrmcal{O}\left(10^9\right)\,\textrm{GeV}$~\cite{Raffelt:2006cw} on $f_a$.
Meanwhile, cosmology restricts the possible range of $f_a$ values from above.
In order to prevent cold axions from overclosing the universe, $f_a$ must be at most of
$\nrmcal{O}\left(10^{12}\right)\,\textrm{GeV}$~\cite{Preskill:1982cy,Sikivie:2006ni},
hence leaving open the following phenomenologically viable window for the axion
decay constant,
\begin{align}
10^9 \,\textrm{GeV} \lesssim f_a \lesssim 10^{12} \,\textrm{GeV} \,.
\label{eq:window}
\end{align}

%%%%%%%%%%%%%%%%%%%%%%%%%%%%%%%%%%%%%%%%%%%%%%%%%%%%%%%%%%%%%%%%%%%%%%%%%%%%%%%%%%%%%%%%%%%%%%%%%%%%

Furthermore, as evident from the effective axion potential in Eq.~\eqref{eq:Vaeff},
the non-perturbative QCD instanton effects break the PQ symmetry to a global
and discrete $Z_{N_{\textrm{DW}}}$ symmetry, where
$N_{\textrm{DW}} = \left|\nrmcal{A}_{\textrm{PQ}}\right| = nk$, commonly referred
to as the domain wall number, counts the number of degenerate axion vacua.
If the breaking of the PQ symmetry occurs after inflation,
this vacuum structure of the axion potential implies the formation of
axion domain walls during the QCD phase transition, thereby leading
to a cosmological disaster~\cite{Sikivie:2006ni,Sikivie:1982qv}.
One obvious solution to this domain wall problem is to impose that inflation
takes place after the spontaneous breaking of the PQ symmetry, such that
the axion field is homogenized across the entire observable universe.%
\footnote{In this case, perturbations in the axion field amplified during inflation
may result in too large isocurvature contributions to the temperature fluctuations
seen in the cosmic microwave background.
A variety of solutions to this isocurvature perturbation problem have however been
proposed in the literature, cf.\ for instance Ref.~\cite{Kawasaki:2013iha} and
references therein, which is why we will not consider it any further.}
Alternatively, one may attempt to construct an axion model with $N_{\textrm{DW}} = 1$,
in which case the axion domain walls collapse under their boundary tension
soon after their formation~\cite{Vilenkin:1982ks}.
In Appendix~\ref{app:domainwallprob}, we present a slight modification of our model that
just yields $N_{\textrm{DW}} = 1$ and which hence allows for a solution of the axion
domain wall problem even if the spontaneous breaking of the PQ symmetry takes place
after inflation.

%%%%%%%%%%%%%%%%%%%%%%%%%%%%%%%%%%%%%%%%%%%%%%%%%%%%%%%%%%%%%%%%%%%%%%%%%%%%%%%%%%%%%%%%%%%%%%%%%%%%

\subsubsection{Mass scale of the extra matter sector}
\label{subsubsec:Qmassscale}

%%%%%%%%%%%%%%%%%%%%%%%%%%%%%%%%%%%%%%%%%%%%%%%%%%%%%%%%%%%%%%%%%%%%%%%%%%%%%%%%%%%%%%%%%%%%%%%%%%%%

As anticipated, the spontaneous breaking of
the PQ symmetry furnishes the extra quarks and anti-quarks with Dirac masses $m_{Q_i}$,
which can be read off from the superpotential $W_Q$ in Eq.~\eqref{eq:WQ}
after expanding the singlet field $P$ around its VEV,
\begin{align}
m_{Q_i} = \frac{\lambda_i}{M_{\textrm{Pl}}^{n-1}}\left(\frac{\Lambda}{\sqrt{2}}\right)^n
\simeq  \left(\frac{\lambda_i}{1}\right)\left(\frac{k}{4}\right)^n
\left(\frac{f_a}{10^{10}\,\textrm{GeV}}\right)^n
\times
\begin{cases}
2.0 \times 10^{10} \,\,\textrm{GeV} & ;\:\: n = 1 \\
6.6 \times 10^{2\phantom{0}}  \,\,\textrm{GeV} & ;\:\: n = 2 \\
3.6 \times 10^{-5} \,\textrm{GeV} & ;\:\: n = 3 \\
\phantom{10}\,\cdots & ; \:\: n = 4
\end{cases}
\,.\label{eq:mQi}
\end{align}
For $n \geq 3$, our model thus predicts $k$ new quark multiplets with
masses below the electroweak scale, which is, of course, inconsistent
with experiments.
Hence, the only viable values for $n$ are $n=1$ and $n=2$.
From a phenomenological point of view, the $n=2$ case is certainly more
interesting as it features new coloured states with masses possibly within
the range of collider experiments.
On the other hand, if no heavy quarks should be found at or above the TeV scale,
our model would not automatically be ruled out.
Falling back to the $n=1$ case, the extra vector-quarks can always be decoupled from
the physics at the TeV scale,
thereby leaving still some room for the realization of our extension of the MSSM.

%%%%%%%%%%%%%%%%%%%%%%%%%%%%%%%%%%%%%%%%%%%%%%%%%%%%%%%%%%%%%%%%%%%%%%%%%%%%%%%%%%%%%%%%%%%%%%%%%%%%

For both viable values of $n$, we can now ask how many new quark flavours we are
allowed to introduce, \textit{i.e.}\ which values $k$ can possibly take.
Recall that in Sec.~\ref{subsubsec:WP} we required the $R$ charge $r_P$
to fulfill all of the seven conditions in Eqs.~\eqref{eq:rPcond5} and \eqref{eq:rPcond2}.
Given the explicit expression for $r_P$ in terms of $n$ and $k$ in Eq.~\eqref{eq:rP},
this requirement then directly translates into a set of $k$ values that, depending on the
values of $n$ and $N$, we are not allowed to employ, cf.\ Tab.~\ref{tab:n12kvalues}.
In Sec.~\ref{subsec:unification}, we will derive further restrictions on the set of
allowed $k$ values based on the requirement that the unification of the
SM gauge couplings ought to occur at the perturbative level.

%%%%%%%%%%%%%%%%%%%%%%%%%%%%%%%%%%%%%%%%%%%%%%%%%%%%%%%%%%%%%%%%%%%%%%%%%%%%%%%%%%%%%%%%%%%%%%%%%%%%

\begin{table}
\begin{center}
\begin{onehalfspacing}
\begin{tabular}{c||ccccccccccc}
 & $Z_3^R$ & $Z_4^R$ & $Z_5^R$ & $Z_6^R$ &  $Z_8^R$ & $Z_9^R$ & $Z_{10}^R$ &
$Z_{12}^R$ & $Z_{18}^R$ \\\hline\hline
$n = 1$ & $2,3,6,9$ & $2,3$ &  & $2,3,6,9$  & $2,3$ &  &  & $3,9$ \\  
$n = 2$ & $2,6$ &  & $2$ & $2,6$ &  & $3$ & $2$ &  & $3$
\end{tabular}
\caption{Values of $k$ leading to unwanted operators in $W_P$,
the superpotential of the new singlet sector,  cf.\ Eq.~\eqref{eq:WPQ}.
This table does not indicate for which $Z_N^R$ symmetries only one extra quark pair
is problematic.
For $n=1$, these are the symmetries with $N = 3,4,5,6,7,8,10,12,14,16,20$;
for $n=2$, it is the symmetries with $N = 5,7,10,11,14,22$.
In addition, independently of $n$, we disregard the possibility of
only one extra quark pair for $N = 3,4,6,8$ in any case,
cf.\ Sec.~\ref{subsubsec:anomaly}.}
\label{tab:n12kvalues}
\end{onehalfspacing}
\end{center}
\end{table}

%%%%%%%%%%%%%%%%%%%%%%%%%%%%%%%%%%%%%%%%%%%%%%%%%%%%%%%%%%%%%%%%%%%%%%%%%%%%%%%%%%%%%%%%%%%%%%%%%%%%

\subsection[Generation of the MSSM $\mu$ term]
{Generation of the MSSM \boldmath{$\mu$} term}
\label{subsec:muterm}

%%%%%%%%%%%%%%%%%%%%%%%%%%%%%%%%%%%%%%%%%%%%%%%%%%%%%%%%%%%%%%%%%%%%%%%%%%%%%%%%%%%%%%%%%%%%%%%%%%%%

In absence of any new physics beyond the MSSM, one might expect the supersymmetric mass of
the MSSM Higgs doublets to be of order the Planck scale, $\mu \sim M_{\textrm{Pl}}$.
Such a large $\mu$ value would then require a miraculous cancellation between the supersymmetric
and the soft SUSY-breaking contributions to the MSSM Higgs scalar potential, given that one
ought to end up with Higgs VEVs $\left<H_{u,d}\right> = v_{u,d}$ close the electroweak scale.
This puzzle, \textit{i.e.}\ the question why $\mu$ should be of the same order as the soft Higgs
masses, represents the infamous $\mu$ problem. 
As we have seen in the previous section, an anomaly-free discrete $R$ symmetry $Z_N^R$ forbids
the $\mu$ term in the MSSM superpotential, thus solving the $\mu$ problem halfway through.
What remains to be done is to demonstrate how the $\mu$ term emerges with
the right order of magnitude once the $Z_N^R$ has been spontaneously broken.

%%%%%%%%%%%%%%%%%%%%%%%%%%%%%%%%%%%%%%%%%%%%%%%%%%%%%%%%%%%%%%%%%%%%%%%%%%%%%%%%%%%%%%%%%%%%%%%%%%%%

\subsubsection[$\mu$ term from spontaneous $R$ symmetry breaking]
{\boldmath{$\mu$} term from spontaneous \boldmath{$R$} symmetry breaking}
\label{subsubsec:GMmechanism}

%%%%%%%%%%%%%%%%%%%%%%%%%%%%%%%%%%%%%%%%%%%%%%%%%%%%%%%%%%%%%%%%%%%%%%%%%%%%%%%%%%%%%%%%%%%%%%%%%%%%

In the special case of a $Z_4^R$ symmetry, the $R$ charges of $H_u$ and $H_d$ sum to zero,
$r_{H_u} + r_{H_d} \modulo{4} 4 \modulo{4} 0$, cf.\ Eq.~\eqref{eq:anomaly-free},
such that a $\mu$ term of the correct magnitude can be easily generated
in the course of spontaneous $R$ symmetry breaking~\cite{Inoue:1991rk}.
This mechanism is based on two ingredients:
(i) the observation that, for $r_{H_u} + r_{H_d} = 0$, the operator $H_uH_d$ can be accommodated
with some $\nrmcal{O}(1)$ coefficient $g_H'$ in the K\"ahler potential, $K \supset g_H' H_u H_d$
as well as (ii) the fact that, during spontaneous $R$ symmetry breaking, the superpotential
acquires a non-zero VEV $\left<W\right> = W_0$,%
\footnote{Since $W$ carries $R$ charge $r_W = 2$, the VEV $\langle W \rangle$
breaks the $Z_N^R$ completely;
$R$ parity, which potentially remains as an unbroken subgroup of the $Z_N^R$,
is not an actual discrete $R$ symmetry, cf.\ Sec.~\ref{subsec:MSSM}.
A possible mechanism to generate a
constant term in the superpotential is the condensation
of hidden gauginos, such that
$W_0 = \langle \nrmcal{W}_\alpha \nrmcal{W}^\alpha \rangle$~\cite{Veneziano:1982ah}.
Alternatively, the VEV of the superpotential might originate from the condensation of
hidden-sector quarks $\widetilde{Q}$, such that
$W_0 = \langle (\widetilde{Q}\widetilde{Q})^n\rangle$.
In Appendix A of Ref.~\cite{Harigaya:2013ns}, we present an exemplary model illustrating how
such a quark condensate could potentially be generated by means of strong gauge dynamics
in some hidden sector.}
where $W_0/M_{\textrm{Pl}}^2$ can be identified
with the gravitino mass, $m_{3/2} = W_0/M_{\textrm{Pl}}^2$.
At low energies, the Higgs operator in the K\"ahler potential
then induces an effective superpotential $W_\mu = g_H' m_{3/2} H_u H_d$,
which is nothing but the desired $\mu$ term with $\mu = g_H' m_{3/2}$.
Besides that, an additional contribution to the $\mu$ term may be generated in the course
of spontaneous SUSY breaking, if the K\"ahler potential should contain a coupling between
the operator $H_uH_d$ and the hidden SUSY breaking sector~\cite{Giudice:1988yz}.
In the remainder of this section, we will now mostly focus on $Z_N^R$ symmetries with $N \neq 4$.

%%%%%%%%%%%%%%%%%%%%%%%%%%%%%%%%%%%%%%%%%%%%%%%%%%%%%%%%%%%%%%%%%%%%%%%%%%%%%%%%%%%%%%%%%%%%%%%%%%%%

\subsubsection[Contributions to the $\mu$ term from spontaneous PQ breaking]
{Contributions to the \boldmath{$\mu$} term from spontaneous PQ breaking}

%%%%%%%%%%%%%%%%%%%%%%%%%%%%%%%%%%%%%%%%%%%%%%%%%%%%%%%%%%%%%%%%%%%%%%%%%%%%%%%%%%%%%%%%%%%%%%%%%%%%

Next, we note that sometimes already the spontaneous breaking of the PQ symmetry
entails the generation of a supersymmetric mass term for the MSSM Higgs doublets
$H_u$ and $H_d$, which, however, turns out to be too small in all viable cases.
The origin for this contribution to the MSSM $\mu$ term
are the following higher-dimensional operators in the
tree-level superpotential,
\begin{align}
W_\mu = \left(C_\mu^{(p)} \frac{P^p}{M_{\textrm{Pl}}^{p-1}} + C_\mu^{(\bar{p})}
\frac{\bar{P}^{\bar{p}}}{M_{\textrm{Pl}}^{\bar{p}-1}}\right) H_u H_d \,,
\label{eq:PQmu}
\end{align}
with $C_\mu^{(p)}$ and $C_\mu^{(\bar{p})}$ denoting dimensionless coupling constants of
$\nrmcal{O}(1)$.
Of course, these couplings are only allowed if they are compatible with the $Z_N^R$
symmetry, which is the case given that
\begin{align}
p \, r_P \modulo{N} - 2 \quad \textrm{and/or} \quad
\bar{p} \, r_{\bar{P}} \modulo{N} - 2 \,.
\label{eq:qcond}
\end{align}
We shall now assume for a moment that at least one of these two conditions can be satisfied.
In case only the first or the second condition can be fulfilled, let $q$ denote
the corresponding value of $p$ or $\bar{p}$.
If both conditions can be satisfied simultaneously,
$q$ shall denote the smaller of the two possible
powers, $q = \min\left\{p,\bar{p}\right\}$.
The spontaneous breaking of the PQ symmetry then induces a
supersymmetric mass $\mu$ for $H_u$ and $H_d$, which looks very similar to
the Dirac masses $m_{Q_i}$ for the extra quarks and anti-quarks in Eq.~\eqref{eq:mQi},
\begin{align}
\mu = \frac{C_\mu^{(q)}}{M_{\textrm{Pl}}^{q-1}}\left(\frac{\Lambda}{\sqrt{2}}\right)^q
\simeq  \bigg(\frac{C_\mu^{(q)}}{1}\bigg)\left(\frac{k}{4}\right)^q
\left(\frac{f_a}{10^{10}\,\textrm{GeV}}\right)^q
\times
\begin{cases}
2.0 \times 10^{10} \,\,\textrm{GeV} & ;\:\: q = 1 \\
6.6 \times 10^{2\phantom{0}}  \,\,\textrm{GeV} & ;\:\: q = 2 \\
3.6 \times 10^{-5} \,\textrm{GeV} & ;\:\: q = 3 \\
\phantom{10}\,\cdots & ; \:\: q = 4
\end{cases}
\,.
\end{align}
For $q = 1$, the generated $\mu$ term is, hence, dangerously large; for $q = 2$
it is of the desired order of magnitude; and for $q \geq 3$ it is drastically too small.
On the other hand, given our restrictions on the $R$ charge $r_P$
in Eqs.~\eqref{eq:rPcond5} and \eqref{eq:rPcond2},
we know that $q$ has to be at least $q=4$.
This means that, for $q = 1,2,3$, \textit{all} possible $R$ charges $r_P$ fulfilling at least one
of the two conditions in Eq.~\eqref{eq:qcond}
lead to an unwanted operator in $W_P$,
the superpotential of the extra scalar sector, cf.\ Eq.~\eqref{eq:WPQ}.
We thus conclude that the spontaneous breaking of the PQ symmetry
does not suffice to generate a $\mu$ term of the right order of magnitude.
For the last time, we are therefore led to extend our model.

%%%%%%%%%%%%%%%%%%%%%%%%%%%%%%%%%%%%%%%%%%%%%%%%%%%%%%%%%%%%%%%%%%%%%%%%%%%%%%%%%%%%%%%%%%%%%%%%%%%%

\subsubsection{Singlet extension of the MSSM Higgs sector}
\label{subsubsec:SMSSM}

%%%%%%%%%%%%%%%%%%%%%%%%%%%%%%%%%%%%%%%%%%%%%%%%%%%%%%%%%%%%%%%%%%%%%%%%%%%%%%%%%%%%%%%%%%%%%%%%%%%%

Extensions of the MSSM aiming at generating the $\mu$ term dynamically usually
couple the MSSM Higgs doublets to another chiral singlet $S$,
which acquires a VEV of order of the soft Higgs masses in the course of
electroweak symmetry breaking.
We will now adopt this approach and introduce a chiral singlet field $S$
with $R$ charge $r_S = - 2$, in order to allow for the operator $S \,H_u H_d$
in the superpotential.
As we will see in the following, this operator usually indeed
yields a $\mu$ term of the right order of magnitude. 

%%%%%%%%%%%%%%%%%%%%%%%%%%%%%%%%%%%%%%%%%%%%%%%%%%%%%%%%%%%%%%%%%%%%%%%%%%%%%%%%%%%%%%%%%%%%%%%%%%%%

Given the fact that the superpotential carries $R$ charge $r_W = 2$, the relation between
the gravitino mass and the VEV of the superpotential, $m_{3/2} = W_0/M_{\textrm{Pl}}^2$,
implies that $m_{3/2}$ should be regarded as a spurious field also carrying
$R$ charge $r_{3/2} = 2$.
After spontaneous $R$ symmetry breaking, the superpotential of the field $S$
hence contains the following terms,%
\begin{align}
N\neq4 \:: \quad W_S \supset g_H H_u H_d \,S + m_{3/2}^2\, S + m_S S^2
\:\:\left(+ \lambda_S S^3\right) \,, \label{eq:WSwoX}
\end{align}
where $m_S$ denotes a supersymmetric mass for the singlet field $S$ and
where the term in parenthesis is only allowed in the case of a discrete $Z_8^R$ symmetry.
In this section, we explicitly exclude the possibility of a $Z_4^R$ symmetry, because in this
case the $\mu$ term is already generated in the course of $R$ symmetry breaking,
cf.\ Sec.~\ref{subsubsec:GMmechanism}.
Besides that, for a $Z_4^R$ symmetry, the $R$ charge of the field $S$ would be
equivalent to $r_S =2 $, such that a tadpole term of order the Planck scale
would be allowed in the superpotential.
Such a large tadpole would then severely destabilize the electroweak scale.
By contrast, all other $Z_N^R$ symmetries successfully prevent the appearance
of a dangerously large tadpole term.
In fact, the only tadpole that we are able to generate for $N \neq 4$ arises
from the spontaneous breaking of $R$ symmetry and is of the size of the gravitino mass,
cf.\ Eq.~\eqref{eq:WSwoX}.

%%%%%%%%%%%%%%%%%%%%%%%%%%%%%%%%%%%%%%%%%%%%%%%%%%%%%%%%%%%%%%%%%%%%%%%%%%%%%%%%%%%%%%%%%%%%%%%%%%%%

Assuming a discrete $Z_3^R$ or $Z_6^R$ symmetry, the $R$ charge of $S^2$
is equivalent to $r_s = 2$ and $m_S$ is expected to be very large,
$m_S = g_S M_{\textrm{Pl}}$, where $g_S$ is a dimensionless constant
of $\nrmcal{O}(1)$ in general.
For all other $Z_N^R$ symmetries, the $S$ mass term is only allowed if, similarly
as for the gravitino mass, $m_S$ is interpreted as a spurious field, now with
$R$ charge $6$ instead of $2$.
On the supposition that only a single dynamical process is responsible for the generation
of $m_{3/2}$ and $m_S$, the $S$ mass then turns out to be heavily suppressed,
\begin{align}
N = 3,6 \:: \quad m_S = g_S M_{\textrm{Pl}} \,, \quad g_S \sim 1 \,; \qquad
N \neq 3,6 \:: \quad m_S \sim \frac{m_{3/2}^3}{M_{\textrm{Pl}}^2} \,.
\end{align}
For $N = 3,6$, the large supersymmetric mass $m_S$ hence
leads to a very small VEV of the field $S$, thereby causing our attempt to
dynamically generate the MSSM $\mu$ term to fail.
Only in case that, for one reason or another, the parameter $g_S$ is severely
suppressed, $g_S \ll 1$, such that $m_S \ll M_{\textrm{Pl}}$,
a $Z_3^R$ or a $Z_6^R$ symmetry may still be considered viable.
Otherwise, $Z_N^R$ symmetries with $N=3,6$ should be regarded disfavoured
within the context of our model.%
\footnote{Interestingly, these are just the two anomaly-free $Z_N^R$ symmetries
of the MSSM.
Now we see that they are most likely not compatible with the generation of the
$\mu$ term by means of an additional singlet field $S$.
This justifies once more
our approach to extend the particle content of the MSSM by a new quark sector
in such a way that the gauge anomalies of the $Z_N^R$ symmetry are always canceled,
independently of the value of $N$.}
By contrast, in the case of all other symmetries, \textit{i.e.}\
$Z_N^R$ symmetries with $N = 5$ or $N\geq 7$,
the mass of the field $S$ is completely
negligible, which is why we will omit from now on.
Thus, as far as the generation of the $\mu$ term in our model is concerned,
we will assume the following terms in the superpotential,
\begin{align}
W_S \supset
\begin{cases}
g_H H_u H_d \,S + m_{3/2}^2\, S + m_S S^2 & ;\:\: N = 3,6 \\
g_H' m_{3/2} H_u H_d & ;\:\: N = 4 \phantom{\Big)}\\
g_H H_u H_d \,S + m_{3/2}^2\, S + \lambda_S S^3 & ;\:\: N = 8 \phantom{\Big)}\\
g_H H_u H_d \,S + m_{3/2}^2\, S & ;\:\: N\neq3,4,6,8 
\end{cases} \,.
\label{eq:WSmu}
\end{align}

%%%%%%%%%%%%%%%%%%%%%%%%%%%%%%%%%%%%%%%%%%%%%%%%%%%%%%%%%%%%%%%%%%%%%%%%%%%%%%%%%%%%%%%%%%%%%%%%%%%%

Together with the scalar masses and couplings in the soft SUSY-breaking
Lagrangian, the interactions in Eq.~\eqref{eq:WSmu} result in a scalar potential
that is minimized for $\left<H_{u,d}\right> = v_{u,d}$ and
$\left<S\right> = \mu / g_H$, whereby our
solution to the $\mu$ problem is completed.
The actual value of the $\mu$ parameter depends in a complicated way
on the couplings in the superpotential $W_S$ as well as on the soft
parameters for the fields $H_{u,d}$ and $S$. 
For our purposes, it will however suffice to treat $\mu$ as an effectively free
parameter that is allowed to vary within some range.

%%%%%%%%%%%%%%%%%%%%%%%%%%%%%%%%%%%%%%%%%%%%%%%%%%%%%%%%%%%%%%%%%%%%%%%%%%%%%%%%%%%%%%%%%%%%%%%%%%%%

Some of the expressions for $W_S$ in Eq.~\eqref{eq:WSmu} are reminiscent of the superpotential
of other extensions of the MSSM that successfully generate the $\mu$ term by means of a singlet
field $S$.
For instance, assuming a discrete $Z_8^R$ symmetry and neglecting the
tadpole term, the superpotential in Eq.~\eqref{eq:WSwoX} corresponds to the Higgs superpotential
of the next-to-minimal supersymmetric standard model (NMSSM).%
\footnote{For reviews of the NMSSM, cf.\ for instance Ref.~\cite{Maniatis:2009re}.}
Conversely, assuming a $Z_N^R$ symmetry with $N\neq3,4,6,8$ 
and taking the tadpole term into account, the superpotential $W_S$ coincides
with the effective Higgs superpotential of the new MSSM
(nMSSM)~\cite{Panagiotakopoulos:1999ah} as well as with
the effective Higgs superpotential of the PQ-invariant
extension of the NMSSM (PQ-NMSSM)~\cite{Jeong:2011jk}.
While in the nMSSM, the shape of the $S$ superpotential is fixed by means
of a discrete $R$ symmetry, similarly as in our model, the PQ-NMSSM invokes
a PQ symmetry by hand in order to ensure the absence of further couplings of the
singlet field $S$.
On the other hand, the PQ-NMSSM features a PQ singlet field, similar to our singlet
fields $P$ and $\bar{P}$, which couples to the field $S$.
By contrast, such a PQ-breaking field is absent in the nMSSM.
But as the mixing between the singlet $S$ and the PQ-breaking sector
is always suppressed by powers of the PQ scale $\Lambda$, this has basically no
effect on the low-energy phenomenology of the Higgs and neutralino sectors.

%%%%%%%%%%%%%%%%%%%%%%%%%%%%%%%%%%%%%%%%%%%%%%%%%%%%%%%%%%%%%%%%%%%%%%%%%%%%%%%%%%%%%%%%%%%%%%%%%%%%

As for the expected low-energy signatures of these two sectors, our model
thus makes the same predictions as the PQ-NMSSM and the nMSSM.
This means, in particular, that our model predicts a fifth neutralino
mostly consisting of the singlino, which only receives a small
mass from mixing with the neutral Higgsinos.
Among all superparticles that either directly belong to the MSSM or
that at least share some renormalizable interaction with it, the singlino-like
neutralino is hence expected to be the lightest.
Furthermore, at small values of $\tan\beta$, the decay of the standard
model-like Higgs boson into two singlino-like neutralinos
might represent the dominant Higgs decay mode.
Such a scenario is already constrained by the search for invisible Higgs
decays by the ATLAS experiment at the LHC~\cite{ATLAS:2013pma} and will be
further tested as data taking at the LHC is resumed.
Another interesting feature of our model is that, independently of $\tan\beta$,
the Higgs boson mass receives positive corrections of the order of a few GeV
from singlino loops, provided that the Higgsinos are lighter than all other
superparticles of the MSSM.
Finally, we mention that our model features a series of interesting
implications for cosmology~\cite{Jeong:2011jk,Menon:2004wv}.

%%%%%%%%%%%%%%%%%%%%%%%%%%%%%%%%%%%%%%%%%%%%%%%%%%%%%%%%%%%%%%%%%%%%%%%%%%%%%%%%%%%%%%%%%%%%%%%%%%%%

The operators on the right-hand side of Eq.~\eqref{eq:WSwoX} are the
only terms in the superpotential of the singlet field $S$ playing a role
in the generation of the $\mu$ term.
Besides that, the field $S$ participates, of course, also in a series of other
interactions.
As the field $X$ carries the same $R$ charge as the gravitino mass,
$r_X = r_{3/2} =2$, the tadpole term in Eq.~\eqref{eq:WSwoX} has, in particular,
to be supplemented by the operators $m_{3/2}XS$ and $X^2S$.
The full superpotential of the field $S$ thus reads
\begin{align}
N\neq4 \:: \quad
W_S = g_H H_u H_d \,S + m_{3/2}^2\, S + g_X \, m_{3/2} \, X S + g_{X^2} X^2 S
\:\:\left(+ m_S S^2\right) \:\left(+ \lambda_S S^3\right) \:\: + .. \,. 
\label{eq:WS}
\end{align}
Here, $g_X$ and $g_{X^2}$ are again dimensionless coupling constants of $\nrmcal{O}(1)$ and
the dots after the plus sign indicate higher-dimensional non-renormalizable terms.
Given the fact that the field $X$ does not carry any PQ charge, the coupling between
$X$ and $S$ immediately implies that the field $S$ also does not transform under PQ
rotations, $q_S = 0$.
This proves in turn our statement in Sec.~\ref{subsec:PQsym} that the
PQ charges of $H_u$ and $H_d$ must sum to zero, $q_{H_u} + q_{H_d} = 0$.
Another important consequence of the operators $m_{3/2}XS$ and $X^2S$ in
Eq.~\eqref{eq:WS} is that, at the supersymmetric level, the scalar field VEVs
in Eq.~\eqref{eq:XPPvevs} no longer represent the unique vacuum configuration.
The PQ-breaking vacuum, in which $\langle P \bar{P}\rangle = \Lambda^2/2$,
is now continuously connected to a family of degenerate vacua,
all of which are characterized by the fact that they fulfill the
condition $\langle P\bar{P}\rangle - g_X/\kappa\,m_{3/2} \langle S\rangle = \Lambda^2/2$.
However, this vacuum degeneracy is fortunately lifted by the soft SUSY breaking masses for
the scalar fields $P$, $\bar{P}$ and $S$, such that, also in the presence of the
operators $m_{3/2}XS$ and $X^2S$, the vacuum configuration of interest, \textit{i.e.}\
$\langle P \bar{P}\rangle = \Lambda^2/2$ together with $\langle S \rangle \sim m_{3/2}$
and $\langle X \rangle \sim m_{3/2}$, corresponds to a local minimum.
Besides that, the new interactions between $S$ and $X$ also lead to a second
local minimum at $\langle X \rangle \sim m_{3/2}^{1/3} \Lambda^{2/3}$, 
$\langle XS \rangle \sim \Lambda^2$ and $\langle P\bar{P}\rangle = 0$.
The energy of this vacuum is, however, much higher than the one of
the PQ-breaking vacuum and hence, we expect the fields $P$, $\bar{P}$, $S$
and $X$ to settle in the PQ-breaking vacuum at low energies,
\begin{align}
\langle P \rangle  = \frac{\Lambda}{\sqrt{2}} e^{A/\Lambda} \,, \quad
\langle \bar{P} \rangle  = \frac{\Lambda}{\sqrt{2}} e^{-A/\Lambda} \,, \quad
\langle S \rangle  \sim m_{3/2} \,, \quad
\langle X \rangle  \sim m_{3/2} \,.
\end{align}

%%%%%%%%%%%%%%%%%%%%%%%%%%%%%%%%%%%%%%%%%%%%%%%%%%%%%%%%%%%%%%%%%%%%%%%%%%%%%%%%%%%%%%%%%%%%%%%%%%%%

\subsubsection{Decay of the extra matter fields into MSSM particles}
\label{subsubsec:Qdecay}

%%%%%%%%%%%%%%%%%%%%%%%%%%%%%%%%%%%%%%%%%%%%%%%%%%%%%%%%%%%%%%%%%%%%%%%%%%%%%%%%%%%%%%%%%%%%%%%%%%%%

The extension of the MSSM Higgs sector by the singlet field $S$
completes the field content of our model.
We are therefore almost ready to turn to the phenomenological constraints on
our model and discuss which values of $N$, $n$ and $k$ allow for a sufficient
protection of the PQ symmetry.
But before we are able to do so, we have to take care of one last detail:
the new quarks and anti-quarks are thermally produced in the early universe,
which potentially results in serious cosmological problems.
If the extra quarks are stable, they might be produced so abundantly that they
overclose the universe.
On the other hand, if they are unstable, their late-time decays might alter
the primordial abundances of the light elements produced during big bang nucleosynthesis (BBN),
so that these are no longer in accordance with the observational data.
To avoid these problems, we require a coupling between the extra quark sector
and the MSSM fields, such that the extra quarks quickly decay after their production.
So far, we only had to fix the $R$ charge $r_{Q\bar{Q}}$ of the quark 
pair operator $Q\bar{Q}$, cf.\ Eq.~\eqref{eq:rQQ}.
The $R$ charges $r_Q$ and $r_{\bar{Q}} = r_{Q\bar{Q}} - r_Q$ of the individual quarks
and anti-quarks have by contrast remained unspecified up to now.
By choosing a particular value for the $R$ charge $r_Q$, we are therefore now able to
pinpoint the operator by means of which the extra quarks shall couple to the MSSM.

%%%%%%%%%%%%%%%%%%%%%%%%%%%%%%%%%%%%%%%%%%%%%%%%%%%%%%%%%%%%%%%%%%%%%%%%%%%%%%%%%%%%%%%%%%%%%%%%%%%%

Under the SM gauge group, the anti-quark fields $\bar{Q}_i$ transform
in the same representation as the MSSM $\mathbf{5}_i^*$ multiplets.
An obvious possibility to couple the new quarks to the MSSM thus is
to allow for the operator $\bar{Q}_i \mathbf{10}_j H_d$ in the superpotential,
in which case the anti-quarks ought to carry the same $R$ charge
as the $\mathbf{5}_i^*$ multiplets, $r_{\bar{Q}} = r_{\mathbf{5}}^*$.
The only way in which the extra anti-quark fields then distinguish themselves
from the MSSM $\mathbf{5}_i^*$ multiplets is their coupling to
the extra quark fields $Q_i$.
More precisely, starting out with a superpotential containing
the operators $P^n Q_i \left(\bar{Q}_i^\prime + \mathbf{5}_i^{*\prime}\right)$
and $\left(\bar{Q}_i^\prime + \mathbf{5}_i^{*\prime}\right)\mathbf{10}_j H_d$,
we can always perform a field transformation
$\left(\bar{Q}_i^\prime,\mathbf{5}_i^{*\prime}\right) \rightarrow
\left(\bar{Q}_i,\mathbf{5}_i^*\right)$, such that, \textit{by definition}, the
MSSM $\mathbf{5}_i^*$ multiplets do not couple to the extra quark fields $Q_i$
and only the operators $P^n Q_i\bar{Q}_i$ and
$\left(\bar{Q}_i + \mathbf{5}_i^*\right)\mathbf{10}_j H_d$
remain in the superpotential.
The operator $\bar{Q}_i \mathbf{10}_j H_d$ then mixes the quarks and
leptons respectively contained in the $\bar{Q}_i$ and $\mathbf{5}_i^*$ multiplets,
which potentially gives rise to dangerous flavour-changing neutral-current (FCNC)
interactions.
In the case of very heavy extra quarks, \textit{i.e.}\ for $n=1$, we however do not have to
worry about FCNC processes as these are
always automatically suppressed by the large quark masses.
Only for $n=2$, we have to pay attention that the mixing between the MSSM fermions
and the new matter fields does not become too large.
For extra quarks with masses around $1\,\textrm{TeV}$, we have for instance
to require that the Yukawa coupling constants
belonging to the operator $\bar{Q}_i \mathbf{10}_j H_d$
are at most of $\nrmcal{O}\left(10^{-2}\right)$~\cite{Cacciapaglia:2011fx}.
This is a rather mild constraint, which may be easily satisfied in a large class of
flavour models.

%%%%%%%%%%%%%%%%%%%%%%%%%%%%%%%%%%%%%%%%%%%%%%%%%%%%%%%%%%%%%%%%%%%%%%%%%%%%%%%%%%%%%%%%%%%%%%%%%%%%

Coupling the new quark sector to the MSSM via the operator
$\bar{Q}_i \mathbf{10}_j H_d$ is therefore certainly a viable option.
The $R$ and PQ charges of the extra quarks and anti-quarks
are then given by
\begin{align}
r_Q \modulo{N} r_{Q\bar{Q}} - r_{\bar{Q}} \,, \quad
r_{\bar{Q}} \modulo{N} r_{\mathbf{5}^*} \,; \quad
q_Q = q_{Q\bar{Q}} - q_{\bar{Q}} \,, \quad q_{\bar{Q}} = q_{\mathbf{5}^*} \,.
\end{align}
Making use of our results for $r_{\mathbf{5}^*}$ and $r_{Q\bar{Q}}$
in Eqs.~\eqref{eq:MSSMRcharges} and \eqref{eq:rQQ}, we find for
$r_Q$ and $r_{\bar{Q}}$,
\begin{align}
r_Q \modulo{N} \frac{13}{5} + \frac{6}{k}
+\left[\frac{\ell_Q}{k} -\frac{\ell}{2}\right]N\,, \quad
r_{\bar{Q}} \modulo{N} -\frac{3}{5} + \ell \frac{N}{2} \,.
\label{eq:rQrQbar}
\end{align}
Likewise, employing our results for $q_{Q\bar{Q}}$ in Eq.~\eqref{eq:qPPQQ}
and setting $q_{\mathbf{5}^*}$ to $0$, we obtain for $q_Q$ and $q_{\bar{Q}}$,
\begin{align}
q_Q = -n \,, \quad q_{\bar{Q}} = 0 \,.
\label{eq:qQqQbar}
\end{align}
Finally, we also note that the operator $\bar{Q}_i \mathbf{10}_j H_d$ explicitly
breaks the vectorial global symmetry in the extra quark sector, such that
the PQ symmetry remains as the only global Abelian symmetry,
\begin{align}
U(1)_{\textrm{PQ}} \times U(1)_Q^V \quad\rightarrow\quad U(1)_{\textrm{PQ}} \,.
\end{align}
Given our choice for the MSSM PQ charges in Sec.~\ref{subsec:PQsym}, we are
now eventually able to determine the relation between the generators of
the three global Abelian symmetries $U(1)_P$, $U(1)_Q^V$ and $U(1)_Q^A$
on the one hand and the PQ generator on the other hand.
Denoting these generators by $P$, $V$, $A$ and $PQ$, respectively, we find
\begin{align}
PQ = P - \frac{n}{2} \left(V+A\right) \,.
\end{align}

%%%%%%%%%%%%%%%%%%%%%%%%%%%%%%%%%%%%%%%%%%%%%%%%%%%%%%%%%%%%%%%%%%%%%%%%%%%%%%%%%%%%%%%%%%%%%%%%%%%%

In order to avoid the above constraint on the Yukawa couplings associated with
$\bar{Q}_i \mathbf{10}_j H_d$ in the case $n=2$, one may alternatively consider
couplings of the new quarks fields to the MSSM via higher-dimensional operators.
Naively, there are three different choices for such an operator, namely
$S \bar{Q}_i \mathbf{10}_j H_d$, $P \bar{Q}_i \mathbf{10}_j H_d$
and $\bar{P}\bar{Q}_i \mathbf{10}_j H_d$.
Replacing the singlet fields $S$, $P$ and $\bar{P}$ in these operators by their respective VEVs,
all of them turn again into $\bar{Q}_i \mathbf{10}_j H_d$, now, however,
with coupling constants that are naturally suppressed compared to unity.
Allowing for any of these operators rather than $\bar{Q}_i \mathbf{10}_j H_d$, we
therefore do not have to fear dangerous FCNC processes
due to the mixing between the $\bar{Q}_i$ and $\mathbf{5}_i^*$
multiplets.
Meanwhile, $S \bar{Q}_i \mathbf{10}_j H_d$ and
$P \bar{Q}_i \mathbf{10}_j H_d$ do not represent viable operators
by means of which the new quarks could couple to the MSSM after all.
In the case of $S \bar{Q}_i \mathbf{10}_j H_d$, the extra quarks
do not decay sufficiently fast in the early universe.
The operator  $S \bar{Q}_i \mathbf{10}_j H_d$ furnishes the new quarks with
two-body and three-body decay channels,
the partial decay rates of which can roughly be estimated as
\begin{align}
\Gamma\left(\bar{Q}_i\rightarrow q_j H_d, e_j^c H_d\right) \sim 
\frac{1}{8\pi} \left(\frac{\mu/g_H}{M_{\textrm{Pl}}}\right)^2 m_{Q_i} \sim & \:
\begin{cases}
10^{2} \,\textrm{s}^{-1} & ; \:\: m_{Q_i} = 10^{10}\,\textrm{GeV} \\
10^{-5} \,\textrm{s}^{-1} & ;\:\: m_{Q_i} = 1\,\textrm{TeV}
\end{cases} \,, \label{eq:Qdec}\\
\Gamma\left(\bar{Q}_i\rightarrow S q_j H_d, S e_j^c H_d\right) \sim
\frac{1}{128\pi^3} \frac{m_{Q_i}^3}{M_{\textrm{Pl}}^2} \sim & \:
\begin{cases}
10^{14} \,\textrm{s}^{-1} & ; \:\: m_{Q_i} = 10^{10}\,\textrm{GeV} \\
10^{-7} \,\textrm{s}^{-1} & ;\:\: m_{Q_i} = 1\,\textrm{TeV}
\end{cases} \,. \nonumber
\end{align}
Here, we have set the VEV of the scalar field $S$ to $\left<S\right> = \mu/g_H = 1\,\textrm{TeV}$.
For $n=2$, the extra quarks thus decay only after
BBN, which begins at a cosmic time of around $1\,\textrm{s}$ and lasts for roughly
$10^3\,\textrm{s}$.
Moreover, if we choose the $R$ charge $r_Q$, such that $P \bar{Q}_i \mathbf{10}_j H_d$ is
contained in the superpotential, also $P^{n-1}Q_i\mathbf{5}_j^*$ is allowed.
Unlike in our first case, in which we considered $\bar{Q}_i \mathbf{10}_j H_d$, this operator
cannot be simply eliminated by a field re-definition.
Together with $P^n\left(Q\bar{Q}\right)_i$, it instead leads to an unacceptably strong
mixing between the $\bar{Q}_i$ and $\mathbf{5}_i^*$ multiplets.

%%%%%%%%%%%%%%%%%%%%%%%%%%%%%%%%%%%%%%%%%%%%%%%%%%%%%%%%%%%%%%%%%%%%%%%%%%%%%%%%%%%%%%%%%%%%%%%%%%%%

\begin{table}[t]
\begin{center}
\begin{doublespacing}
\begin{tabular}{r||cccc}
 & $SU(5)$ & $P_M$ & $Z_N^R$ & $U(1)_{\textrm{PQ}}$ \\\hline\hline
$(q,u^c,e^c)$ & $\mathbf{10}$ & $-$ & $\frac{1}{5} + \ell\frac{N}{2}\phantom{\Big]}$ & $0$ \\
$(d^c,\ell)$ & $\mathbf{5}^*$ & $-$ & $-\frac{3}{5} + \ell\frac{N}{2}\phantom{\Big]}$ & $0$  \\
$(n^c)$ & $\mathbf{1}$ & $-$ & $1 + \ell\frac{N}{2}\phantom{\Big]}$ & $0$ \\
$H_u$ &  $\mathbf{2}_L$ & $+$ & $2-\frac{2}{5}\phantom{\Big]}$ & $0$ \\
$H_d$ & $\mathbf{2}_L$ & $+$ & $2+\frac{2}{5}\phantom{\Big]}$ & $0$ \\
$Q$ & $\mathbf{5}$ & $-$ & $\frac{13}{5} + \frac{(n+1)6}{nk}
-\left[\frac{\ell}{2} + \frac{\ell_P}{n} - \frac{(n+1)\ell_Q}{nk}\right]N$ & $-n-1$ \\
$\bar{Q}$ & $\mathbf{5}^*$ & $-$ & $-\frac{3}{5} - \frac{6}{nk} +
\left[\frac{\ell}{2} + \frac{\ell_P}{n} - \frac{\ell_Q}{nk}\right]N$ & $1$ \\
$P$ & $\mathbf{1}$ &  $+$ & $-\frac{6}{nk} + (k \, \ell_P - \ell_Q)\frac{N}{nk}$ & $1$ \\
$\bar{P}$ & $\mathbf{1}$ & $+$ & $\frac{6}{nk} - (k \, \ell_P - \ell_Q)\frac{N}{nk}$ & $-1$ \\
$X$ & $\mathbf{1}$ & $+$ & $2\phantom{\Big]}$ & $0$ \\
$S$ & $\mathbf{1}$ & $+$ & $-2\phantom{\Big]}$ & $0$ 
\end{tabular}
\end{doublespacing}
\begin{onehalfspacing}
\caption{Summary of the possible charge assignments in our model assuming
that the extra quarks couple to the MSSM via the operator
$\bar{P} \bar{Q}_i \mathbf{10}_j H_d$.
If the extra quarks should instead couple to the MSSM via
$\bar{Q}_i \mathbf{10}_j H_d$, the values given in Eqs.~\eqref{eq:rQrQbar}
and \eqref{eq:qQqQbar} must be used for the $R$ and PQ charges of the fields
$Q_i$ and $\bar{Q}_i$.
The $\mathbf{2}_L$ in the column indicating the $SU(5)$ representations
denote $SU(2)_L$ doublets.
All $R$ charges are only defined up to the addition of integer multiples of $N$.
The MSSM $R$ charges can additionally be changed by acting on them with $Z_5$
transformations.
$N\geq 3$; $n=1,2$; $k\geq1$; $\ell$; $\ell_P$ and $\ell_Q$ are all integers.}
\label{tab:charges} 
\end{onehalfspacing}
\end{center}
\end{table}

%%%%%%%%%%%%%%%%%%%%%%%%%%%%%%%%%%%%%%%%%%%%%%%%%%%%%%%%%%%%%%%%%%%%%%%%%%%%%%%%%%%%%%%%%%%%%%%%%%%%

The only remaining option therefore is to allow for
$\bar{P} \bar{Q}_i \mathbf{10}_j H_d$.
In this case, the superpotential also features $P^{n+1}Q_i\mathbf{5}_j^*$,
which cannot be transformed away as well, but which fortunately results in the mixing between
the $\bar{Q}_i$ and $\mathbf{5}_j^*$ multiplets being suppressed by a factor of
$\nrmcal{O}\left(\Lambda/M_{\textrm{Pl}}\right)$.
Furthermore, $\bar{P} \bar{Q}_i \mathbf{10}_j H_d$ gives rise to two-body decays
of the extra quarks at a fast rate.
After replacing the scalar field $\bar{P}$ by $\Lambda/\sqrt{2}$, we obtain
\begin{align}
\Gamma\left(\bar{Q}_i\rightarrow q_j H_d, e_j^c H_d\right) \sim 
\frac{1}{16\pi} \left(\frac{\Lambda}{M_{\textrm{Pl}}}\right)^2 m_{Q_i} \sim
\begin{cases}
10^{16} \,\textrm{s}^{-1} & ; \:\: m_{Q_i} = 10^{10}\,\textrm{GeV} \\
10^{10} \,\textrm{s}^{-1} & ;\:\: m_{Q_i} = 1\,\textrm{TeV}
\end{cases} \,,
\end{align}
where we have chosen the PQ-breaking scale $\Lambda$ such that
it respectively results in $m_{Q_i} = 10^{10}\,\textrm{GeV}$ or
$m_{Q_i} = 1\,\textrm{TeV}$, if $n$ is set to $1$ or $2$, cf.\ Eq.~\eqref{eq:mQi}.
Similarly to $S \bar{Q}_i \mathbf{10}_j H_d$, the operator
$\bar{P} \bar{Q}_i \mathbf{10}_j H_d$ also entails three-body decays,
which, however, always proceed at a slower rate than the corresponding
two-body decays, cf.\ Eq.~\eqref{eq:Qdec}.
A coupling of the extra quarks to the MSSM via $\bar{P} \bar{Q}_i \mathbf{10}_j H_d$
is hence a viable alternative to the coupling via $\bar{Q}_i \mathbf{10}_j H_d$.
A particular advantage of this coupling is that we do not have to require
suppressed Yukawa couplings, if $n=2$.
On the other hand, the charges of the extra anti-quarks now do not coincide
any more with the charges of the MSSM $\mathbf{5}_i^*$ multiplets.
The $R$ and PQ charges of the new quark and anti-quark fields are instead given by
\begin{align}
r_Q \modulo{N} r_{Q\bar{Q}} - r_{\bar{Q}} \,, \quad
r_{\bar{Q}} \modulo{N} r_{\mathbf{5}^*} - r_{\bar{P}} \,; \quad
q_Q = q_{Q\bar{Q}} - q_{\bar{Q}} \,, \quad q_{\bar{Q}} = - q_{\bar{P}} \,.
\end{align}
Our results for $r_{\mathbf{5}^*}$, $r_{Q\bar{Q}}$ and $r_{\bar{P}}$
in Eqs.~\eqref{eq:MSSMRcharges}, \eqref{eq:rQQ} and \eqref{eq:rPbar} therefore provide us with
\begin{align}
r_Q \modulo{N} \frac{13}{5} + \frac{(n+1)6}{nk}
-\left[\frac{\ell}{2} + \frac{\ell_P}{n} - \frac{(n+1)\ell_Q}{nk}\right]N\,, \quad
r_{\bar{Q}} \modulo{N} -\frac{3}{5} - \frac{6}{nk} +
\left[\frac{\ell}{2} + \frac{\ell_P}{n} - \frac{\ell_Q}{nk}\right]N
\end{align}
Similarly, making use of the fact that $q_{\bar{P}} = -1$ and $q_{Q\bar{Q}} = -n$,
cf.\ Eq.~\eqref{eq:qPPQQ}, we find for $q_Q$ and $q_{\bar{Q}}$,
\begin{align}
q_Q = -n -1 \,, \quad q_{\bar{Q}} = 1 \,.
\end{align}
Combining this result with our choice for the MSSM PQ charges in Sec.~\ref{subsec:PQsym},
the relation between the four Abelian generators $P$, $V$, $A$ and $PQ$ now turns out
to be
\begin{align}
PQ = P - V - \frac{n}{2} \left(V+A\right) \,.
\end{align}

%%%%%%%%%%%%%%%%%%%%%%%%%%%%%%%%%%%%%%%%%%%%%%%%%%%%%%%%%%%%%%%%%%%%%%%%%%%%%%%%%%%%%%%%%%%%%%%%%%%%

\medskip

These findings complete the construction of our model.
To sum up, in this section, we have introduced (i) the field content of the MSSM along with
three generations of right-handed neutrinos, (ii) $k$ pairs of extra quarks and anti-quarks
in order to render the discrete $R$ symmetry anomaly-free, (iii) an additional singlet
sector in order to provide masses to the new quarks and anti-quarks
and (iv) a singlet field $S$ in order to dynamically generate the MSSM $\mu$ term.
The charges of all these fields are summarized in Tab.~\ref{tab:charges}.

%%%%%%%%%%%%%%%%%%%%%%%%%%%%%%%%%%%%%%%%%%%%%%%%%%%%%%%%%%%%%%%%%%%%%%%%%%%%%%%%%%%%%%%%%%%%%%%%%%%%

\section{Phenomenological constraints}
\label{sec:constraints}

%%%%%%%%%%%%%%%%%%%%%%%%%%%%%%%%%%%%%%%%%%%%%%%%%%%%%%%%%%%%%%%%%%%%%%%%%%%%%%%%%%%%%%%%%%%%%%%%%%%%

The MSSM extension presented in the previous section is subject to a variety
of phenomenological constraints.
As we have already seen in Sec.~\ref{subsubsec:Qmassscale}, the positive integer $n$
can, for instance, only be $1$ or $2$, since otherwise the extra quarks
would always have masses below the electroweak scale.
Besides that, \textit{i.e.}\ besides the lower
bound on the masses of the new quarks, we also have to ensure
(i) that, despite our extension of the MSSM particle
content, the unification of the SM gauge coupling constants still occurs
at the perturbative level, (ii) that operators explicitly breaking the PQ symmetry do not
induce shifts in the QCD vacuum angle larger than $10^{-10}$ as well as (iii) that
the axion decay constant takes a value
within the experimentally allowed window, cf. Eq.~\eqref{eq:window}.
In the next two subsections, we will now discuss these constraints in turn
and show how they allow us to single out the phenomenologically viable combinations
of $N$, $n$ and $k$ along with corresponding upper and lower bounds on $f_a$.

%%%%%%%%%%%%%%%%%%%%%%%%%%%%%%%%%%%%%%%%%%%%%%%%%%%%%%%%%%%%%%%%%%%%%%%%%%%%%%%%%%%%%%%%%%%%%%%%%%%%

\subsection{Gauge coupling unification}
\label{subsec:unification}

%%%%%%%%%%%%%%%%%%%%%%%%%%%%%%%%%%%%%%%%%%%%%%%%%%%%%%%%%%%%%%%%%%%%%%%%%%%%%%%%%%%%%%%%%%%%%%%%%%%%

The new quark and anti-quark fields contribute to the beta functions of the
SM gauge coupling constants and thus cause a change in the value
$g_{\textrm{GUT}}$ at which these coupling constants unify at high energies.
The more extra quark pairs we add to the MSSM particle content, the higher
$g_{\textrm{GUT}}$ turns out to be, which provides us with a means to constrain
the allowed number of extra quark pairs $k$ from above.
For given masses $m_{Q_i}$ of the new quarks, we define the maximal viable
number of extra quark pairs $k_{\textrm{max}}$ such that
\begin{align}
g_{\textrm{GUT}} \left(m_{Q_i}, k=k_{\textrm{max}}\right) \leq \sqrt{4\pi} \,,\quad
g_{\textrm{GUT}} \left(m_{Q_i}, k=k_{\textrm{max}} + 1\right) > \sqrt{4\pi} \,,\quad
k_{\textrm{max}} = k_{\textrm{max}}\left(m_{Q_i}\right) \,.
\label{eq:kmax}
\end{align}

%%%%%%%%%%%%%%%%%%%%%%%%%%%%%%%%%%%%%%%%%%%%%%%%%%%%%%%%%%%%%%%%%%%%%%%%%%%%%%%%%%%%%%%%%%%%%%%%%%%%

\begin{figure}
\begin{center}
\includegraphics[width=0.7\textwidth]{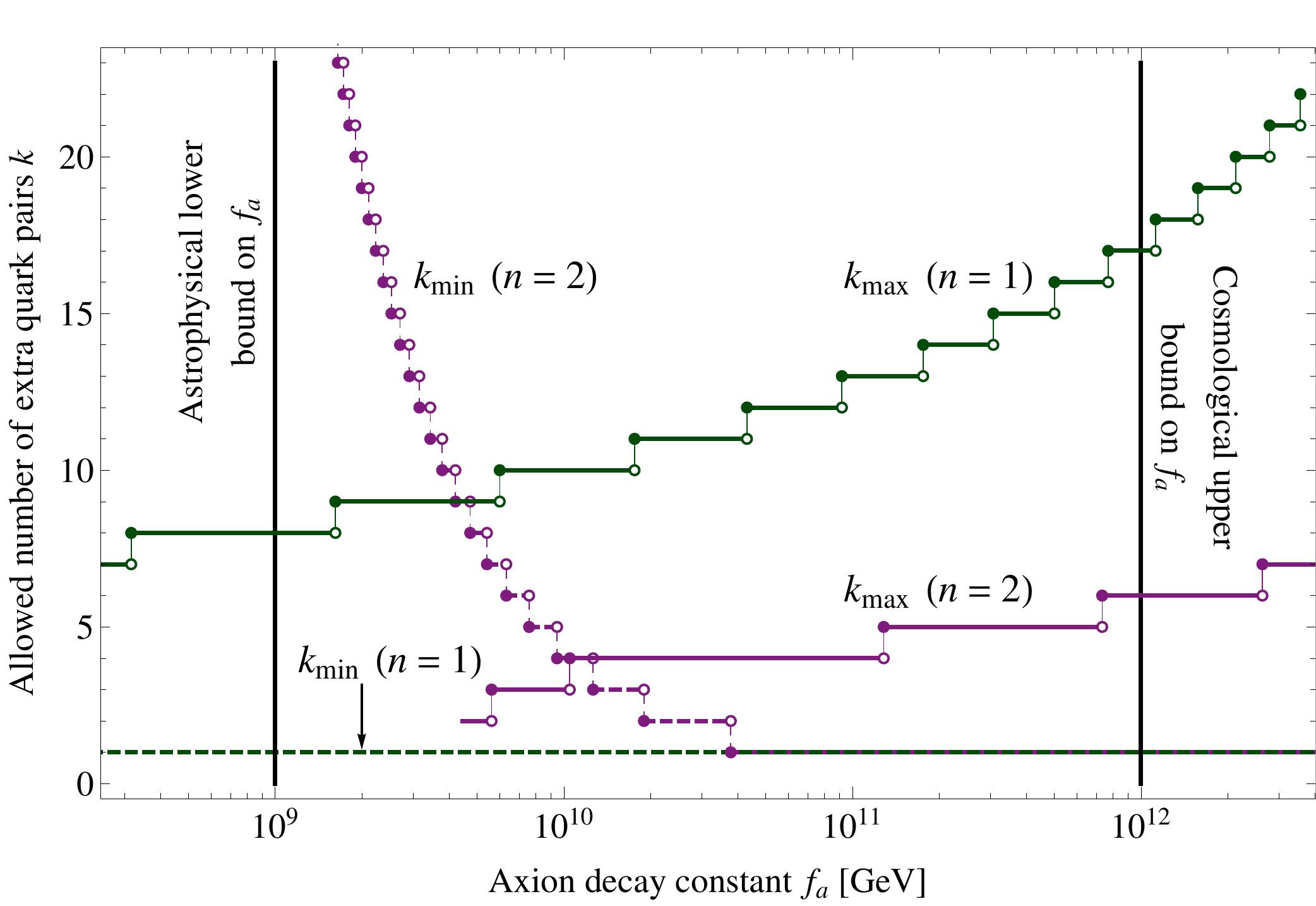}
\caption{Constraints on the number of extra quark pairs $k$ for $n=1$ and $n=2$, respectively.
The lower bounds are due to the experimental lower bound on the mass of new heavy down-type quarks;
the upper bounds derive from the requirement of perturbative gauge coupling unification.}
\label{fig:kbounds}
\end{center}
\end{figure}

In order to determine $k_{\textrm{max}}$ in dependence of the heavy quark mass spectrum,
we make the simplifying approximation that
all new quark flavours have the same mass, $M_Q = m_{Q_i}$, where
$M_Q = \left(\Lambda/\sqrt{2}/M_{\textrm{Pl}}\right)^n M_{\textrm{Pl}}$,
cf.\ Eq.~\eqref{eq:mQi}.
At the same time, we assume that all superparticles share a common soft SUSY breaking mass
$M_{\textrm{SUSY}}$ of $1\,\textrm{TeV}$.
When solving the renormalization group equations of the SM gauge
couplings for energy scales $\mu$ ranging from the $Z$ boson mass $M_Z = 91.2\,\textrm{GeV}$
to the GUT scale $M_{\textrm{GUT}} = 2\times 10^{16}\,\textrm{GeV}$,
we then have to distinguish between two different scenarios:
\begin{itemize}
\item If $M_Q > M_{\textrm{SUSY}}$, we use the SM one-loop beta functions for
$M_Z \leq \mu < M_{\textrm{SUSY}}$, the MSSM one-loop beta functions for
$M_{\textrm{SUSY}} \leq \mu < M_Q$ and the two-loop beta functions of the MSSM plus
the extra quark multiplets in the NSVZ scheme~\cite{Novikov:1983uc} for
$M_Q \leq \mu \leq M_{\textrm{GUT}}$.
\item If $M_Q \leq M_{\textrm{SUSY}}$, we use the SM one-loop beta functions for
$M_Z \leq \mu < M_Q$, the one-loop beta functions of the standard model plus
the extra fermionic quarks for $M_Q \leq \mu < M_{\textrm{SUSY}}$ and the two-loop beta
functions of the MSSM plus the extra quark multiplets, \textit{i.e.}\ plus the extra
fermionic \textit{and scalar} quarks, in the NSVZ scheme for
$M_{\textrm{SUSY}} \leq \mu \leq M_{\textrm{GUT}}$.
\end{itemize}
Given the solutions of the renormalization group equations, we are able to determine
$k_{\textrm{max}}$ as a function of $M_Q$ according to Eq.~\eqref{eq:kmax}.
The relation between the PQ scale $\Lambda$ and the axion decay constant $f_a$
in Eq.~\eqref{eq:LQCDeff} then provides us with $k_{\textrm{max}}$ as a
function of $f_a$.
The result of our calculation is presented in Fig.~\ref{fig:kbounds},
which displays $k_{\textrm{max}}$ as a function of $f_a$ for $n=1$ and $n=2$, respectively.

%%%%%%%%%%%%%%%%%%%%%%%%%%%%%%%%%%%%%%%%%%%%%%%%%%%%%%%%%%%%%%%%%%%%%%%%%%%%%%%%%%%%%%%%%%%%%%%%%%%%

%%%%%%%%%%%%%%%%%%%%%%%%%%%%%%%%%%%%%%%%%%%%%%%%%%%%%%%%%%%%%%%%%%%%%%%%%%%%%%%%%%%%%%%%%%%%%%%%%%%%

Moreover, we note that collider searches for heavy down-type quarks are capable of placing
a lower bound $M_Q^{\textrm{min}}$ on the quark mass scale $M_Q$.
As $M_Q$ decreases with $f_a$ and $k$, cf.\ Eq.~\eqref{eq:mQi}, this lower bound on $M_Q$
readily translates into a lower bound $k_{\textrm{min}}$ on $k$,
\begin{align}
M_Q \left(f_a, n, k=k_{\textrm{min}}\right) \geq M_Q^{\textrm{min}} \,, \quad
M_Q \left(f_a, n, k=k_{\textrm{min}} - 1\right) < M_Q^{\textrm{min}} \, \quad
k_{\textrm{min}} = k_{\textrm{min}}\left(f_a,n\right) \,,
\label{eq:kmin}
\end{align}
Assuming that the new quarks primarily couple to the SM quarks of the third generation
via the operator $\bar{Q}_i \mathbf{10}_j H_d$, such as in the model discussed
in Ref.~\cite{AguilarSaavedra:2009es}, the ATLAS experiment at the LHC has
recently reported a lower bound of $590 \,\textrm{GeV}$ on the heavy
quark mass scale~\cite{ATLAS:2013-051}.
In the following, we will adopt this value for $M_Q^{\textrm{min}}$, although we
remark that smaller values of $M_Q$ might still be viable, if the new quarks
should predominantly couple to the first or second generation of the SM quarks rather than
to the third generation.
Conversely, an even larger mass range could in principle be excluded using the present data,
if the new quarks should couple to the MSSM via the operator
$\bar{P}\bar{Q}_i \mathbf{10}_j H_d$ rather than via the operator
$\bar{Q}_i \mathbf{10}_j H_d$.
In this case, the new quarks would be long-lived, thereby leaving
very distinct signatures in collider experiments.
In this section, we, however, assume a coupling via the operator
$\bar{Q}_i \mathbf{10}_j H_d$ and set $M_Q^{\textrm{min}}$ to $590\,\textrm{GeV}$.
Solving Eq.~\eqref{eq:kmin} for $k_{\textrm{min}}$, we then find $k_{\textrm{min}}$
as a function of $f_a$, cf.\ Fig.~\ref{fig:kbounds}.
For $n=1$ and all values of $f_a$ of interest, $k_{\textrm{min}}$ is always $1$.
On the other hand, for $n=2$ and $f_a \lesssim 3 \times 10^{10}\,\textrm{GeV}$,
the minimal possible number of quark pairs rapidly grows as we go to smaller and
smaller values of $f_a$.

%%%%%%%%%%%%%%%%%%%%%%%%%%%%%%%%%%%%%%%%%%%%%%%%%%%%%%%%%%%%%%%%%%%%%%%%%%%%%%%%%%%%%%%%%%%%%%%%%%%%

In summary, we conclude that, for each value of $f_a$, the requirements
of perturbative gauge coupling unification as well as the
lower bound on the mass of heavy down-type quarks provide us with a range of
possible $k$ values, cf.\ Eqs.~\eqref{eq:kmax} and \eqref{eq:kmin},
\begin{align}
k_{\textrm{min}}\left(f_a,n\right) \leq k \leq k_{\textrm{max}}\left(f_a,n\right) \,.
\end{align}
Turning this statement around, we can say that, for given values of $n$ and $k$,
our two phenomenological constraints imply a lower bound on $f_a$,
\begin{align}
f_a \geq \max\left\{f_a^\textrm{min,p}, f_a^\textrm{min,m}\right\} \,,
\end{align}
where $f_a^\textrm{min,p}$ and $f_a^\textrm{min,m}$ are defined such
\begin{align}
g_{\textrm{GUT}} \left(f_a^\textrm{min,p},n,k\right) = \sqrt{4\pi} \,, \quad
M_Q\left(f_a^\textrm{min,m},n,k\right) = M_Q^{\textrm{min}} \,, \quad
f_a^\textrm{min,i} = f_a^\textrm{min,i}\left(k,n\right) \,, \quad
i = p,m \,.
\end{align}
In addition to that, we know from astrophysical and cosmological observations that
the axion decay constant must not be smaller than $\nrmcal{O}\left(10^9\right)\,\textrm{GeV}$
and not be larger than
$\nrmcal{O}\left(10^{12}\right)\,\textrm{GeV}$, cf.\ Eq.~\eqref{eq:window},
so that we are eventually led to imposing the following lower and upper bounds on $f_a$,
\begin{align}
f_a^\textrm{min} \leq f_a \leq 10^{12}\,\textrm{GeV} \,, \quad
f_a^\textrm{min} = \max\left\{10^9\,\textrm{GeV},
f_a^\textrm{min,p}, f_a^\textrm{min,m}\right\} \,.
\label{eq:famin}
\end{align}

%%%%%%%%%%%%%%%%%%%%%%%%%%%%%%%%%%%%%%%%%%%%%%%%%%%%%%%%%%%%%%%%%%%%%%%%%%%%%%%%%%%%%%%%%%%%%%%%%%%%

\subsection{Shifts in the QCD vacuum angle}
\label{subsec:fabounds}

%%%%%%%%%%%%%%%%%%%%%%%%%%%%%%%%%%%%%%%%%%%%%%%%%%%%%%%%%%%%%%%%%%%%%%%%%%%%%%%%%%%%%%%%%%%%%%%%%%%%

Given the particle content and charge assignments of our model, it is easy
to construct operators that explicitly break the PQ symmetry.
Instead of an exact symmetry, the PQ symmetry therefore merely ends up being an
approximate symmetry, which poses a threat to the PQ solution of the strong $CP$ problem.
Most PQ-breaking operators induce a shift in the VEV of the axion field,
such that the $\bar{\theta}$ term in the QCD Lagrangian is no longer completely canceled.
The magnitudes of these shifts in $\left<a\right>$ differ from operator to operator
and depend in addition on the axion decay constant $f_a$, the gravitino mass $m_{3/2}$
as well as on the scalar VEVs $\langle S \rangle$ and $\langle X \rangle$ in some cases.
In this section, we will now investigate for which $Z_N^R$ symmetries, which
choices of $n$ and $k$ as well as which values of $f_a$
the total shift in the axion VEV remains small enough, such that the shifted
$\bar{\theta}$ angle does not exceed the upper experimental bound,
$\bar{\theta} \lesssim 10^{-10}$.

%%%%%%%%%%%%%%%%%%%%%%%%%%%%%%%%%%%%%%%%%%%%%%%%%%%%%%%%%%%%%%%%%%%%%%%%%%%%%%%%%%%%%%%%%%%%%%%%%%%%

\subsubsection{PQ-breaking operators in the superpotential}

%%%%%%%%%%%%%%%%%%%%%%%%%%%%%%%%%%%%%%%%%%%%%%%%%%%%%%%%%%%%%%%%%%%%%%%%%%%%%%%%%%%%%%%%%%%%%%%%%%%%

All PQ-breaking operators in the superpotential inducing a shift in
$\langle a \rangle$ are of the following form%
\footnote{PQ-breaking operators that do not involve any power of
$P$ or $\bar{P}$ (for instance, $Q^5$ or $\bar{Q}^5$) do not induce a shift
in the axion VEV and are therefore irrelevant for our purposes.}
\begin{align}
W \supset \frac{C\,P^p \,\bar{P}^{\bar{p}}}{p!\,\bar{p}!\,h!\,s!\,x!\,M_{\textrm{Pl}}^c} \,
\left(H_u H_d\right)^h \,m_{3/2}^m  \,S^s \,X^x  \,,\quad
c = p+\bar{p}+h+m+s+x-3 \,, \quad p \neq \bar{p} \,, 
\label{eq:WOdang}
\end{align}
where $C$ is a $\nrmcal{O}(1)$ constant and where the powers
of the various fields have to be chosen such that,
\begin{align}
r_P\left(p - \bar{p}\right) + 4 h + 2 m -2s +2x \modulo{N} 2 \,.
\end{align}
Our intention behind explicitly dividing the operators in Eq.~\eqref{eq:WOdang}
by the factorials of the powers $p$, $\bar{p}$, $h$, $s$ and $x$ is to eventually obtain
maximally conservative bounds on the axion decay constant.
Fortunately, we do not have to consider all possible combinations of $p$, $\bar{p}$,
$h$, $m$, $s$ and $x$ in the following.
For instance, if some operator involving powers $(p,\bar{p})$ with
$\max\{p,\bar{p}\} > \min\{p,\bar{p}\}>0$ is allowed in the superpotential,
the same operator with $(p,\bar{p})$ being either replaced by $(p-\bar{p},0)$ or
$(0,\bar{p}-p)$ is also allowed.
The shift in $\left<a\right>$ induced by this second operator is then enhanced
compared to the shift induced by the original operator by a factor of
$\nrmcal{O}\left(M_{\textrm{Pl}}^q/\Lambda^q\right)$, where $q = 2 \min\{p,\bar{p}\}$.
Consequently, we are allowed to solely focus on PQ-breaking operators in the following
that either involve some power of $P$ \textit{or} some power of $\bar{P}$.
For a similar reason, we do not have to care about operators involving some power of
$H_u H_d$.
Given an operator with powers $h\geq 1$ and $m \geq 0$, we can always write down
a similar operator in which $(h,m)$ is replaced by $(0,m+2h)$.
This is possible because $\left(H_uH_d\right)^h$ and $m_{3/2}^{2h}$ have the same $R$
charge up to an integer multiple of $N$.
Now assuming that $m_{3/2}^2$ is larger than $\left<H_uH_d\right> = v_u v_d$, the operator
with powers $(0,m+2h)$ always yields a larger shift in $\left<a\right>$ than the operator
with powers $(h,m)$.
Furthermore, the same game as with the
fields $P$ and $\bar{P}$ can also be played with $m_{3/2}$ and
the fields $S$ and $X$.
Operators with powers $(m,s,x)$ satisfying the
relation $s>m+x\geq0$ can always be traded for operators
with powers $(0,s-m-x,0)$.
The shift in $\left<a\right>$ due to these alternative operators is then enhanced compared
to the shift due to the original operators by a factor of
$\nrmcal{O}\big(M_{\textrm{Pl}}^{2(m+x)}/\big(m_{3/2}^{m}
\left<S\right>^{m+x}\left<X\right>^x\big)\big)$.
In the end, we therefore only have to consider the following set of PQ-breaking operators,
\begin{align}
W \supset \frac{C\,P^p}{p!\,M_{\textrm{Pl}}^c}\left[\frac{1}{s!} S^s \,,\:
\frac{1}{x!} m_{3/2}^m X^x \right]
\quad|\:\: (P,p) \leftrightarrow (\bar{P},\bar{p}) \,.
\label{eq:dangOpW}
\end{align}

%%%%%%%%%%%%%%%%%%%%%%%%%%%%%%%%%%%%%%%%%%%%%%%%%%%%%%%%%%%%%%%%%%%%%%%%%%%%%%%%%%%%%%%%%%%%%%%%%%%%

Each of the operators in Eq.~\eqref{eq:dangOpW}
results in PQ-breaking terms in the scalar potential.
Among these PQ-breaking contributions to the scalar potential, one class of terms derives from
the $F$-terms of the fields $S$ and $X$,
\begin{align}
F_S = & \: \frac{C}{M_{\textrm{Pl}}^c}\left[ \frac{s}{p!\,s!} P^p S^{s-1} \,,\:
\frac{s}{\bar{p}!\,s!} \bar{P}^{\bar{p}} S^{s-1}\right] +
F_S^0 \,, \\ \label{eq:FSFX}
F_X = & \: \frac{C}{M_{\textrm{Pl}}^c}\left[
\frac{x}{p!\,x!} P^p m_{3/2}^m X^{x-1} \,,\:
\frac{x}{\bar{p}!\,x!} \bar{P}^{\bar{p}} m_{3/2}^m X^{x-1} \right]
+ F_X^0 \,, \nonumber
\end{align}
where we have introduced $F_S^0$ and $F_X^0$ to denote the contributions to $F_X$ and
$F_S$ deriving from PQ-invariant operators in the superpotential.
Given the superpotential in Eq.~\eqref{eq:WS}
and taking into account the various supergravity effects
induced by the constant term in the superpotential, $W_0 = m_{3/2}M_{\textrm{Pl}}$,
we are able to estimate of what order of magnitude we expect $F_S^0$ and $F_X^0$ to be,
\begin{align}
F_S^0 = & \: \nrmcal{O}\left(m_{3/2}^2,v_uv_d,m_{3/2}
\left<X\right>,\left<X^2\right>,..\right)  \,, \\
F_X^0 = & \: \nrmcal{O}\left(m_{3/2}^2,m_{3/2}
\left<S\right>,\left<XS\right>,..\right) \nonumber \,,
\end{align}
with the dots denoting further contributions to $F_S^0$ and $F_X^0$ that only arise in the case
of certain $Z_N^R$ symmetries.
The VEVs of the fields $S$ and $X$ are both of the order of the gravitino mass, such that
the leading contributions to $F_S^0$ and $F_X^0$ can eventually be estimated as
\begin{align}
F_S^0 = \nrmcal{O}\left(m_{3/2}^2\right)  \,, \quad F_X^0 = \nrmcal{O}\left(m_{3/2}^2\right) \,.
\end{align}
The mixing between $F_S^0$ and $F_X^0$ and the PQ-breaking contributions
to $F_S$ and $F_X$ in Eq.~\eqref{eq:FSFX} then gives rise to the following PQ-breaking
terms in the scalar potential,
\begin{align}
V \supset m_{3/2}^2 \frac{C\,P^p}{p!\,M_{\textrm{Pl}}^c} \left[
\frac{s}{s!}S^{s-1}  \,,\: \frac{x}{x!} m_{3/2}^m X^{x-1} \right]
+ \textrm{h.c.} \quad|\:\: (P,p) \leftrightarrow (\bar{P},\bar{p}) \,.
\label{eq:VPQF}
\end{align}

%%%%%%%%%%%%%%%%%%%%%%%%%%%%%%%%%%%%%%%%%%%%%%%%%%%%%%%%%%%%%%%%%%%%%%%%%%%%%%%%%%%%%%%%%%%%%%%%%%%%

A second important class of PQ-breaking terms in the scalar potential are the $A$-terms
which derive from  the mixing between the operators in Eq.~\eqref{eq:dangOpW} and the VEV
of the superpotential $W_0$,
\begin{align}
V \supset \frac{W_0}{M_{\textrm{Pl}}^2}\frac{C\,P^p}{p!\,M_{\textrm{Pl}}^c} \left[
\frac{(p+s-3)}{s!}S^s  \,,\: \frac{(p+x-3)}{x!} m_{3/2}^m  X^x \right]
+ \textrm{h.c.} \quad|\:\: (P,p) \leftrightarrow (\bar{P},\bar{p}) \,.
\label{eq:VPQA}
\end{align}
For a given operator in the superpotential with powers $(p,s)$ or
$(\bar{p},s)$, the largest PQ-breaking term in the scalar potential
hence corresponds to
\begin{align}
V \supset & \: m_{3/2}\frac{C\,P^p}{p!\,s!\,M_{\textrm{Pl}}^c}
\max\left\{s\,m_{3/2}, \left|p+s-3\right|S\right\} S^{s-1}
+ \textrm{h.c.}  \quad|\:\: (P,p) \leftrightarrow (\bar{P},\bar{p})\,.
\label{eq:VPS}
\end{align}
Similarly, the largest term induced by an operator with powers $(m,p,x)$
or $(m,\bar{p},x)$ is given by%
\footnote{Note that, in Eqs.~\eqref{eq:VPS} and \eqref{eq:VPX}, we have implicitly
absorbed the sign of $(p+s-3)$ and $(p+x-3)$ in $C$.\smallskip}
\begin{align}
V \supset & \: m_{3/2}\frac{C\,P^p}{p!\,x!\,M_{\textrm{Pl}}^c}
\max\left\{x\,m_{3/2}, \left|p+x-3\right|X\right\} m_{3/2}^m X^{x-1}
+ \textrm{h.c.} \quad|\:\: (P,p) \leftrightarrow (\bar{P},\bar{p})\,.
\label{eq:VPX}
\end{align}
Next, we replace all scalar fields in these two operators by their VEVs,
\begin{align}
P\rightarrow\frac{\Lambda}{\sqrt{2}} \,
\exp\left(i\frac{a}{\sqrt{2}\,\Lambda}\right) \,,\quad
\bar{P}\rightarrow\frac{\Lambda}{\sqrt{2}} \,
\exp\left(-i\frac{a}{\sqrt{2}\,\Lambda}\right) \,, \quad
S \rightarrow \langle S\rangle \,, \quad
X \rightarrow \langle X\rangle \,.
\end{align}
This provides us with contributions to
the axion potential all of which are of the following form,
\begin{align}
\Delta V_a = \frac{1}{2} M^4 \left[\exp\left(i\frac{p\,a}{\sqrt{2}\,\Lambda}\right)
+ \textrm{h.c.}\right] = 
M^4 \cos\left(p \,\frac{a}{\sqrt{2}\,\Lambda}\right) \,,
\label{eq:DeltaVa}
\end{align}
where, for the terms in the scalar potential in Eqs.~\eqref{eq:VPS} and \eqref{eq:VPX},
the mass scale $M$ is respectively to be identified as%
\footnote{The expressions for $M^4$ corresponding
to the operators $\bar{P}^{\bar{p}}S^s$ and $\bar{P}^{\bar{p}}m_{3/2}^m X^x$ look
exactly the same.}
\begin{align}
P^p S^s \:: \quad &
M^4 \rightarrow \frac{2\,C\,m_{3/2}}{p!\,s!\,M_{\textrm{Pl}}^c}
\left(\frac{\Lambda}{\sqrt{2}}\right)^p
M_S \,\langle S\rangle^{s-1} \,, \quad M_S = \max\left\{s\,m_{3/2},
\left|p+s-3\right|\langle S\rangle\right\} \,,
\label{eq:Mscale}\\
P^p m_{3/2}^m X^x \:: \quad &
M^4 \rightarrow \frac{2\,C\,m_{3/2}}{p!\,x!\,M_{\textrm{Pl}}^c}
\left(\frac{\Lambda}{\sqrt{2}}\right)^p
M_X \,m_{3/2}^m \,\langle X\rangle^{s-1} \,, \quad
M_X = \max\left\{x\,m_{3/2}, \left|p+x-3\right|\langle X\rangle\right\} \,, \nonumber
\end{align}

%%%%%%%%%%%%%%%%%%%%%%%%%%%%%%%%%%%%%%%%%%%%%%%%%%%%%%%%%%%%%%%%%%%%%%%%%%%%%%%%%%%%%%%%%%%%%%%%%%%%

\subsubsection{PQ-breaking operators in the K\"ahler potential and the effective potential}

%%%%%%%%%%%%%%%%%%%%%%%%%%%%%%%%%%%%%%%%%%%%%%%%%%%%%%%%%%%%%%%%%%%%%%%%%%%%%%%%%%%%%%%%%%%%%%%%%%%%

Next to the PQ-breaking operators in the superpotential, we also have to take into
account the PQ-breaking contributions to the K\"ahler potential $K$.
It is, however, easy to show that the PQ-breaking terms in the scalar potential
induced by the K\"ahler potential can at most be as large as the terms induced
by the superpotential.
Given some PQ-breaking term $K_{\cancel{\textrm{PQ}}} \subset K$, its largest
contribution to the scalar potential is given by
\begin{align}
V \supset \frac{C'}{M_{\textrm{Pl}}^2} \left|W_0\right|^2  K_{\cancel{\textrm{PQ}}}
= C' \, m_{3/2}^2 K_{\cancel{\textrm{PQ}}} \,, \quad
C' \sim \nrmcal{O}(1) \,.
\end{align}
The operator $K_{\cancel{\textrm{PQ}}}$ is either holomorphic
from the outset or it is accompanied by a holomorphic term in the K\"ahler
potential $K_{\cancel{\textrm{PQ}}}'$ that follows from
$K_{\cancel{\textrm{PQ}}}$ by performing the following replacements,
\begin{align}
P^\dagger \rightarrow \bar{P} \,, \quad \bar{P}^\dagger \rightarrow  P \,, \quad
S^\dagger \rightarrow X \,, \quad X^\dagger \rightarrow S \,, \quad
\left(H_u H_d\right)^\dagger \rightarrow S^2 \,.
\end{align}
Furthermore, we know that, in order to be consistent with the $Z_N^R$ symmetry,
the $R$ charge of $K_{\cancel{\textrm{PQ}}}$ must be zero.
As the gravitino mass carries $R$ charge $2$, the holomorphicity of
$K_{\cancel{\textrm{PQ}}}^{(\prime)}$ in combination
with its vanishing $R$ charge thus directly implies that
$m_{3/2} K_{\cancel{\textrm{PQ}}}^{(\prime)}$
is one of the allowed operators in the superpotential.
The $A$-term deriving from $m_{3/2} K_{\cancel{\textrm{PQ}}}^{(\prime)}$ is then exactly
of the same order of magnitude as the term in the scalar potential induced by
$K_{\cancel{\textrm{PQ}}}$, cf.\ Eq.~\eqref{eq:VPQA}.
We therefore do not have to take care of the PQ-breaking terms in the
K\"ahler potential explicitly.
By studying the effects on the axion VEV related to the PQ-breaking operators
in the superpotential, we automatically cover all relevant effects on
the axion VEV related to the K\"ahler potential.

%%%%%%%%%%%%%%%%%%%%%%%%%%%%%%%%%%%%%%%%%%%%%%%%%%%%%%%%%%%%%%%%%%%%%%%%%%%%%%%%%%%%%%%%%%%%%%%%%%%%

So far, we have only discussed PQ-breaking terms in the tree-level scalar potential.
Below the heavy quark mass threshold, interactions at the loop level  give rise
to further PQ-breaking terms in the \textit{effective} scalar potential.
These higher-dimensional terms are then no longer solely suppressed by the Planck
scale, but partly also by the heavy quark mass scale $M_Q$,
\begin{align}
V_{\textrm{eff}} \supset \frac{1}{M_{\textrm{Pl}}^{c} M_Q^d}
\frac{C\,P^p}{p!\,s!\,x!} m_{3/2}^m S^s X^x \,, \quad c+d = p + m + s + x - 4 \,, \quad d > 0 \, 
\quad|\:\: (P,p) \leftrightarrow (\bar{P},\bar{p}) \,,
\label{eq:Veff}
\end{align}
where the coupling constant $C$ is in general now also field-dependent.
We might therefore worry that some of these effective operators could yield larger shifts in
the axion VEV than the actual tree-level operators that we have considered up to now.
By imposing the requirement that the radiatively induced terms in the scalar potential
must vanish in the limit $M_Q \rightarrow 0$,
\begin{align}
M_Q \rightarrow 0 \:: \quad
V_{\textrm{eff}} \supset \frac{1}{M_{\textrm{Pl}}^c M_Q^d}
\frac{C\,P^p}{p!\,s!\,x!} m_{3/2}^m S^s X^x \rightarrow 0 \,,
\quad|\:\: (P,p) \leftrightarrow (\bar{P},\bar{p}) \,, 
\end{align}
one can however show that
the factor $M_Q^{-d}$ in Eq.~\eqref{eq:Veff} is always canceled by factors contained
in the coupling constant $C$, such that all of the effective terms can eventually
be rewritten as
\begin{align}
V_{\textrm{eff}} \supset \frac{1}{M_{\textrm{Pl}}^{c'}}
\frac{C'\,P^{p'}}{p'!\,s'!\,x'!} m_{3/2}^{m'} S^{s'} X^{x'} \,,
\:\: c' = p' + m' + s' + x' - 4 \,,
\:\: C' \in \mathbb{R} \,,
\quad|\:\: (P,p') \leftrightarrow (\bar{P},\bar{p}') \,.
\end{align}
The radiative corrections in the effective potential are hence \textit{not} enhanced
with respect to the terms in the tree-level scalar potential.
We conclude that, for our purposes, it will suffice to only consider the
PQ-breaking terms in the scalar potential induced by the superpotential.
A separate treatment of K\"ahler-induced effects or radiative corrections is not necessary.

%%%%%%%%%%%%%%%%%%%%%%%%%%%%%%%%%%%%%%%%%%%%%%%%%%%%%%%%%%%%%%%%%%%%%%%%%%%%%%%%%%%%%%%%%%%%%%%%%%%%

\subsubsection{Upper bounds on the axion decay constant}

%%%%%%%%%%%%%%%%%%%%%%%%%%%%%%%%%%%%%%%%%%%%%%%%%%%%%%%%%%%%%%%%%%%%%%%%%%%%%%%%%%%%%%%%%%%%%%%%%%%%

The $\Delta V_a$ terms in the scalar potential, cf.\ Eq.~\eqref{eq:DeltaVa},
disturb the effective QCD instanton-induced
potential $V_a^{\textrm{eff}}$, cf.\ Eq.~\eqref{eq:Vaeff}, such that the axion
potential is no longer minimized by $f_a \bar{\theta}$,
\begin{align}
\left.\frac{d \left(V_a + \Delta V_a\right)}{da}\right|_{a = \langle a \rangle} = 0 \,, \quad
\langle a \rangle = f_a \left(\bar{\theta} + \Delta \bar{\theta}\right) \,.
\end{align}
This shift in the axion VEV directly translates into a non-zero value $\Delta\bar{\theta}$
of the QCD vacuum angle.
Making use of our results for $V_a^{\textrm{eff}}$ and $\Delta V_a$ in Eqs.~\eqref{eq:Vaeff}
and \eqref{eq:DeltaVa}, we obtain for $\Delta\bar{\theta}$
\begin{align}
\Delta\bar{\theta} = \Delta\bar{\theta}_0 \,
\sin\Big(\frac{p}{\left|\nrmcal{A}_{\textrm{PQ}}\right|}\,\bar{\theta}\Big) +
\nrmcal{O}\left(\left(\Delta\bar{\theta}_0\right)^2\right) \,, \quad
\Delta\bar{\theta}_0 = \frac{p}{\left|\nrmcal{A}_{\textrm{PQ}}\right|}
\frac{M^4}{\Lambda_{\textrm{QCD}}^4} \,.
\label{eq:Deltatheta}
\end{align}
According to the experimental upper bound on the QCD vacuum angle,
$\Delta \bar{\theta}_0$ must not be larger than $10^{-10}$,
\begin{align}
\Delta \bar{\theta}_0 \leq \Delta \bar{\theta}_0^{\textrm{max}} =  10^{-10} \,, \quad 
M^4 \leq \frac{\left|\nrmcal{A}_{\textrm{PQ}}\right|}{p} \,\Delta\bar{\theta}_0^{\textrm{max}}\,
\Lambda_{\textrm{QCD}}^4 \,,
\end{align}
which results in an upper bound on the mass scale $M$.
Combining this constraint with our expressions for $M$ in Eq.~\eqref{eq:Mscale},
we are able to derive an upper bound on the axion decay constant $f_a$
for each of the PQ-breaking operators in the superpotential,
\begin{align}
P^p S^s \:: \quad & f_a^{\textrm{max},S} =
\left[2^{p-1}p!\,s!\,\frac{\Delta\bar{\theta}_0^{\textrm{max}}}{C} 
\frac{\left|\nrmcal{A}_{\textrm{PQ}}\right|^{1-p}}{p}
\frac{\Lambda_{\textrm{QCD}}^4 \, M_{\textrm{Pl}}^c}
{M_S \,\langle S\rangle^{s-1} \,m_{3/2}}\right]^{1/p} \,, \\
P^p m_{3/2}^m X^x\:: \quad & f_a^{\textrm{max},X} =
\left[2^{p-1}p!\,x!\,\frac{\Delta\bar{\theta}_0^{\textrm{max}}}{C} 
\frac{\left|\nrmcal{A}_{\textrm{PQ}}\right|^{1-p}}{p}
\frac{\Lambda_{\textrm{QCD}}^4 \, M_{\textrm{Pl}}^c}
{M_X \,\langle X\rangle^{x-1} \,m_{3/2}^{m+1}}\right]^{1/p} \,, \nonumber
\end{align}
with the bounds corresponding to $\bar{P}^{\bar{p}}S^s$ and
$\bar{P}^{\bar{p}}m_{3/2}^m X^x$ being of exactly the same form.

%%%%%%%%%%%%%%%%%%%%%%%%%%%%%%%%%%%%%%%%%%%%%%%%%%%%%%%%%%%%%%%%%%%%%%%%%%%%%%%%%%%%%%%%%%%%%%%%%%%%

\begin{table}
\begin{center}
\begin{doublespacing}
\begin{tabular}{c||ccccccccccccccc}
$k$ & $3$ & $4$ & $5$ & $6$ & $7$ & $8$ & $9$ & $10$ &
$11$ & $12$ & $13$ & $14$ & $15$ & $16$ & $17$\\\hline\hline
\# & $(0,4)$ & $(0,38)$ & $(16,54)$ & $(4,36)$ &
$49$ & $54$ & $36$ & $62$ & $101$ & $36$ & $120$ & $110$ & $56$ & $132$ & $160$
\end{tabular}
\end{doublespacing}
\begin{onehalfspacing}
\caption{Numbers of viable scenarios for individual values of $k$, including as well
all scenarios with a $Z_3^R$ or a $Z_6^R$ symmetry.
For $k\leq 6$, the respective numbers of solutions
for the two cases $n=1$ and $n=2$ are indicated in the format
$\left(\left.\#\right|_{n=1},\left.\#\right|_{n=2}\right)$.}
\label{tab:kfix} 
\end{onehalfspacing}
\end{center}
\end{table}

%%%%%%%%%%%%%%%%%%%%%%%%%%%%%%%%%%%%%%%%%%%%%%%%%%%%%%%%%%%%%%%%%%%%%%%%%%%%%%%%%%%%%%%%%%%%%%%%%%%%

For given values of $N$, $n$, $k$ and $r_P$, a multitude of different
PQ-breaking operators might be allowed in the superpotential, all of
which imply an upper limit on $f_a$.
Let us denote the most restrictive among these upper limits by
$f_a^{\textrm{max},\bar{\theta}}$,
\begin{align}
f_a^{\textrm{max},\bar{\theta}} = \min \left\{\textrm{all }f_a^{\textrm{max},S},
\textrm{all }f_a^{\textrm{max},X} \right\} \,.
\label{eq:famaxth}
\end{align}
Together with the constraints on $f_a$ in Eq.~\eqref{eq:famin}, we thus find
the following total lower and upper limits on the axion decay constant $f_a$,
\begin{align}
f_a^\textrm{min} \leq f_a \leq f_a^\textrm{max} \,, \:\:
f_a^\textrm{min} = \max\left\{10^9\,\textrm{GeV},
f_a^\textrm{min,p}, f_a^\textrm{min,m}\right\} \,, \:\:
f_a^\textrm{max} = \min\big\{10^{12}\,\textrm{GeV},
f_a^{\textrm{max},\bar{\theta}}\big\} \,.
\end{align}
By virtue of this result, we are now able to identify the phenomenologically viable
combinations of $N$, $n$, $k$ and $r_P$.
The corresponding criterion is nothing but the requirement that
there has to be an allowed window of possible values for $f_a$,
\begin{align}
f_a^{\textrm{min}} < f_a^{\textrm{max}} \quad\Rightarrow\quad
\left(N,n,k,r_p\right) \textrm{ viable} \,.
\label{eq:viable}
\end{align}
To determine the allowed combinations of $N$, $n$, $k$ and $r_P$, we compute
$f_a^{\textrm{min}}$ and $f_a^{\textrm{max}}$ for 
\begin{align}
N = 3,4,.., 12\,; \quad n =1,2 \,; \quad k = 1,2,..,
k_{\textrm{max}}\left(10^{12}\,\textrm{GeV},n\right)
\,; \quad r_P = r_P \left(N,n,k,\ell_Q,\ell_P\right) \,,
\label{eq:NnKrtanges}
\end{align}
where $k_{\textrm{max}}\left(10^{12}\,\textrm{GeV},1\right) = 17$ and
$k_{\textrm{max}}\left(10^{12}\,\textrm{GeV},2\right) = 6$ and where
$r_P$ as a function of $N$, $n$, $k$, $\ell_Q$ and $\ell_P$ is given
in Eq.~\eqref{eq:rP},%
\footnote{In total, we thus scan $1950$ different combinations of $N$, $n$, $k$ and $r_P$.
Out of these combinations, $1530$ belong
to the case $n=1$, whereas $430$ belong to the case $n=2$.}
and check whether or not the criterion in
Eq.~\eqref{eq:viable} is fulfilled.
In doing so, we set all dimensionless coupling constants to $1$ and
use a common value of $1\,\textrm{TeV}$ for the gravitino mass
and the scalar VEVs,
\begin{align}
m_{3/2} = 1\,\textrm{TeV} \,, \quad \langle S \rangle = 1\,\textrm{TeV} \,, \quad
\langle X \rangle = 1\,\textrm{TeV} \,.
\label{eq:m32SXvals}
\end{align}
Larger values of $m_{3/2}$, $\langle S\rangle$ and $\langle X \rangle$ would lead
to more stringent bounds on $f_a$, which means that the bounds that we obtain
should be regarded as conservative.
For the non-perturbative scale of QCD, we employ the $\overline{\textrm{\footnotesize MS}}$
value above the bottom-quark mass threshold,
$\Lambda_{\textrm{QCD}} \simeq 213 \,\textrm{MeV}$~\cite{Beringer:1900zz}.

%%%%%%%%%%%%%%%%%%%%%%%%%%%%%%%%%%%%%%%%%%%%%%%%%%%%%%%%%%%%%%%%%%%%%%%%%%%%%%%%%%%%%%%%%%%%%%%%%%%%

\subsubsection{Phenomenologically viable scenarios}

%%%%%%%%%%%%%%%%%%%%%%%%%%%%%%%%%%%%%%%%%%%%%%%%%%%%%%%%%%%%%%%%%%%%%%%%%%%%%%%%%%%%%%%%%%%%%%%%%%%%

Restricting ourselves to the parameter values specified in Eqs.~\eqref{eq:NnKrtanges}
and \eqref{eq:m32SXvals}, we find in total $1068$ viable combinations of $N$, $n$, $k$
and $r_P$, where $936$ of these solutions belong to the case $n=1$
and $132$ solutions to the case $n=2$.
In Tabs.~\ref{tab:kfix} and \ref{tab:Nfix}, we indicate how many solutions we
respectively obtain for the individual values of $k$ and $N$ under study.
In Tab.~\ref{tab:viableNnkrP}, we list all viable combinations of $N$, $n$, $k$
and $r_P$ for all $k$ values up to $k=6$.
In summary, we conclude that our minimal extension of the MSSM apparently gives rise
to a large landscape of viable scenarios.
It is in particular surprising and intriguing that the order $N$ of the $Z_N^R$ symmetry
can take any value, as long as the number of extra quark pairs $k$ is chosen appropriately.
A comprehensive phenomenological study of this landscape of possible solutions is beyond
the scope of this paper.
In the following, we shall thus restrict ourselves to a few interesting observations,
illustrating what kind of questions one might be able to answer based on the full
numerical data describing the landscape.

%%%%%%%%%%%%%%%%%%%%%%%%%%%%%%%%%%%%%%%%%%%%%%%%%%%%%%%%%%%%%%%%%%%%%%%%%%%%%%%%%%%%%%%%%%%%%%%%%%%%

\begin{table}
\begin{center}
\begin{doublespacing}
\begin{tabular}{c||cccccccc}
$N$ & $3$ & $4$ & $5$ & $6$ & $7$ & $8$ & $9$ & $10$ \\\hline\hline
\# & $(76, 8)$ & $(104, 16)$ & $(91, 13)$ & $(76, 8)$ & $(103, 14)$ & $(104, 16)$ & $(90,6)$ &
$(91, 13)$ \\\hline\hline
$N$ & $11$ & $12$ \\\hline\hline
\# & $(109, 22)$ & $(92, 16)$
\end{tabular}
\end{doublespacing}
\begin{onehalfspacing}
\caption{Numbers of viable scenarios for individual values of $N$
 in the format $\left(\left.\#\right|_{n=1},\left.\#\right|_{n=2}\right)$.}
\label{tab:Nfix} 
\end{onehalfspacing}
\end{center}
\end{table}

%%%%%%%%%%%%%%%%%%%%%%%%%%%%%%%%%%%%%%%%%%%%%%%%%%%%%%%%%%%%%%%%%%%%%%%%%%%%%%%%%%%%%%%%%%%%%%%%%%%%

\begin{table}
\begin{center}
\begin{small}
\begin{doublespacing}
\begin{tabular}{r||llll}
$\left(n,k\right)$ & $Z_3^R$ & $Z_4^R$ & $Z_5^R$ & $Z_6^R$\\\hline\hline
$(2,4)$ & $\{3, 9, 15, 21\}$ & $\{2, 6, 10, 14, 18, 22, 26, 30\}$ &
$\{9, 19, 29, 39\}$ & $\{6, 18, 30, 42\}$ \\
$(2,5)$ & $\{3, 9, 21, 27\}$ & $\{2, 6, 14, 18, 22, 26, 34, 38\}$ &
$\{9, 19, 29, 39, 49\}$ & $\{6, 18, 42, 54\}$ \\
$(2,6)$ &   & $\{2, 10, 14, 22, 26, 34, 38, 46\}$ & $\{19, 29, 49, 59\}$ \\\hline\hline
$\left(n,k\right)$ & $Z_7^R$ & $Z_8^R$ & $Z_9^R$ & $Z_{10}^R$\\\hline\hline
$(1,5)$ & $\{1, 8, 22, 29\}$ &    & $\{3, 12, 21, 39\}$ &    \\
$(1,6)$ & $\{1, 29\}$ &    &    &    \\
$(2,3)$ & $\{1, 29\}$ &    &    &    \\
$(2,4)$ & $\{1, 15, 29, 43\}$ & $\{2, 10, 18, 26, 34, 42, 50, 58\}$ &
$\{3, 21, 39, 57\}$ & $\{14, 34, 54, 74\}$ \\
$(2,5)$ & $\{1, 29, 43, 57\}$ & $\{2, 18, 26, 34, 42, 58, 66, 74\}$ &
$\{3, 12, 21, 39, 48, 57, 66, 84\}$ & $\{14, 34, 54, 74, 94\}$ \\
$(2,6)$ & $\{1, 29, 43, 71\}$ & $\{2, 10, 26, 34, 50, 58, 74, 82\}$ &
$\{3, 57\}$ & $\{14, 34, 74, 94\}$ \\\hline\hline
$\left(n,k\right)$ & \multicolumn{2}{l}{$Z_{11}^R$} &
\multicolumn{2}{l}{$Z_{12}^R$} \\\hline\hline
$(1,5)$ & \multicolumn{2}{l}{$\{16, 27, 38, 49\}$} &
\multicolumn{2}{l}{$\{6, 18, 42, 54\}$} \\
$(1,6)$ & \multicolumn{2}{l}{$\{5, 49\}$} \\
$(2,3)$ & \multicolumn{2}{l}{$\{5, 49\}$} \\
$(2,4)$ & \multicolumn{2}{l}{$\{5, 27, 38, 49, 71, 82\}$} &
\multicolumn{2}{l}{$\{6, 18, 30, 42, 54, 66, 78, 90\}$} \\
$(2,5)$ & \multicolumn{2}{l}{$\{16, 27, 38, 49, 71, 82, 93, 104\}$} &
\multicolumn{2}{l}{$\{6, 18, 42, 54, 66, 78, 102, 114\}$} \\
$(2,6)$ & \multicolumn{2}{l}{$\{5, 38, 49, 71, 82, 115\}$} 
\end{tabular}
\end{doublespacing}
\begin{onehalfspacing}
\caption{Viable values of $r_P$ in units of $1/(nk)$ and
in dependence of $N$, $n$ and $k$ for all $k$ values up to $k=6$.
We also include the $r_P$ values for $(N,n,k) = (4,2,4), (8,2,4),(9,2,5)$,
which are actually phenomenologically unviable if we believe in the perturbative
unification of the gauge coupling constants.
For these combinations of $N$, $n$ and $k$, we namely find
$10^9\,\textrm{GeV} \leq f_a^{\textrm{max},0}\leq f_a^{\textrm{min}}$,
cf.\ Eq.~\eqref{eq:fminfmax} and Fig.~\ref{fig:kbounds}.}
\label{tab:viableNnkrP} 
\end{onehalfspacing}
\end{small}
\end{center}
\end{table}

%%%%%%%%%%%%%%%%%%%%%%%%%%%%%%%%%%%%%%%%%%%%%%%%%%%%%%%%%%%%%%%%%%%%%%%%%%%%%%%%%%%%%%%%%%%%%%%%%%%%

We observe for instance that, for all possible
combinations of $N$, $n$ and $k$, there exists either
no viable $r_P$ value at all or at least two different values.
It is therefore interesting to ask which of the various possible $r_P$ values for
given $N$, $n$ and $k$ yields the least stringent upper bound on $f_a$,
\begin{align}
f_a^{\textrm{max},0}(N,n,k) = \max_{r_P}
\left\{f_a^{\textrm{max},\bar{\theta}}(N,n,k,r_P)\right\} \,.
\label{eq:famax0}
\end{align}
This maximal upper bound can then be regarded as the most conservative constraint
on $f_a$ for the respective combinations of $N$, $n$ and $k$.
The two panels of Fig.~\ref{fig:kbounds} present $f_a^{\textrm{max},0}$ as a function
of $N$ and $k$ for $n=1$ and $n=2$, respectively.
Apart from four exceptions, $f_a^{\textrm{max},0}$ interestingly
always exceeds $f_a^{\textrm{min}}$ as long as it is larger than $10^9\,\textrm{GeV}$,
\begin{align}
(N,n,k) \neq (4,2,4), (8,2,4), (9,2,5), (12,1,15) \:: \quad
f_a^{\textrm{max},0} \geq 10^9\,\textrm{GeV}
\:\:\Rightarrow\:\: f_a^{\textrm{max},0} > f_a^{\textrm{min}} \,.
\label{eq:fminfmax}
\end{align}
Only for $(N,n,k) = (4,2,4), (8,2,4), (9,2,5), (12,1,15)$, $f_a^{\textrm{max},0}$
is smaller than $f_a^{\textrm{min}}$, which
renders these four cases phenomenologically unviable.
This is indicated in Fig.~\ref{fig:kbounds} by the diagonal black lines
crossing out the respective squares.

%%%%%%%%%%%%%%%%%%%%%%%%%%%%%%%%%%%%%%%%%%%%%%%%%%%%%%%%%%%%%%%%%%%%%%%%%%%%%%%%%%%%%%%%%%%%%%%%%%%%

\begin{figure}
\begin{center}
\includegraphics[width=0.7\textwidth]{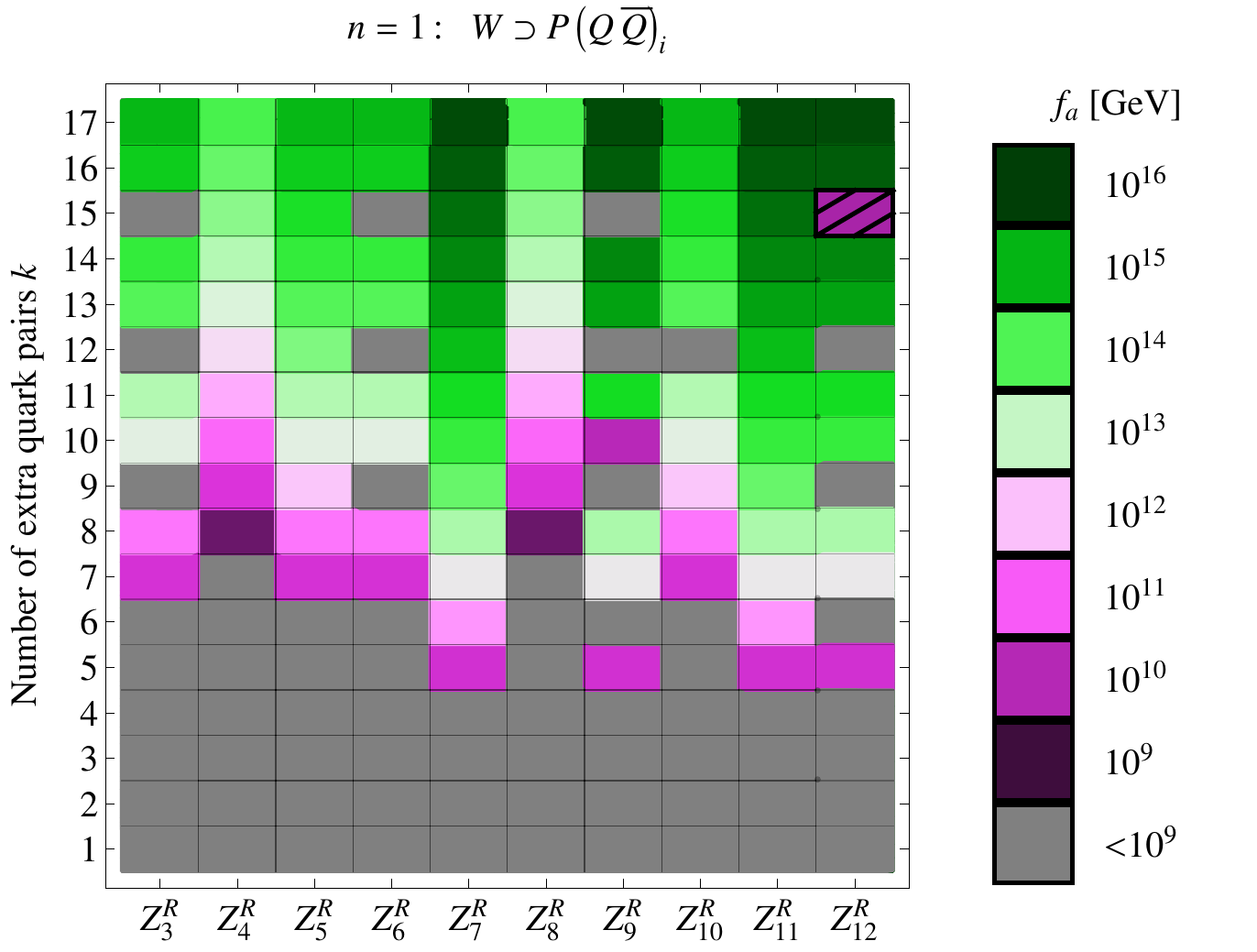}

\vspace{0.7cm}
\includegraphics[width=0.7\textwidth]{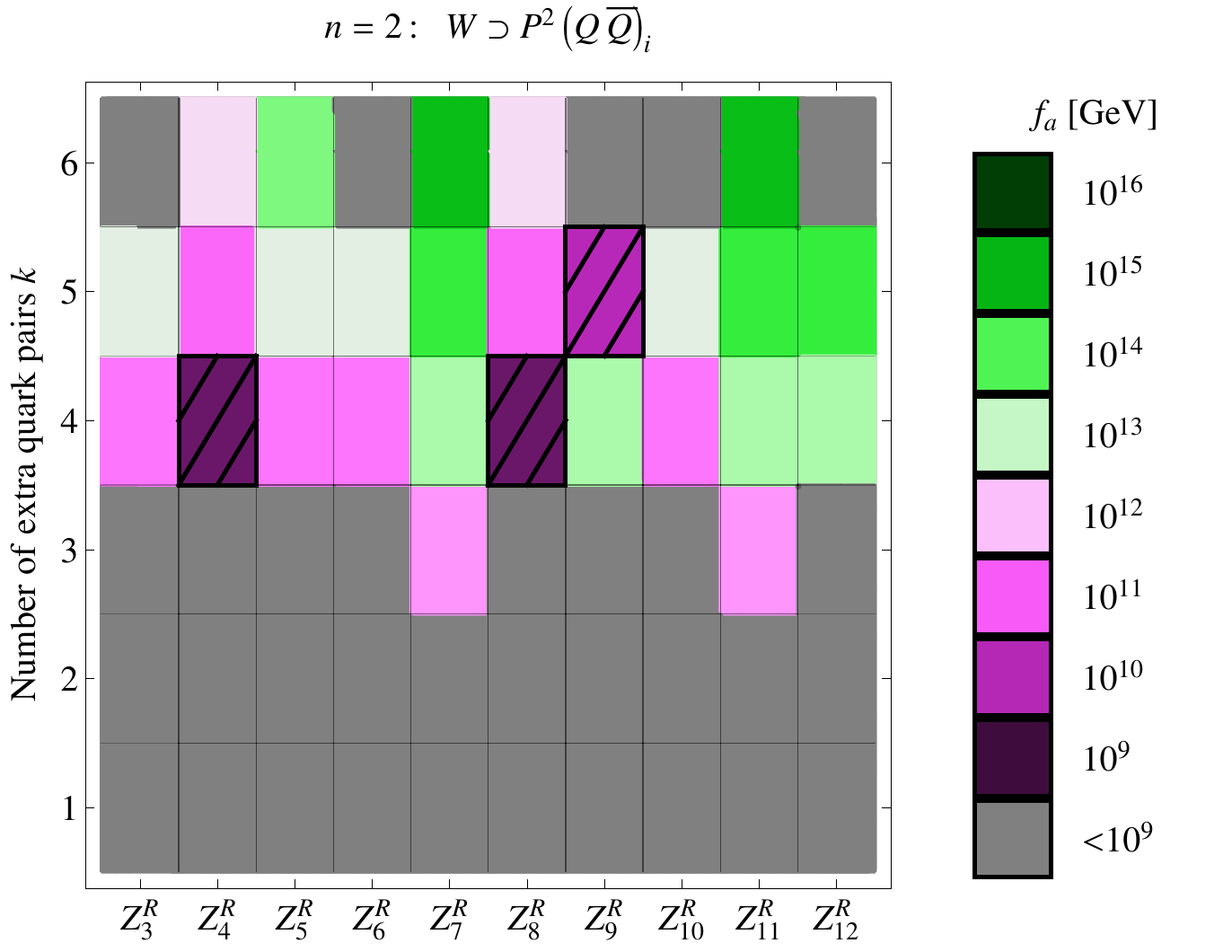}
\caption{Upper bounds $f_a^{\textrm{max},0}$ on the
axion decay constant $f_a$ according to the
requirement that the shift in the QCD vacuum angle $\bar{\theta}$ induced by
PQ-breaking operators not be larger
than $10^{-10}$, cf.\ Eqs.~\eqref{eq:famaxth} and \eqref{eq:famax0}.
Both plots are based on $m_{3/2} = 1\,\textrm{TeV}$,
$\left<S\right> = \mu/g_H = 1\,\textrm{TeV}$ and $\left<X\right> = 1\,\textrm{TeV}$.
At the same time, all dimensionless coupling constants have been set to $1$.
The black diagonal lines indicate that
$10^9\,\textrm{GeV} \leq f_a^{\textrm{max},0}\leq f_a^{\textrm{min}}$, cf.\ Eq.~\eqref{eq:famin}.}
\end{center}
\end{figure}

%%%%%%%%%%%%%%%%%%%%%%%%%%%%%%%%%%%%%%%%%%%%%%%%%%%%%%%%%%%%%%%%%%%%%%%%%%%%%%%%%%%%%%%%%%%%%%%%%%%%

A further question that one might be interested in is which of the viable scenarios are
compatible with the assumption of axion dark matter.
In case inflation takes place after the spontaneous breaking of the PQ
symmetry, the only contribution to the relic axion density
stems from the vacuum realignment of the zero-momentum mode of
the axion field during the QCD phase transition~\cite{Preskill:1982cy}.
The present value of the axion density parameter $\Omega_a^0 h^2$ can then
be estimated as~\cite{Sikivie:2006ni}
\begin{align}
\Omega_a^0 h^2 \sim 0.50 \left(\frac{\bar{\theta}_i^2}{\pi^2/3}\right)
\left(\frac{f_a}{10^{12}\,\textrm{GeV}}\right)^{7/6} \,,
\label{eq:Omegash2}
\end{align}
where $\bar{\theta}_i \in (-\pi,\pi]$ denotes the initial misalignment angle of the axion field
before the onset of the QCD phase transition, $\bar{\theta}_i = a\left(t_i\right) /f_a$.
In the derivation of Eq.~\eqref{eq:Omegash2}, it is assumed that $\bar{\theta}_i$
is constant across the entire observable universe as well as
that the axion relic density is not diluted
after its generation by some form of late-time entropy production.
If all possible values of $\bar{\theta}_i$ are equally likely, we expect
that $\left< \bar{\theta}_i^2 \right> = \pi^2/3$.
By comparing the expression for $\Omega_a^0 h^2$ in Eq.~\eqref{eq:Omegash2}
with the density parameter of cold dark matter (CDM),
which has recently been determined very precisely by the PLANCK
satellite, $\Omega_{\textrm{CDM}}^0 h^2 \simeq 0.1199$~\cite{Ade:2013zuv},
we see that, for an axion decay constant $f_a$ of $\nrmcal{O}(10^{12})\,\textrm{GeV}$,
cold axions may completely account for the relic density of dark matter.
For $f_a \gtrsim 10^{12}\,\textrm{GeV}$, the axion density exceeds the measured
abundance of dark matter, which is nothing but the cosmological upper bound on $f_a$
which we introduced in Sec.~\ref{subsec:PQsym}.

%%%%%%%%%%%%%%%%%%%%%%%%%%%%%%%%%%%%%%%%%%%%%%%%%%%%%%%%%%%%%%%%%%%%%%%%%%%%%%%%%%%%%%%%%%%%%%%%%%%%

\begin{table}
\begin{center}
\begin{doublespacing}
\begin{tabular}{c||cccccccc}
$(n,N,k)$ & $(1,4,12)$ & $(1,5,9)$ & $(1,7,7)$ & $(1,8,12)$ & $(1,9,7)$
\\\hline\hline
$10^{10}\,\bar{\theta}^0$ & $1 \times 10^{-4}$ & $3 \times 10^{-1}$
& $2 \times 10^{-4}$ & $1 \times 10^{-4}$
& $2 \times 10^{-4}$\\\hline\hline
$(n,N,k)$ & $(1,10,9)$ & $(1,11,7)$ & $(1,12,7)$ & $(2,4,6)$ & $(2,8,6)$ \\\hline\hline
$10^{10}\,\bar{\theta}^0$ & $3 \times 10^{-1}$ & $2 \times 10^{-4}$ & $2 \times 10^{-4}$
& $1 \times 10^{-4}$ & $1 \times 10^{-4}$
\end{tabular}
\end{doublespacing}
\begin{onehalfspacing}
\caption{All combinations of $N$, $n$ and $k$ which, in the case of axion dark matter,
\textit{i.e.}\ for $f_a = 10^{12}\,\textrm{GeV}$,
result in a lower bound $\bar{\theta}^0$ on the theta angle 
between $10^{-15}$ and $10^{-10}$, cf.\ Eq.~\eqref{eq:bartheta0}.}
\label{tab:theta0} 
\end{onehalfspacing}
\end{center}
\end{table}

%%%%%%%%%%%%%%%%%%%%%%%%%%%%%%%%%%%%%%%%%%%%%%%%%%%%%%%%%%%%%%%%%%%%%%%%%%%%%%%%%%%%%%%%%%%%%%%%%%%%

Setting $f_a$ to $10^{12}\,\textrm{GeV}$, we can now ask how large a QCD vacuum
angle $\bar{\theta}$ we expect to be induced by the PQ-breaking operators in the respective
viable scenarios.
In the case of those scenarios for which we found that
$f_a^{\textrm{min}} < f_a^{\textrm{max},\bar{\theta}} < 10^{12}\,\textrm{GeV}$,
the induced QCD vacuum angle, of course, turns out to be larger than $10^{-10}$,
\textit{i.e.}\ only scenarios in which
$f_a^{\textrm{min}} < 10^{12}\,\textrm{GeV} < f_a^{\textrm{max},\bar{\theta}}$
are compatible with the requirement of axion dark matter.
In total, we find $861$ of such scenarios.
Among these, $763$ belong to the case $n=1$ and $98$ to the case $n=2$.
Analogously to the upper bounds on the axion decay constant, for which we introduced
$f_a^{\textrm{max},0}$, cf.\ Eq.~\eqref{eq:famax0}, we would also
like to know which $R$ charge $r_P$ for given
$N$, $n$ and $k$ yields the smallest QCD vacuum angle,
\begin{align}
f_a = 10^{12}\,\textrm{GeV} \: : \quad \bar{\theta}^0(N,n,k) =
\min_{r_P} \left\{\bar{\theta}(N,n,k,r_P)\right\} \,, \quad
\bar{\theta}(N,n,k,r_P) = \max\{\textrm{all }\Delta\bar{\theta}_0\}
\label{eq:bartheta0}
\end{align}
where the shifts in the QCD vacuum angle $\Delta\bar{\theta}_0$ are to be calculated
according to Eq.~\eqref{eq:Deltatheta}.
The angles $\bar{\theta}^0$ then represent the most conservative
lower bounds on $\bar{\theta}$ for the respective combinations of $N$, $n$ and $k$.
For $96$ combinations of $N$, $n$ and $k$, splitting into $79$ combinations
corresponding to $n=1$ and $17$ combinations corresponding to $n=2$,
the angle $\bar{\theta}^0$ does not exceed $10^{-10}$.
But only for a few of these solutions, $\bar{\theta}^0$ falls into a range that might
be experimentally accessible in the not so far future.
For instance, only for $10$ solutions we find values of $\bar{\theta}^0$ between
$10^{-15}$ and $10^{-10}$, cf.\ Tab.~\ref{tab:theta0}.
Provided that dark matter is really composed out of axions, these $10$ scenarios
can then be tested in experiments aiming at measuring a non-zero value of
the QCD vacuum angle.

%%%%%%%%%%%%%%%%%%%%%%%%%%%%%%%%%%%%%%%%%%%%%%%%%%%%%%%%%%%%%%%%%%%%%%%%%%%%%%%%%%%%%%%%%%%%%%%%%%%%

\section{Conclusions and discussion}
\label{sec:conclusions}

%%%%%%%%%%%%%%%%%%%%%%%%%%%%%%%%%%%%%%%%%%%%%%%%%%%%%%%%%%%%%%%%%%%%%%%%%%%%%%%%%%%%%%%%%%%%%%%%%%%%

The PQ solution of the strong $CP$ problem requires an anomalous global Abelian symmetry,
$U(1)_{\textrm{PQ}}$.
On the other hand, any global symmetry is expected to be explicitly broken by
quantum gravity effects.
In this paper, we have pointed out that imposing a gauged and
discrete $R$ symmetry, $Z_N^R$, one is able to retain a PQ symmetry of high enough quality
as an approximate and accidental symmetry in the low-energy effective theory.
The reasoning behind the construction of our model was the following:
In order to render the $Z_N^R$ symmetry anomaly-free, it is, in general,
necessary to extend the particle content of the MSSM by new matter multiplets.
Except for some special cases, these new particles are \textit{a priori} massless,
which calls for a further extension of the spectrum by an extra singlet sector that is
capable of generating masses for the new particles.
As we were able to show, the new matter and singlet sectors then exhibit several global
Abelian symmetries, a linear combination of which can be
identified as the PQ symmetry.
In addition to that, for all $Z_N^R$ symmetries
apart from $Z_4^R$, we supplemented the MSSM Higgs sector
by an additional chiral singlet $S$, so as to allow
for a dynamical generation of the MSSM $\mu$ term.

%%%%%%%%%%%%%%%%%%%%%%%%%%%%%%%%%%%%%%%%%%%%%%%%%%%%%%%%%%%%%%%%%%%%%%%%%%%%%%%%%%%%%%%%%%%%%%%%%%%%

The presence of the extra matter multiplets and the singlet $S$ in our model
entail a potentially rich phenomenology in collider experiments.
Depending on the nature of the coupling between the extra matter and singlet
sectors, the new particles might either have masses in the TeV or multi-TeV range 
or they might be very heavy, with their masses being close to the scale of PQ symmetry breaking.
In the former case, our model is being directly probed by searches
for heavy vector-like quarks at the LHC.
At the same time, the phenomenology of the Higgs sector of our model is similar to
the one in the PQ-NMSSM or in the nMSSM.
Next to the four ordinary neutralinos, we expect a fifth, very light neutralino,
the singlino $\tilde{S}$, which receives its mass only from mixing with the
neutral Higgsinos.
The singlino may play an important role in the decay of the standard model-like
Higgs boson and contribute to the relic density of dark matter.

%%%%%%%%%%%%%%%%%%%%%%%%%%%%%%%%%%%%%%%%%%%%%%%%%%%%%%%%%%%%%%%%%%%%%%%%%%%%%%%%%%%%%%%%%%%%%%%%%%%%

In order to single out the phenomenologically viable variants of our model,
we imposed four phenomenological constraints.
We required
(i) the masses of the new quarks to exceed the lower experimental
bound on the mass of heavy down-type quarks, $M_Q^{\textrm{min}} = 590 \,\textrm{GeV}$,
(ii) the unification of the standard model gauge couplings to still
occur at the perturbative level, $g_{\textrm{GUT}} \leq \sqrt{4\pi}$,
(iii) the shift in the QCD vacuum angle induced by higher-dimensional PQ-breaking
operators to remain below the upper experimental bound, $\bar{\theta}< 10^{-10}$,
as well as (iv) the axion decay constant to take a value within the experimentally
allowed window, $10^9 \,\textrm{GeV} \lesssim f_a \lesssim 10^{12}\,\textrm{GeV}$.
To our surprise, we found a large landscape of possible scenarios, all compatible
with these four constraints.
In particular, we showed that, for an appropriately chosen number of extra matter
multiplets, the order $N$ of the $Z_N^R$ symmetry can take any integer value larger than $2$.
Besides that, for each viable scenario, we derived an upper bound on the axion decay constant
based on the requirement that QCD vacuum angle must not exceed $10^{-10}$.
In many cases, these upper bounds turned out to be larger than $10^{12}\,\textrm{GeV}$, thereby
rendering the corresponding scenarios compatible with the assumption of axion dark matter.
For these scenarios, we then estimated the expected value of the QCD vacuum angle,
in case dark matter should really be composed out of axions.
A measurement of a non-zero theta angle in combination with a confirmation
of axion dark matter would therefore allow for a highly non-trivial
experimental test of our model.

%%%%%%%%%%%%%%%%%%%%%%%%%%%%%%%%%%%%%%%%%%%%%%%%%%%%%%%%%%%%%%%%%%%%%%%%%%%%%%%%%%%%%%%%%%%%%%%%%%%%

We also emphasized the virtues of the special case of a $Z_4^R$ symmetry.
In the case of a $Z_4^R$ symmetry, the MSSM $\mu$ term can be easily generated in
the course of spontaneous $R$ symmetry breaking, such that there is no need to
introduce an additional chiral singlet.
As a consequence of that, the scalar potential does not exhibit a flat direction
in the supersymmetric limit, so that we do not have to rely on the soft SUSY breaking
masses to stabilize the PQ-breaking vacuum, as is the case for all other $Z_N^R$
symmetries.
Moreover, a $Z_4^R$ is the only discrete $R$ symmetry that allows for MSSM
$R$ charges consistent with the assumption of $SO(10)$ unification, cf.\ Appendix~\ref{app:MSSMRcharges}.

%%%%%%%%%%%%%%%%%%%%%%%%%%%%%%%%%%%%%%%%%%%%%%%%%%%%%%%%%%%%%%%%%%%%%%%%%%%%%%%%%%%%%%%%%%%%%%%%%%%%

Finally, we mention that our study needs be extended into several directions.
First of all, it is necessary to embed our extension of the MSSM into a grander model
that explains the origin of the $Z_N^R$ symmetry and provides some guidance as to
the number of extra matter multiplets and the exact nature of their couplings.
Likewise, it is important to further explore the cosmological implications of our model.
One open question, for instance, is the generation and composition of dark
matter in terms of axions, saxions, neutralinos and/or gravitinos in dependence
of our model parameters.
Besides that, it would be interesting to make contact between our model
and $R$-invariant scenarios of inflation~\cite{Kumekawa:1994gx}.
%
%%%%%%%%%%%%%%%%%%%%%%%%%%%%%%%%%%%%%%%%%%%%%%%%%%%%%%%%%%%%%%%%%%%%%%%%%%%%%%%%%%%%%%%%%%%%%%%%%%%%
%
Altogether, our model promises to give rise to a rich phenomenology that can
be probed at colliders and in astrophysical and cosmological observations.
Future experiments will thus be able to test the intriguing possibility
that the PQ symmetry, required for the PQ solution of the strong $CP$ problem,
is indeed an accidental consequence of a gauged and discrete $R$ symmetry.
   
%%%%%%%%%%%%%%%%%%%%%%%%%%%%%%%%%%%%%%%%%%%%%%%%%%%%%%%%%%%%%%%%%%%%%%%%%%%%%%%%%%%%%%%%%%%%%%%%%%%%

\appendix

%%%%%%%%%%%%%%%%%%%%%%%%%%%%%%%%%%%%%%%%%%%%%%%%%%%%%%%%%%%%%%%%%%%%%%%%%%%%%%%%%%%%%%%%%%%%%%%%%%%%

\section{Possible \boldmath{$R$} charges of the MSSM fields}
\label{app:MSSMRcharges}

%%%%%%%%%%%%%%%%%%%%%%%%%%%%%%%%%%%%%%%%%%%%%%%%%%%%%%%%%%%%%%%%%%%%%%%%%%%%%%%%%%%%%%%%%%%%%%%%%%%%

In Sec.~\ref{subsec:MSSM}, we derive five constraints on
$r_{\mathbf{10}}$, $r_{\mathbf{5}^*}$, $r_{\mathbf{1}}$,
$r_{H_u}$ and $r_{H_d}$, the $R$ charges of the MSSM matter and
Higgs multiplets, cf.\ Eqs.~\eqref{eq:anomaly-free},
\eqref{eq:yukawacond} and \eqref{eq:neutrinocond}.
As these conditions only hold up to the addition of integer multiples
of $N$, they do not suffice to fix the values of the MSSM $R$ charges uniquely.
In this appendix, we now show that, for each value of $N$, there
exist exactly ten different $R$ charge assignments for
the MSSM fields that comply with all constraints.
Moreover, we also discuss under which circumstances these solutions
are equivalent to each other.

%%%%%%%%%%%%%%%%%%%%%%%%%%%%%%%%%%%%%%%%%%%%%%%%%%%%%%%%%%%%%%%%%%%%%%%%%%%%%%%%%%%%%%%%%%%%%%%%%%%%

\subsection*%[$R$ charge assignments consistent with all constraints]
{\boldmath{$R$} charge assignments consistent with all constraints}

%%%%%%%%%%%%%%%%%%%%%%%%%%%%%%%%%%%%%%%%%%%%%%%%%%%%%%%%%%%%%%%%%%%%%%%%%%%%%%%%%%%%%%%%%%%%%%%%%%%%

To begin with, let us rewrite the five conditions in Eqs.~\eqref{eq:anomaly-free},
\eqref{eq:yukawacond} and \eqref{eq:neutrinocond} as follows,
\begin{align}
\label{eq:Xconds}
r_{H_u} + r_{H_d} = & \: 4 + \ell_1 N \,, \quad 
2 r_{\mathbf{10}} + r_{H_u} = 2 + \ell_2 N \,, \quad
r_{\mathbf{5}^*} + r_{\mathbf{10}} + r_{H_d} = 2 + \ell_3 N \,,\\
r_{\mathbf{5}*} + r_{\mathbf{1}} + r_{H_u} = & \: 2 + \ell_4 N \,, \quad
2  r_{\mathbf{1}} = 2 + \ell_5 N \,, \nonumber
\end{align}
where we have made use of the relation in Eq.~\eqref{eq:modulo}
and with $\ell_i \in \mathbb{Z}$ for all $i=1,..,5$.
Solving this system of linear equations for the $R$ charges
$\boldsymbol{r} = \left(r_{\mathbf{10}}, r_{\mathbf{5}^*}, r_{\mathbf{1}}, r_{H_u},
r_{H_d}\right)^T$ yields
\noindent\begin{align}
\begin{pmatrix}
r_{\mathbf{10}} \\ r_{\mathbf{5}^*} \\ r_{\mathbf{1}} \\ r_{H_u} \\ r_{H_d}
\end{pmatrix} \modulo{N}
\begin{pmatrix} \frac{1}{5} \\ -\frac{3}{5} \\ 1 \\
2 - \frac{2}{5} \\ 2 + \frac{2}{5}\end{pmatrix} + 
\tilde{\ell} \,\frac{N}{10}
\begin{pmatrix} 1 \\ -3 \\ 5 \\ -2 \\ 2\end{pmatrix} + 
\begin{pmatrix}
0 & 0 & 0 & 0 & 0\\
-1 & 1 & 1 & 0 & 0\\
1 & -2 & -1 & 1 & 0\\
0 & 1 & 0 & 0 & 0\\
1 & -1 & 0 & 0 & 0
\end{pmatrix}
\begin{pmatrix} \ell_1 \\ \ell_2 \\ \ell_3 \\ \ell_4 \\ \ell_5 \end{pmatrix} N\,,
\label{eq:RMSSMX}
\end{align}
with $\tilde{\ell} = -2 \ell_1 + 4\ell_2 + 2\ell_3 - 2\ell_4 + \ell_5 \in \mathbb{Z}$.
As indicated by the $\modulo{N}$ symbol in Eq.~\eqref{eq:RMSSMX},
all $R$ charges are only defined modulo $N$.
Thus, after picking explicit values for the $\ell_i$, we always have to take all
$R$ charges modulo $N$, such that $0\leq r_i < N$ for all fields $i$.
At the same time, the last summand on the right-hand side of Eq.~\eqref{eq:RMSSMX}
does nothing but shifting the charges $r_{\mathbf{5}^*}$, $r_{\mathbf{1}}$,
$r_{H_u}$ and $r_{H_d}$ by integer multiples of $N$.
Its effect is hence always nullified by the modulo $N$ operation,
allowing us to omit it in the following.
Furthermore, we observe that the entries of the second column vector
on the right-hand side of Eq.~\eqref{eq:RMSSMX} correspond to the
$X$ charges of the MSSM multiplets.%
\footnote{$X$ denotes the charge corresponding to the Abelian symmetry
$U(1)_X$, which is the subgroup of $U(1)_{B-L} \times U(1)_Y$ that commutes
with $SU(5)$.
In terms of $B$$-$$L$ and the weak hypercharge $Y$, it is given as $X = 5(B$$-$$L) - 4Y$.}
We shall therefore denote this vector by $\boldsymbol{X}$, such that
\begin{align}
\boldsymbol{r} \modulo{N}
\boldsymbol{r}_0 + \tilde{\ell} \,\frac{N}{10}\,\boldsymbol{X} \,,\quad
\boldsymbol{r}_0 = \left(\frac{1}{5}, -\frac{3}{5}, 1,
2 - \frac{2}{5}, 2 + \frac{2}{5}\right)^T\,,\quad
\boldsymbol{X} =  \left(1,-3,5,-2,2\right)^T \,.
\label{eq:tensol}
\end{align}
This result illustrates that, for any given $N$, there are indeed ten different possible
$R$ charge assignments $\boldsymbol{r}$.
Independently of the concrete value of $N$, the assignment $\boldsymbol{r}_0$
always represents a solution to the conditions in Eqs.~\eqref{eq:anomaly-free},
\eqref{eq:yukawacond} and \eqref{eq:neutrinocond}.
All other solutions can be constructed from $\boldsymbol{r}_0$
by adding multiples of $\frac{N}{10}\boldsymbol{X}$ to it.
Here, the fact that the $R$ charges $r_i$ are only defined modulo $N$
implies that all $R$ charge assignments corresponding to values of
$\tilde{\ell}$ that differ from each other by
integer multiples of $10$ are equivalent to each other.
The ten possible solutions for the MSSM $R$ charges then follow from
Eq.~\eqref{eq:tensol} by setting $\tilde{\ell}$ to $\tilde{\ell} = 0,1,2,..,9$.

%%%%%%%%%%%%%%%%%%%%%%%%%%%%%%%%%%%%%%%%%%%%%%%%%%%%%%%%%%%%%%%%%%%%%%%%%%%%%%%%%%%%%%%%%%%%%%%%%%%%

Among all viable $R$ charge assignments that can be obtained
from Eq.~\eqref{eq:tensol}, there are several which are particularly interesting.
For instance, for $N = 4$, it is possible to assign $R$ charges to the MSSM
fields in such a way that they are consistent with the assumption of $SO(10)$
unification.
In this case,  the GUT gauge group contains $SO(10)$ as a subgroup,
$G_{\textrm{GUT}} \supset SO(10) \supset SU(5)$, and the MSSM matter and Higgs
fields are unified in $SO(10)$ multiplets, such that
$r_{\mathbf{10}} = r_{\mathbf{5}^*} = r_{\mathbf{1}}$ and $r_{H_u}= r_{H_d}$.
For $N=4$ and $\tilde{\ell} = 2,7$, these two relations can indeed be realized,
\begin{align}
N = 4\,: \quad \tilde{\ell} = 2\,: \boldsymbol{r} = \left(1,1,1,0,0\right) \,, \quad
\tilde{\ell} = 7\,: \boldsymbol{r} = \left(3,3,3,0,0\right) \,.
\label{eq:rSO10}
\end{align}
In the case of $N=4$, the $R$ charge of the superpotential is equivalent to $-2$.
Hence, given any viable $R$ charge assignment, reversing the signs of all $R$ charges
and applying the modulo $N$ operation, so that all $R$ charges lie again in the interval
$[0,N)$, provides one with another viable $R$ charge assignment. 
The two solutions for $\boldsymbol{r}$ in Eq.~\eqref{eq:rSO10} are related to
each other in just this way, implying that they are in fact equivalent.
In Refs.~\cite{Lee:2011dya,Lee:2010gv},
the discrete $Z_4^R$ symmetry with $R$ charges
$\boldsymbol{r} = \left(1,1,1,0,0\right)$ has been discussed in more detail.
Allowing  for anomaly cancellation via the Green-Schwarz
mechanism, this symmetry has in particular been identified as
the \textit{unique} discrete $R$ symmetry of
the MSSM that may be rendered  anomaly-free without introducing any new particles
and which, at the same time, commutes with $SO(10)$ and forbids
the $\mu$ term in the superpotential.
Finally, we point out that the two $R$ charge assignments in Eq.~\eqref{eq:rSO10}
only feature integer-valued $R$ charges.
We mention in passing that, in fact, for each value of $N$ that is not an integer multiple of $5$
there is at least one viable $R$ charge assignment that only involves integer-valued $R$ charges.
This is a direct consequence of our result for $\boldsymbol{r}$ in Eq.~\eqref{eq:tensol}
and the fact that all $R$ charges in $\boldsymbol{r}_0$ are integer multiples of $\frac{1}{5}$.

%%%%%%%%%%%%%%%%%%%%%%%%%%%%%%%%%%%%%%%%%%%%%%%%%%%%%%%%%%%%%%%%%%%%%%%%%%%%%%%%%%%%%%%%%%%%%%%%%%%%

\subsection*%[Relationship between the different $R$ charge assignments]
{Relationship between the different \boldmath{$R$} charge assignments}

%%%%%%%%%%%%%%%%%%%%%%%%%%%%%%%%%%%%%%%%%%%%%%%%%%%%%%%%%%%%%%%%%%%%%%%%%%%%%%%%%%%%%%%%%%%%%%%%%%%%

The form of our result for $\boldsymbol{r}$ in Eq.~\eqref{eq:tensol}
reflects the symmetries of the MSSM superpotential that commute with $SU(5)$.
Among these symmetries, there is in particular a $Z_{10}$ subgroup of $U(1)_X$.
To see this, notice that the MSSM superpotential without the Majorana mass term
for the neutrino singlets $\mathbf{1}_i$ is invariant under $U(1)_X$ transformations.
The Majorana mass term, however, carries $X$ charge $10$ and thus breaks the $U(1)_X$
symmetry to its $Z_{10}$ subgroup.
Our solutions for the MSSM $R$ charges are therefore related to each other by
$Z_{10}$ transformations, which also explains why we have found exactly ten different solutions
for each value of $N$.
This result is independent of the question of whether or not we assume the $U(1)_X$ symmetry
to be part of the gauge group above some high energy scale.
We will address this question shortly, but before we do that, we remark that the
$Z_{10}$ subgroup of $U(1)_X$ is not the only symmetry of the MSSM superpotential that
commutes with $SU(5)$.
By definition, the center of $SU(5)$, a discrete $Z_5$ symmetry, also commutes with
all $SU(5)$ elements.
Under this $Z_5$ symmetry, the MSSM multiplets $\mathbf{10}_i$, $\mathbf{5}_i^*$,
$H_u$ and $H_d$ carry charges $1$, $2$, $3$ and $2$, while all SM singlets
have zero charge.
At the same time, all SM singlets of our model transform
trivially under the $Z_5$ subgroup of the $Z_{10}$ contained in $U(1)_X$.
The $Z_5$ center of $SU(5)$ is hence equivalent to this $U(1)_X$ subgroup,
\begin{align}
SU(5) \supset Z_5 \cong Z_5 \subset Z_{10} \subset U(1)_X.
\label{eq:Z5}
\end{align}
Therefore, independently of whether $U(1)_X$ is gauged or not, the $Z_5$
subgroup of $Z_{10}$ always has to be treated as a gauge symmetry, as it
is also contained in $SU(5)$.
Under the action of this gauged $Z_5$ symmetry, the $R$ charge assignments
in Eq.~\eqref{eq:tensol} split into two equivalence classes
of respectively five solutions.
The $R$ charge assignments corresponding to $\tilde{\ell} = 2,4,6,8$
can all be generated by acting with $Z_5$ transformations on the
$R$ charge assignment corresponding to $\tilde{\ell} = 0$.
Similarly, the $R$ charge assignments corresponding to $\tilde{\ell} = 1,3,7,9$
can all be generated by acting with $Z_5$ transformations on the
$R$ charge assignment corresponding to $\tilde{\ell} = 5$.
All viable $R$ charge assignments are hence physically equivalent to one
of the following two solutions, cf.\ Eq.~\eqref{eq:MSSMRcharges},
\begin{align}
r_{\mathbf{10}} \modulo{N} \frac{1}{5} + \ell \,\frac{N}{2} \,,\quad
r_{\mathbf{5}^*} \modulo{N} -\frac{3}{5} + \ell \,\frac{N}{2}\,, \quad
r_{\mathbf{1}} \modulo{N} 1 + \ell \,\frac{N}{2} \,, \quad
r_{H_u} \modulo{N} 2 - \frac{2}{5} \,, \quad r_{H_d} \modulo{N} 2 + \frac{2}{5} \,,
\label{eq:requiv}
\end{align}
where $\ell=0,1$.
These two remaining $R$ charge assignments are related to each other by
transformations under the quotient group $Z_{10}/Z_5$, which is nothing
but a simple $Z_2$ parity.

%%%%%%%%%%%%%%%%%%%%%%%%%%%%%%%%%%%%%%%%%%%%%%%%%%%%%%%%%%%%%%%%%%%%%%%%%%%%%%%%%%%%%%%%%%%%%%%%%%%%

Whether the two solutions in Eq.~\eqref{eq:requiv} are also equivalent to each other
depends on the nature of this $Z_2$ parity.
If $U(1)_X$ is part of the gauge group at high energies, its $Z_{10}$ subgroup
is a gauge symmetry at low energies.
Dividing the center of $SU(5)$ out of this $Z_{10}$, we are then left with a
gauged $Z_2$ parity, which can be identified as matter parity, $P_M = Z_{10}/Z_5$.
The transformations relating the two solutions in Eq.~\eqref{eq:requiv} to each
other are then gauge transformations and both solutions end up being equivalent.
On the other hand, if $U(1)_X$ is not gauged and matter parity
is contained in the $Z_N^R$ symmetry, $P_M \subset Z_N^R$, the $Z_{10}$ subgroup
of $U(1)_X$ is also only a global symmetry.
The $Z_2$ parity transformations relating the two solutions in Eq.~\eqref{eq:requiv} to each
other are then global transformations, rendering these two $R$ charge assignments 
physically inequivalent.
In conclusion, we hence arrive at the following picture,
\begin{align}
P_M = Z_{10} / Z_5 \subset U(1)_X \,: & \quad \textrm{all 10 solutions equivalent} \,, \\
P_M \subset Z_N^R \,: & \quad \textrm{2 equivalence classes containing
respectively 5 solutions} \,. \nonumber
\end{align}

%%%%%%%%%%%%%%%%%%%%%%%%%%%%%%%%%%%%%%%%%%%%%%%%%%%%%%%%%%%%%%%%%%%%%%%%%%%%%%%%%%%%%%%%%%%%%%%%%%%%

\subsection*%[$R$ charges in a $U(1)_X$-invariant extension of the MSSM]
{\boldmath{$R$} charges in a \boldmath{$U(1)_X$}-invariant extension of the MSSM}

%%%%%%%%%%%%%%%%%%%%%%%%%%%%%%%%%%%%%%%%%%%%%%%%%%%%%%%%%%%%%%%%%%%%%%%%%%%%%%%%%%%%%%%%%%%%%%%%%%%%

If matter parity is a subgroup of the $U(1)_X$, our model as presented
in Sec.~\ref{sec:model} is not yet complete, as it still lacks an
explanation for the spontaneous breaking of $U(1)_X$ at some high energy scale.
In the last subsection of this appendix, we shall thus illustrate by means
of a minimal example how our model could be embedded into a $U(1)_X$-invariant
extension of the MSSM.

%%%%%%%%%%%%%%%%%%%%%%%%%%%%%%%%%%%%%%%%%%%%%%%%%%%%%%%%%%%%%%%%%%%%%%%%%%%%%%%%%%%%%%%%%%%%%%%%%%%%

The seesaw extension of the MSSM is not invariant under $U(1)_X$ transformations
because of the lepton number-violating Majorana mass term in the superpotential
$W_{\textrm{MSSM}}$, cf.\ Eq.~\eqref{eq:WMSSM}.
Imagine, however, that this Majorana mass term  derives from the
Yukawa interaction of the neutrino singlets $\mathbf{1}_i$ with
some chiral singlet $\Phi$
carrying $B$$-$$L$ charge $-2$ that acquires a non-vanishing VEV
$\Lambda_{B-L}/\sqrt{2}$ at the GUT scale,
\begin{align}
W_{\textrm{MSSM}} \supset \frac{1}{2} M_i \mathbf{1}_i \mathbf{1}_i \quad \rightarrow \quad
\frac{1}{\sqrt{2}} h_i^n \, \Phi \mathbf{1}_i \mathbf{1}_i +
\lambda\,  T \left(\frac{\Lambda_{B-L}^2}{2} - \Phi\bar{\Phi}\right) \,.
\label{eq:WBL}
\end{align}
Here, $T$ and $\bar{\Phi}$ are two further SM singlets with $B$$-$$L$ charges
$0$ and $2$, respectively.
The field $T$ carries $R$ charge $r_T = 2$, while $\Phi$ and $\bar{\Phi}$ have opposite
$R$ charges, $r_\Phi = - r_{\bar{\Phi}}$.
The diagonal matrix $h^n$ denotes a fourth Yukawa matrix and $\lambda$ is
a dimensionless coupling constant.
This replacement of the Majorana mass term evidently renders the superpotential
$U(1)_X$-invariant, which allows us to enlarge the gauge group of our model by a
$U(1)_X$ factor.
Above the GUT scale, the gauge group hence contains the following subgroup,
\begin{align}
G_{\textrm{GUT}} \supset \left[SU(5) \times U(1)_X \times Z_N^R\right] / Z_5 \,,
\end{align}
where the $Z_5$ symmetry dividing $SU(5)\times U(1)_X$
corresponds to the center of $SU(5)$ and, at the same time,
to a subgroup of $U(1)_X$, cf.\ Eq.~\eqref{eq:Z5}.
To prevent it from appearing twice in the gauge group, it has to be divided out once.
At energies around $\Lambda_{B-L}$, the Higgs fields $\Phi$ and $\bar{\Phi}$
acquire non-vanishing VEVs, whereby they spontaneously break $U(1)_X / Z_5$
to matter parity $P_M$,
\begin{align}
\left<\Phi\right>, \langle\bar{\Phi}\rangle \rightarrow\frac{\Lambda_{B-L}}{2}
\,, \quad U(1)_X\rightarrow Z_{10} \,, \quad P_M = Z_{10} / Z_5 \,.
\end{align}

%%%%%%%%%%%%%%%%%%%%%%%%%%%%%%%%%%%%%%%%%%%%%%%%%%%%%%%%%%%%%%%%%%%%%%%%%%%%%%%%%%%%%%%%%%%%%%%%%%%%

Having introduced the fields $\Phi$, $\bar{\Phi}$ and $T$ and
modified the superpotential as in Eq.~\eqref{eq:WBL}, we have successfully
embedded our model into a $U(1)_X$-invariant extension of the MSSM.
Let us now discuss the set of possible $R$ charge assignments in this extended model.
Next to the five $R$ charges of the MSSM fields, $r_\Phi$, 
the $R$ charge of the Higgs field $\Phi$, now represents a
further, sixth independent $R$ charge.
All six $R$ charges are again subject to five constraints,
which are almost identical to those in Eq.~\eqref{eq:Xconds}.
The only difference now is that the condition deriving from
the neutrino Majorana mass term has to modified, so as to
account for the presence of the field $\Phi$,
\begin{align}
2  r_{\mathbf{1}} = 2 + \ell_5 N \quad\rightarrow\quad
2  r_{\mathbf{1}} +  r_{\Phi} = 2 + \ell_5 N \,.
\end{align}
This replacement entails a shift of all viable $R$ charge assignments,
cf.\ Eq.~\eqref{eq:tensol}, proportional to $r_\Phi$, which itself
remains undetermined, in the direction of the vector $\boldsymbol{X}$,
\begin{align}
\boldsymbol{r} \modulo{N}
\boldsymbol{r}_0 + \tilde{\ell} \,\frac{N}{10}\,\boldsymbol{X}
\quad\rightarrow\quad
\boldsymbol{r} \modulo{N}
\boldsymbol{r}_0 + \frac{1}{10}\left(-r_\Phi + \tilde{\ell}\, N\right)\boldsymbol{X} \,.
\label{eq:tensolX}
\end{align}
For $r_\Phi = 0$, we hence recover exactly the same solutions as in Eq.~\eqref{eq:tensol}.
On the other hand, for $r_\Phi \neq 0$,
all solutions are shifted by $- \frac{r_\Phi}{10}\boldsymbol{X}$.
The \textit{universal} solution $\boldsymbol{r}_0$, which always satisfies the conditions in
Eq.~\eqref{eq:Xconds}, irrespectively of the value of $N$, turns in particular into
$\boldsymbol{r}_0 - \frac{r_\Phi}{10}\boldsymbol{X}$.
Given the fact that the field $\Phi$ carries $X$ charge $-10$, these
shifts are readily identified as $U(1)_X$ gauge transformations
acting on the MSSM $R$ charges as well as on the $R$ charge $r_\Phi$.
The form of our result in Eq.~\eqref{eq:tensolX} is hence a direct consequence of
the $U(1)_X$ invariance of our extended model.
As expected, any $R$ charge assignment is only uniquely
defined up to arbitrary $U(1)_X$ gauge transformations.
Before closing this section, we remark that we are able to use
this observation to render all $R$ charges of the universal solution
integer-valued.
Performing a $U(1)_X$ transformation such that $r_\Phi = -8$, we obtain for the
MSSM $R$ charges
\begin{align}
r_\Phi = - 8 \: : \quad \boldsymbol{r}_0 - \frac{r_\Phi}{10} \boldsymbol{X} =
\left(1,-3,5,0,4\right)^T \,.
\end{align}

%%%%%%%%%%%%%%%%%%%%%%%%%%%%%%%%%%%%%%%%%%%%%%%%%%%%%%%%%%%%%%%%%%%%%%%%%%%%%%%%%%%%%%%%%%%%%%%%%%%%

\section{Solution to the axion domain wall problem}
\label{app:domainwallprob}

%%%%%%%%%%%%%%%%%%%%%%%%%%%%%%%%%%%%%%%%%%%%%%%%%%%%%%%%%%%%%%%%%%%%%%%%%%%%%%%%%%%%%%%%%%%%%%%%%%%%

If the PQ-breaking sector only exhibits a single vacuum, \textit{i.e.}\ if $N_{\textrm{DW}} = 1$,
the axion domain wall problem~\cite{Sikivie:1982qv} does not exist from the outset, whatever
the thermal history of the universe is~\cite{Vilenkin:1982ks}.
In this appendix, we now illustrate how our model may be easily modified in such a way that
it ends up having a unique PQ-breaking vacuum.

%%%%%%%%%%%%%%%%%%%%%%%%%%%%%%%%%%%%%%%%%%%%%%%%%%%%%%%%%%%%%%%%%%%%%%%%%%%%%%%%%%%%%%%%%%%%%%%%%%%%

The simplest way to have a unique vacuum is to couple only one pair of
additional quarks to the singlet field $P$, as is done in the original KSVZ axion model.
For $n=1$, we may for instance impose the following superpotential, cf.\ Eq.~\eqref{eq:WQ},
\begin{align}
W_Q = \lambda_1 P \left(Q \bar{Q}\right)_1 \,, \quad \lambda_1 \sim \nrmcal{O}(1)\,.
\label{eq:WQalt}
\end{align}
The other $k-1$ quark pairs are then supposed to obtain masses in consequence of the
spontaneous breaking of $R$ symmetry~\cite{Inoue:1991rk}.
Given the coupling in Eq.~\eqref{eq:WQalt} and requiring vanishing $R$ charges for
all quark pairs that do not couple to the singlet field $P$,
the $R$ and PQ charges of $P$, $\bar{P}$ as well as of the
new quarks and anti-quarks can be fixed as listed in Table~\ref{tab:charge-nodomailwall}.

%%%%%%%%%%%%%%%%%%%%%%%%%%%%%%%%%%%%%%%%%%%%%%%%%%%%%%%%%%%%%%%%%%%%%%%%%%%%%%%%%%%%%%%%%%%%%%%%%%%%

Let us now discuss whether the PQ symmetry can be a \textit{good} accidental symmetry.
First of all, in the continuous $R$ symmetry limit, the global $U(1)_P$, $U(1)_Q^V$
and $U(1)_Q^A$ symmetries are \textit{almost} exact accidental symmetries of the extra singlet and
extra quark sectors, respectively.
The $U(1)_P$ symmetry is, however, always explicitly broken by the operator
$P m_{3/2}^{k+1}$ in the superpotential.
Likewise, given the charge assignments in Tab.~\ref{tab:charge-nodomailwall},
we see that the operator $\bar{P} S^{k+1}$ is always allowed in the
superpotential, even in the continuous $R$ symmetry limit.
Therefore, the number of extra quark pairs $k$ should be large enough
in order to ensure that the PQ symmetry is not broken too severely.
After performing an analysis similar to the one Sec.~\ref{sec:constraints},
we find that
\begin{align}
k \gtrsim 3.3 + 0.12 \ln \left(\frac{\langle S \rangle}{1\,\textrm{TeV}}\right) + 0.028
\ln \left(\frac{m_{3/2}}{1\,\textrm{TeV}}\frac{\Lambda}{10^{12}\,\textrm{GeV}}\right)
\end{align}
is required in order to keep the QCD vacuum angle below $10^{-10}$.
Consequently, at least $k=4$ extra quark pairs are needed.
With this setup, the PQ
symmetry becomes a good accidental symmetry for sufficiently large $N$,
as is the case in the model discussed in the main body of this paper.

%%%%%%%%%%%%%%%%%%%%%%%%%%%%%%%%%%%%%%%%%%%%%%%%%%%%%%%%%%%%%%%%%%%%%%%%%%%%%%%%%%%%%%%%%%%%%%%%%%%%

\begin{table}
\begin{center}
\begin{doublespacing}
\begin{tabular}{c||cccccc}
        & $Q_1$       & $\bar{Q}_1$    & $Q_i~(i>1)$    & $\bar{Q}_i~(i>1)$ & $P$
        & $\bar{P}$ \\\hline\hline
$Z_N^R$ & $r_{Q_1}$   & $6+2k-r_{Q_1}$ & $r_{Q_i}$      & $-r_{Q_i}$ &
	     $-4-2k$ & $4+2k$ \\
$U(1)_{\rm PQ}$ & $q_Q$ & $-1-q_Q$  & $0$ & $0$ & $1$ & $-1$ 
\end{tabular}
\end{doublespacing}
\begin{onehalfspacing}
\caption{$R$ and PQ charge assignments in a model with a single PQ-breaking vacuum.
All $R$ charges are only defined up to the addition of integer multiples of $N$.
As far as the PQ mechanism is concerned, the $R$ charges of the extra quark fields $Q_i$
can be chosen arbitrarily.
They may, however, be further constrained by requiring appropriate couplings between
the new quarks and anti-quarks and the fields of the MSSM, cf.\ Sec.~\ref{subsubsec:Qdecay}.}
\label{tab:charge-nodomailwall}
\end{onehalfspacing}
\end{center}
\end{table}

%%%%%%%%%%%%%%%%%%%%%%%%%%%%%%%%%%%%%%%%%%%%%%%%%%%%%%%%%%%%%%%%%%%%%%%%%%%%%%%%%%%%%%%%%%%%%%%%%%%%

\section*{Acknowledgments}
This work is supported by Grant-in-Aid for Scientific Research from the
Ministry of Education, Science, Sports, and Culture (MEXT), Japan,
No.\ 22244021 (T.T.Y.), No.\ 24740151 (M.I.), and by the World Premier
International Research Center Initiative (WPI Initiative), MEXT, Japan.
The work of K.H.\ is supported in part by a JSPS Research Fellowship
for Young Scientists. K.S.\ would like to thank Patrick Vaudrevange and
Taizan Watari for helpful discussions pertaining
Appendix~\ref{app:MSSMRcharges}.

%%%%%%%%%%%%%%%%%%%%%%%%%%%%%%%%%%%%%%%%%%%%%%%%%%%%%%%%%%%%%%%%%%%%%%%%%%%%%%%%%%%%%%%%%%%%%%%%%%%%

%%%%%%%%%%%%%%%%%%%%%%%%%%%%%%%%%%%%%%%%%%%%%%%%%%%%%%%%%%%%%%%%%%%%%%%%%%%%%%%%%%%%%%%%%%%%%%%%%%%%

\end{document}